\documentclass[12pt]{JHEP3}

\usepackage{amssymb}
\usepackage{amsmath}
\usepackage{graphicx}
\usepackage{latexsym}

\def\mylimit#1{\mathrel{\mathop{\kern0pt\longrightarrow}\limits_{#1}}}
\def\myequiv#1{\mathrel{\mathop{\kern0pt\longleftrightarrow}\limits_{#1}}}

\newcommand{\bequ}{\begin{equation}}
\newcommand{\eequ}{\end{equation}}
\newcommand{\beqn}{\begin{eqnarray}}
\newcommand{\eeqn}{\end{eqnarray}}
\newcommand{\bctr}{\begin{center}}
\newcommand{\ectr}{\end{center}}
\newcommand{\bit}{\begin{itemize}}
\newcommand{\eit}{\end{itemize}}

\newcommand{\e}[1]{{\rm e}^{#1}}

\def\e6{{\rm E}_6}
\def\sgn{\mathop{\rm sgn}}

\newcommand{\psla}{p\kern-.45em/}
\newcommand{\bfpsla}{{\bf p}\kern-.45em/}
\newcommand{\esla}{E\kern-.50em/}
\newcommand{\ebsla}{\bar{\epsilon}\kern-.45em/}
\newcommand{\qsla}{q\kern-.45em/}
\newcommand{\pbsla}{\bar{p}\kern-.45em/}
\newcommand{\qbsla}{\bar{q}\kern-.45em/}

\def\ti    {\widetilde}
\def\sf    {{\ti f}}

\def\ch    {\ti {\chi}}

\def\cpl    {\ti {\chi}^+}
\def\cm    {\ti {\chi}^-}

\def\none  {\ti \chi^0_1}
\def\ntwo  {\ti \chi^0_2}
\def\nthre {\ti \chi^0_3}
\def\nfour {\ti \chi^0_4}

\def\gluino{\ti g}

\def\gsim  {\hspace{0.3em}\raisebox{0.4ex}{$>$}\hspace{-0.75em}\raisebox{-.7ex}{$\sim$}\hspace{0.3em}}
\def\lsim  {\hspace{0.3em}\raisebox{0.4ex}{$<$}\hspace{-0.75em}\raisebox{-.7ex}{$\sim$}\hspace{0.3em}}

\def\bra {\langle}
\def\ket {\rangle}

\def\over {\overline}

\def\dg {\dagger}

\def\5b{\bf {\overline{5}}}
\def\1{\bf 1}

\title{LHC signature of supersymmetric models
with non-universal sfermion masses}

\author{Sung-Gi Kim\\
Department of Physics, Tohoku University,
Sendai 980-8578, Japan}

\author{Nobuhiro Maekawa, Keiko I. Nagao\\
Department of Physics, Nagoya University, Nagoya 464-8602, Japan}

\author{Mihoko M. Nojiri\\
KEK Theory Center, IPNS, KEK,  
The Graduate University for Advanced Studies (Sokendai), 
1-1 Oho, Tsukuba, 305-0801, Japan \\
IPMU, Tokyo University, Kashiwa, Chiba, 277-8568, Japan}
\author{Kazuki Sakurai\\
KEK Theory Center, IPNS, KEK,  1-1 Oho, Tsukuba, 305-0801, Japan}

\preprint{KEK-TH-1324}

\abstract{
We study the LHC signature of the minimal supersymmetric standard model
with non-universal sfermion masses.
In the model, soft masses of gauginos and the 3rd generation of $\bf 10$ of SU(5) are around the weak scale,
while other sfermion soft mass is universal and around a few TeV.
Such sfermion mass spectrum is motivated not only from
flavor, CP and naturalness constraints but also from $E_6$ grand unified model with non-Abelian
horizontal (flavor) symmetry.
The characteristic signature of the model at the LHC is the dominance of the
events with 4 $b$ partons in the final state together with high
rate of mildly boosted top quark arising from gluino decay.
The prominent high $p_T$ jet also arises from squark decay.
We show it is possible to find the characteristic signature in the early stage of the LHC.
The discrimination of our scenario from some CMSSM model points with similar signature may be possible
with large integrated luminosity.  
The result of sparticle mass measurement using
exclusive channel  with the help of hemisphere analysis, and
inclusive measurement of gluino and squark masses using $M_{T2}$ and $M_{T2}^{\rm min}$  
in some representative model points are presented.
}


\begin{document}




\section{Introduction}

Supersymmetry (SUSY) is one of the most promising candidates 
for physics beyond the Standard Model (SM).
The minimal supersymmetric standard model (MSSM) has some attractive features, for instance
 improvement of the gauge coupling unification, the radiative electroweak symmetry breaking, 
and providing a dark matter candidate as the lightest superparticle (LSP)
\cite{Nilles:1983ge, Haber:1984rc, Martin:1997ns}.
Moreover, one of the most attractive features of the MSSM is 
the stabilization of the weak scale in the case that the
SUSY breaking parameters and the higgsino mass parameter ($\mu$) are around the 
weak scale. 
The signatures of the supersymmetry may be discovered
 and some properties of the MSSM will be revealed at ATLAS and CMS experiments
 at the CERN Large Hadron Collider (LHC). 
The LHC signatures of various SUSY models, such as 
minimal supergravity (mSUGRA) model 
\cite{Hinchliffe:1996iu,Hinchliffe:1999zc,Bachacou:1999zb,Allanach:2000kt},
gauge mediated SUSY breaking (GMSB) model \cite{Feng:1997zr,Hinchliffe:1998ys,Kawagoe:2003jv}, 
anomaly mediated SUSY breaking (AMSB) model 
\cite{Paige:1999ui,Barr:2002ex,Datta:2002vy,Asai:2007sw,Asai:2008sk},
mixed modulas anomaly mediation (MMAM) model 
\cite{Baer:2006tb,Cho:2007fg},
 have been studied.

The sparticle mass measurement is an important physics target at the LHC.
Various methods have been developed for sparticle mass determination from event kinematics
[4\,-\,7,\,18\,-\,35].
Endpoint methods of the various leptonic exclusive channels is known to be very successful
\cite{Hinchliffe:1996iu,Hinchliffe:1999zc,Bachacou:1999zb,Allanach:2000kt,Abdullin:1998pm,Atlas}.
By combining the measured endpoints of the invariant mass distributions
of the jets and leptons from relatively clean and long cascade decay channels involving 
neutralinos ($\ti \chi^0_i$) and sleptons ($\ti l$),
one can determine not only the masses of the squark and gluino, 
but also the masses of neutralinos and sleptons arising from their cascade decays.
Recently, progress has been made in the use of $M_{T2}$ distributions
for the sparticle mass measurement
\cite{Lester:1999tx,Barr:2003rg,Cho:2007qv,Barr:2007hy,
Cho:2007dh,Nojiri:2008hy,Nojiri:2008vq,Burns:2008va,Cho:2008tj}.
It has been pointed out that the endpoint of the $M_{T2}$ distributions as a function 
of the test LSP mass ($\chi$) may exhibit a kink, 
which indicates the initially produced sparticle mass and the true LSP mass simultaneously
\cite{Cho:2007qv,Barr:2007hy,Cho:2007dh}. 
Moreover if we define the $M_{T2}$ and the sub-system $M_{T2}$ 
\cite{Nojiri:2008vq,Burns:2008va} inclusively,
we can roughly measure the squark and gluino masses even in the very early stage at the LHC
\cite{Nojiri:2008hy,Nojiri:2008vq,Alwall:2009zu}.

The purpose of this paper is investigating 
the LHC signature of SUSY models with non-universal sfermion masses.
Most of the SUSY models that have been studied so far have
universality of sfermion soft masses in the flavor space 
except for some literature \cite{Bityukov:2001yf,Bhattacharya:2008dk,Barbieri:2009ev}.
Introduction of the sfermion non-universality may induce unacceptably large 
flavor changing neutral currents (FCNCs) and electric dipole moments (EDMs)
\cite{Ellis:1981ts, Barbieri:1981gn, Hagelin:1992tc, Gabbiani:1996hi}.
However, the sfermion non-universality may be partially introduced for the 3rd generation sfermions
because constraints from the 3rd generation FCNCs are not so severe
\cite{Dimopoulos:1995mi,Pomarol:1995xc,Cohen:1996vb}
(See also Refs. \cite{Kim:2006ab,Kim:2008yta,Ishiduki:2009gr}).   
If soft masses of gluino and right and left-handed stops are around  the weak scale 
while the other sfermion masses are around a few TeV,
some of FCNC and EDM constraints are relaxed 
with keeping 
weak scale stabilization.
In grand unified theories (GUTs) such as SU(5) GUT, 
both the left and right-handed stops are involved in 
the 3rd generation of ${\bf 10}$-plet of SU(5).
In this paper, 
we consider a minimal non-univeral  models in which only the 3rd generation sfermions involved in ${\bf 10}$
have a different soft mass ($m_{30}$) from the other universal soft sfermion mass ($m_0$) at the cutoff scale. 
The non-universal sfermion mass scenario is motivated both on 
phenomenological and theoretical grounds.
There are models that predict the non-universal structure adopted in this paper
along with realistic fermion masses and mixing matrices \cite{E6Hori, E6Hori2}.

This paper is organized as follows.
In Section 2, we introduce a non-universal sfermion mass scenario and explain motivations.
We identify the motivated region of parameter space from some phenomenological constraints.
In Section 3, we investigate the LHC signature at some representative model points.
 The charactaristic signature of the model is the 4 $b$ partons in the final state and mildly boosted 
top quark with high rate for gluino gluino production. 
The prominent high $p_T$ jet also arises for squark gluino co-production.  
We demonstrate that it is possible to find the characteristic signature of the model
through particle level Monte Carlo Simulation with detector smearing.
In Section 4, we search constrained MSSM (CMSSM) parameter space \cite{Ellis:2001msa},
where all scalar fields have common SUSY breaking mass at the cutoff scale,
 and find a model point
whose signature is similar to that of our scenario. 
We propose a key measurement to  discriminate our scenario from the CMSSM model point.
In Section 5, we discuss sparticle mass measurement in our scenario.
We study leptonic exclusive analysis, the top reconstruction from gluino decay,  
and measurement of the $tb$ endpoint. 
We also study  inclusive $M_{T2}$ and  $M_{T2}^{\rm min }$ distributions 
\cite{Nojiri:2008hy,Nojiri:2008vq,Alwall:2009zu}
 and demonstrate that gluino and the first two generation squark masses can be measured
in a stage of the LHC such as $\int {\cal L} dt = 5 - 20 \,{\rm fb^{-1}}$.
Section 6 is devoted to the conclusion. 

\section{Modified universal sfermion mass scenario}
\label{sec_nusm}

In supersymmetric models
there is no quadratic divergence in the Higgs sector.
Therefore the naturalness problem in the SM is significantly relaxed.
However the supersymmetry is softly broken and the scale of the quantum
correction to the Higgs mass is {\rm of the order of } the soft SUSY breaking parameters.
If the SUSY breaking scale is much lager than the weak scale,
unnatural tuning among the soft masses and higgsino mass, $\mu$, is required.

Let us look this issue more closely.
In the MSSM, the condition for the electroweak symmetry breaking (EWSB) is given as
\begin{equation}
\frac{m_Z^2}{2} = -|\mu|^2 - m^2_{H_u}(\Lambda) - \Delta m^2_{H_u} 
+ {\cal O} \Big( \frac{m^2_{H_{u,d}}}{\tan^2\beta} \Big),
\label{tuning_rough}
\end{equation}
where $m^2_{H_u}(\Lambda)$ is a soft mass of the up-type Higgs boson  at the cutoff scale $\Lambda$, and $\Delta m^2_{H_u}$ is a quantum correction to the $m^2_{H_u}$ at the weak scale, 
which is roughly given as 
\begin{equation}
\Delta m^2_{H_u} \sim -\frac{6|Y_t|^2}{(4\pi)^2} m^2_{\ti t} \ln \Big( \frac{\Lambda^2}{m_{\ti t}^2} \Big),
\label{del_mhu}
\end{equation}
where $Y_t$ is the top Yukawa coupling and $m_{\ti t}$ is the averaged stop mass.
The $\Delta m^2_{H_u}$ is large if $m_{\ti t}$ is much larger than the weak scale.
In that case, a relatively large cancellation is required among the terms
in the right hand side of Eq.\,(\ref{tuning_rough}).

Unlike stop masses,
the other squark and slepton masses do not affect the Higgs potetial
because their Yukawa couplings are small 
unless the bottom Yukawa coupling is as large as the top Yukawa coupling.
As long as both left and right-handed stop masses are around the weak scale, 
we can take their masses much larger than the weak scale.
In $SU(5)$ GUT,
two stops are unified into a single ${\bf 10}$-plet field at the GUT scale.
We consider a model in which a soft mass of the 3rd generation of ${\bf 10}$ (${\bf 10}_3$) is independent of 
 the other  universal sfermion soft masses  at the GUT scale. 
We parameterize sfermion soft mass matrices at the cutoff scale as follows:
\begin{equation}
m^2_{\bf 10} =
\begin{pmatrix}
m_0^2 & & \\
& m_0^2 & \\
& & m_{30}^2
\end{pmatrix},~~~~~
m^2_{\overline {\bf 5}} =
\begin{pmatrix}
m_0^2 & & \\
& m_0^2 & \\
& & m_{0}^2
\end{pmatrix}.
\label{nusm_matrix}
\end{equation} 
Here, ${\bf 10}=(Q,U^c,E^c)$, ${\bf {\overline 5}}=(D^c,L)$.
In this paper, 
we take the cutoff $\Lambda$ to be the unification scale of the three gauge couplings
($\Lambda \simeq 2 \times 10^{16}$\,GeV).

In the super-CKM basis, sfermion mass matrices for ${\bf 10}$ sector
are given as $m^2_{\ti f}=V^\dagger_{f} m^2_{\bf 10} V_f$ $(f=u_L,d_L,u_R,e_R)$
at the cutoffs scale.
Here $V_f$ are unitary matrices that diagonalize the Yukawa matrices as
$V_{u_L}^T Y_u V_{u_R}^* = Y_u^{\rm diag}$.
The mixing induced by $V_f$ should be sufficiently small to avoid large FCNCs.
Thus in addition to Eq.\,(\ref{nusm_matrix}),
we assume \cite{Kim:2006ab,Kim:2008yta,Ishiduki:2009gr},
\begin{eqnarray}
(V_{u_L})_{ij},~ (V_{d_L})_{ij},~ (V_{u_R})_{ij},~ (V_{e_R})_{ij} ~\lsim~ (V_{\rm CKM})_{ij}, ~
~~~(i \neq j).
\label{assump}
\end{eqnarray}

In $E_6$ SUSY GUT models with $SU(2) \times U(1)$ horizontal symmetry $H$, 
Eqs.\,(\ref{nusm_matrix}) and (\ref{assump}) are derived 
along with realistic fermion masses and mixing matrices
\cite{E6Hori,E6Hori2}.
In the models,  all the  ${\bf \over 5}$-plets of three families  
and the first two generations of ${\bf 10}$-plets (${\bf 10}_1$ and ${\bf 10}_2$) 
are involved in a $SU(2)_H$ doublet, $\Psi({\bf 27,2})_a=(\Psi_1,\Psi_2)$, 
while ${\bf 10}_3$ and MSSM Higgs fields
are involved in $SU(2)_H$ singlets, $\Psi({\bf 27,1})_3$ and $\Phi({\bf 27,1})$, respectively.\footnote{
The numbers in the parenthesis denote representations under $E_6 \times SU(2)_H$}
The Yukawa terms for ${\Psi({\bf 27,2})}_a$ are forbidden by SUSY and the horizontal symmetry
of the superpotential. 
On the other hand, a term $W \ni \Psi({\bf 27,1})_3 \Psi({\bf 27,1})_3 \Phi({\bf 27,1})$ is allowed.
Thus, the Yukawa coupling for 
${\bf 10}_3$ can be of order one,
which is identified as the top Yukawa coupling.

Other Yukawa couplings arise though the higher dimensional operators
after breaking the horizontal symmetry.
They are suppressed by factor $(\langle F \rangle / \Lambda)^n$,
where $\langle F \rangle$ is a breaking scale of $H$ and $n$ is some integer.
Thus, the Yukawa hierarchy in ${\bf 10}$ sector is larger than that in ${\bf \over 5}$ residing in 
$\Psi({\bf 27},{\bf 2})$.
This explains why the mass hierarchy of the up-quark sector ($Q,U^c \subset {\bf 10}$) is the largest and 
the lepton flavor mixing ($L \subset {\bf \over 5}$) is larger than the quark flavor mixing.

The model predicts the non-universal sfermion masses Eq.\,(\ref{nusm_matrix}) at the leading order.
The renormalizable sfermion soft mass terms can be written as
\begin{equation}
V_{\rm soft}^{\rm renorm} = m_0^2 \phi({\bf 27, 2})^{\dg a} \phi({\bf 27, 2})_a 
+ m^2_{30} |\phi({\bf 27, 1})_3|^2,
\end{equation} 
where $\phi$ represents the scalar components of the superfield ${\Psi}$.
Again, the model has not only a partial (1st-2nd) universality of ${\bf 10}$-plet sfermions
but also a full (1st-3rd) universality of ${\bf \over 5}$-plet sfermions.  
The full universality of ${\bf \over 5}$ sfermions is crucial to satisfy the FCNC and EDM constraints
because the unitary matrices that diagonalize the Yukawa matrices for ${\bf \over 5}$s 
are expected to have large off-diagonal entries like the MNS (Maki-Nakagawa Sakita)  matrix.

In the following, we identify a region of parameter space where 
both the naturalness and the FCNC constraints are satisfied.
In this model, 
the condition Eq.\,(\ref{tuning_rough}) can be expressed numerically 
in terms of the fundamental parameters defined at the cutoff scale as
\cite{Dermisek:2006qj}
\begin{equation}
m^2_Z \simeq -1.9 |\mu(\Lambda)|^2 -1.2 m^2_{H_u}(\Lambda) + 1.5 m^2_{30} + 5.9 m^2_{1/2} + \cdots .
\label{tuning_num} 
\end{equation}
The term proportional to $m_{1/2}$ arises through the stop mass dependence of the RGE.
The large cancellation is not required in the right hand side of Eq.\,(\ref{tuning_num})
as long as parameters $\mu$, $m_{H_u}$, $m_{30}$ and $m_{1/2}$ are 
around the $Z$ boson mass scale. 
Therefore we study the region of the parameter space where
\begin{equation}
m_{1/2} \lsim 300\,{\rm GeV},~~~~m_{30}, ~m_{H_u}(\Lambda) \lsim 500\,{\rm GeV}.
\label{m1/2_m30_mHu}
\end{equation}
In this case, $\mu$ is typically 200 to 500\,GeV.

The other sfermion masses are given by $m_0$.
If $m_0$ is also around the $Z$ boson mass scale,
FCNCs and EDMs severely constrain the flavor off-diagonal terms 
of the sfermion mass matrices and CP violating phases of the various SUSY breaking parameters.
Such constraints are sometimes problematic to construct explicit models,
because various sources that violate the universality
are expected.\footnote{
The RG effect or the effect of the gravity mediation  
is one of the sources to produce the non-universality.
In models with the horizontal symmetry,
the effect of the horizontal symmetry breaking is also the source
to produce the non-universality.
}
On the other hand, if $m_0$ is much larger than the weak scale,
the constraints can be relaxed.\footnote{
The constraint from the up-quark (C)EDM is still severe
because some contributions do not decouple with increading $m_0$.
A spontaneous CP violation mechanism may solve this issue \cite{Ishiduki:2009gr}.
}

However, there is upper bound on $m_0$.
A large mass splitting
between the first two generations and the 3rd generation sfermions tends to make 
the 3rd generation sfermion mass squared negative through the 
2-loop RG effects,
and cause the color and charge breaking (CCB) \cite{ArkaniHamed:1997ab}.
In addition, large $\tan\beta$ ($\tan\beta \equiv \bra H_u \ket / \bra H_d \ket$) potentially cause
the CCB problem in our scenario.
The  ${\bf 10}_3$ couple to the ${\bf \over 5}_3$ through the bottom Yukawa coupling,
$Y_b$.
The negative correction to $(m^2_{\ti q})_{33}$ is roughly given as
\begin{equation}
\Delta (m^2_{\ti q})_{33} \sim -\frac{|Y_b|^2}{(4\pi)^2} m^2_{\ti b_R} 
\ln \Big( \frac{\Lambda^2}{m^2_{\ti b_R}} \Big).
\label{del_mhd}
\end{equation}
Note $m_{\ti b_R}=m_0$ at the cutoff scale.
This contribution would be as large as the 2-loop RG effect unless $Y_b$
is sufficiently small.
Small $\tan\beta$ ($\tan\beta \sim 10$) is also preferable  
in $E_6$ SUSY GUT model with horizontal symmetry.
In the models the bottom Yukawa coupling, 
which is originated from the Yukawa couplings for $\Psi({\bf 27,2})$,
is suppressed because it is forbidden under the horizontal symmetry.

\begin{figure}[t!]
\begin{center}
\includegraphics[width=8cm,clip]{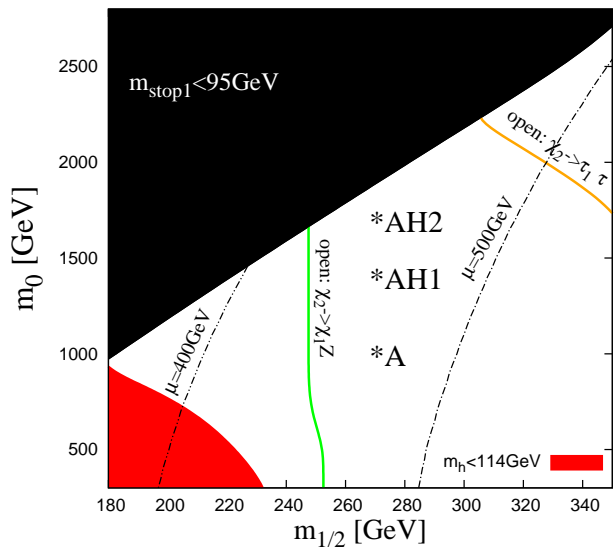}
\caption{\footnotesize{
The allowed region on ($m_{1/2} - m_0$) plane.
The other parameters are taken 
$m_{30}=m_{H_u}(\Lambda)=m_{H_d}(\Lambda)=300$\,GeV, 
$A_0=-600$\,GeV, $\tan\beta=10$, $\sgn(\mu)=+$.
}}
\label{m0var}
\end{center}
\end{figure}

In Fig.\,\ref{m0var}, we show the allowed region on ($m_{1/2}-m_0$) plane.
Here, the low energy particle spectra are calculated using ${\tt ISAJET\,7.75}$ \cite{isajetref}.
We fix the other parameters as
$m_{30}=m_{H_u}(\Lambda)=m_{H_d}(\Lambda)=300\,$GeV, $A_0=-600$\,GeV, $\tan\beta=10$ and $\sgn(\mu)=+$,
where $A_0$ is the universal trilinear coupling.
In the black region the lighter mass eigenstate of stops, $\ti t_1$, is
unacceptably light due to the 1 and 2-loop RG effects.
We find that $m_0$ cannot exceed 2\,TeV 
in the region where $m_{30}, m_{1/2} \lsim 300$\,GeV and $-A_0 \gsim 600$\,GeV.
Thus, in this paper we study the parameter region where\footnote{
This scenario however cannot explain the anomaly of the muon $g-2$ \cite{Hagiwara:2006jt},
because $\ti \mu$ and $\ti \nu_{\mu}$ are heavy due to the large $m_0$.
}
\begin{equation}
m_0 \sim 1-2\,{\rm TeV}, ~~~~~~\tan\beta =10.
\label{m0_tanb_region}
\end{equation}
In the following, we call the scenario characterized by Eqs.\,(\ref{nusm_matrix}),  (\ref{m1/2_m30_mHu}) 
and (\ref{m0_tanb_region})   {\it modified universal sfermion mass} (MUSM) {\it scenario}.

\begin{figure}[t!]
\begin{tabular}{lr}
\begin{minipage}{0.45\hsize}
\begin{flushleft}
\includegraphics[width=7.5cm,clip]{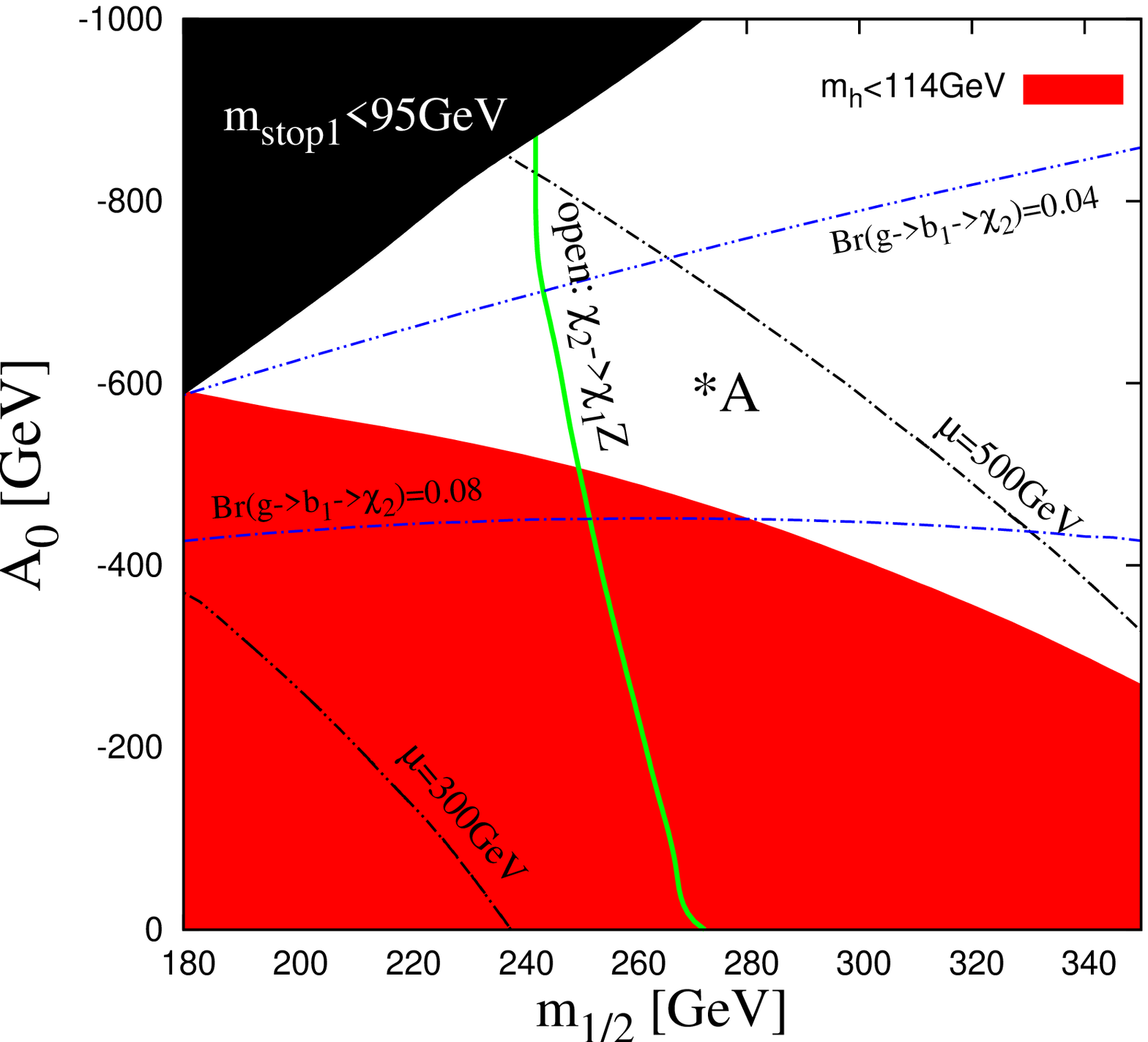}
\caption{\footnotesize{
The allowed region on ($m_{1/2}-A_0$) plane.
The other parameters are taken as
$m_0=1$\,TeV, $m_{30}=m_{H_u}(\Lambda)=m_{H_d}(\Lambda)=300$\,GeV,
$\tan\beta=10$ and $\sgn(\mu)=+$.
}}
\label{mh300}
\end{flushleft}
\end{minipage}
\hspace{9mm}
\begin{minipage}{0.45\hsize}
\begin{flushright}
\includegraphics[width=7.5cm,clip]{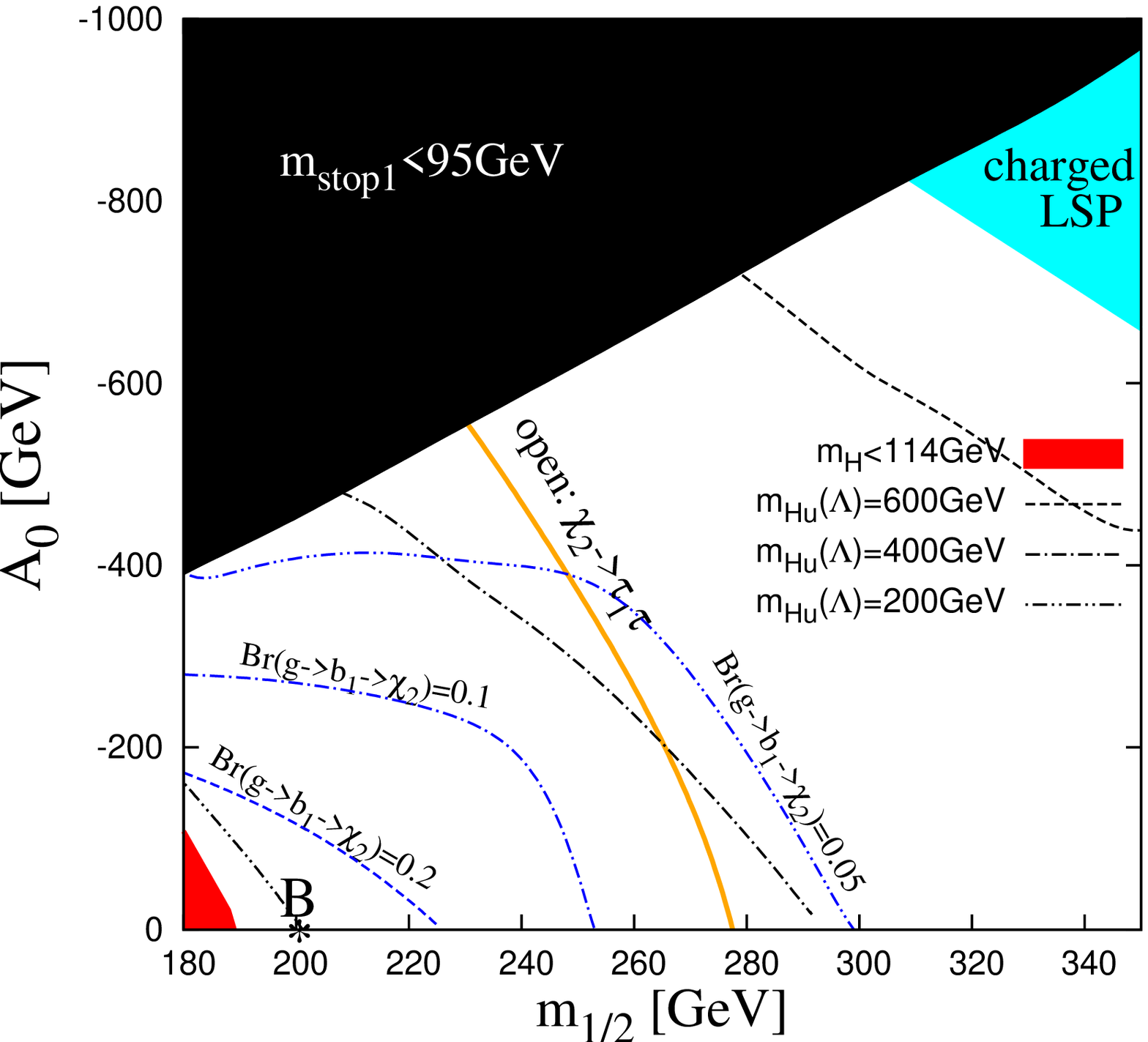}
\caption{\footnotesize{
The allowed region on ($m_{1/2}-A_0$) plane in the light Higgs scenario.
The other parameters are taken as
$m_0=1$\,TeV, $m_{30}=200$\,GeV, $\tan\beta=10$, 
$\mu=250$\,GeV and $m_A=105$\,GeV.
}}
\label{mu250}
\end{flushright}
\end{minipage}
\end{tabular}
\end{figure}

The mass of the lightest CP-even Higgs boson, $m_h$, tends to be lighter than LEP II 
SM Higgs mass bound $m_{\varphi_{\rm SM}} > 114.4$\,GeV \cite{LEPhiggs}.
To push up $m_h$ above the bound,
the quantum correction to the Higgs quartic coupling is crucial \cite{Okada:1990vk}.
This typically requires large stop masses or a trilinear coupling.
Since large $m_{30}$ and $m_{1/2}$ are not preferable in view of the naturalness,
we search an allowed region in the direction of large $|A_0|$. 
Fig.\,\ref{mh300} shows the allowed region on ($m_{1/2}-A_0$) plane.
The red region is excluded by the condition $m_h>114$\,GeV.
We find that the Higgs mass bound requires $|A_0| \gsim 400-600$\,GeV 
in the region where $m_{30}, m_{1/2} \sim 200-300$\,GeV.
The two black dashed lines represent
$\mu=300$\,GeV and 500\,GeV at the stop mass scale.
The $\mu$ value is not sensitive to $A_0$ compared with its dependence on $m_{1/2}$ in Fig.\,\ref{mh300}.
This means the weak scale $m_{H_u}$ is not so sensitive to $A_0$,
and we can take large value of $A_0$ without making naturalness worse.
However a large $|A_0|$ may also cause the CCB problem,
because it may lead one of the stop masses squared to be negative.
In the black region of Fig.\,\ref{mh300}, the lighter stop becomes unacceptably light, 
$m_{\ti t_1}<95$\,GeV \cite{Amsler:2008zzb},
due to the large $|A_0|$ value.


\begin{table}[t!]
~~
\begin{center}
\renewcommand{\arraystretch}{1.1}
\begin{tabular}{|r|c|c|c|c|c|}
\hline
 ~~~      & \multicolumn{5}{|c|}{ Values } \\
\cline{2-6}
 Parameters         & A &  AH1 &  AH2 &  B & U  \\
\hline \hline
$m_0$             & 1000 & 1400 & 1700 & 1000 & 1000 \\
$m_{30}$           & 300  & 300  & 300  & 200  & 1000 \\
$m_{1/2}$         & 270  & 270  & 270  &  200 &  270 \\
$A_0$          &$-$600&$-$600&$-$600& 0    &$-1600$ \\
$\tan\beta$              & 10   &  10  &    10 &   10 &  20  \\
$m^2_{H_d}(\Lambda)$  & $(300)^2$  & $(300)^2$  &  $(300)^2$ &$-(216.0)^2$& $(1000)^2$ \\
$m_{H_u}(\Lambda)$ & 300  & 300  &  300 &  196.7& 1000 \\
\hline
\end{tabular} 
\caption{\footnotesize{
The parameters of each model point. The unit of mass parameters is GeV .   
}}\label{point}
\end{center}
%
\vspace{2mm}
\begin{center}
\renewcommand{\arraystretch}{1.1}
\begin{tabular}{|c|c|c|c|c|c|}
\hline
 ~~~      & \multicolumn{5}{|c|}{ Masses (GeV)} \\
\cline{2-6}
 Particles         & A &  AH1 &  AH2 &  B & U  \\
\hline \hline
$\ti q$             & 1150 & 1500 & 1780 & 1080 & 1145 \\
$\ti t_1$           & 321  & 262  & 187  &  296 & 281 \\
$\ti b_1$          & 540  & 499  & 456  &  400 &  856 \\
$\ti g$             &697    &711   & 721  &   537  & 706 \\
$\none$          & 110   &  111  & 111 &   77 &  114  \\
\hline
\end{tabular} 
\caption{\footnotesize{
The masses of some sparticles at each model point.
}}
\end{center}\label{mass}
\end{table}

To perform Monte Carlo simulation studies
we choose a representative parameter set, Point A with $m_0=1000$\,GeV, $m_{30}=300$\,GeV and 
$m_{1/2}=270$\,GeV. 
The other parameters are listed in Table\,\ref{point}. 
In order to see the $m_0$ dependence of the collider signature,
we also choose model points AH1 and AH2 with $m_0=1400$ and 1700\,GeV, respectively.
These model points are shown in Figs\,\ref{m0var}, \ref{mh300} and Table\,\ref{point}. 
Point U shown in Table\,\ref{point} is a CMSSM model point 
which will be investigated in Section\,\ref{alike_msugra}.
The masses of some sparticles are shown in Table\,\ref{mass}.

If Higgs masses can be non-universal,
there is another allowed region where
the heaviest CP-even Higgs boson, 
$H$, becomes the SM-like Higgs boson \cite{Kane:2004tk,lhp_bsg}.  
This scenario is referred as 
{\it a light Higgs scenario} or {\it an inverted hierarchy scenario}.
This can be realized if the CP-odd Higgs boson mass, $m_A$, satisfies
$m_A \sim 100$\,GeV, which also prefers small $|\mu|$.
Because of the small $m_A$ and $\mu$ parameter, 
the thermal relic density of the $\none$ tends to be the same or 
small compared with the observed dark matter density 
\cite{Asano:2007gv,Kim:2008uh}.
In this scenario all Higgs bosons, $h$, $H$, $A$, $H^{\pm}$, have small masses
around the weak scale, 
therefore they would contribute to various rare $B$ decay processes.
Actually, constraints from $Br(B_u^+ \to \tau^+ \nu_{\tau})$ \cite{Hou:1992sy} and 
$Br(B_s \to \mu^+ \mu^-)$ \cite{Babu:1999hn} 
exclude $\tan\beta \gsim 15$ region \cite{Isidori:2006pk}.
And $Br(b \to s \gamma)$ requires ${\rm sign}(\mu)=+$ and small masses of both $\ti t_i$ and $\ti \chi^{\pm}_i$
\cite{lhp_bsg}, so that the charged Higgs\,-\,top quark contribution can cancel with 
the chargino\,-\,stop contribution.
The prediction of $Br(b \to s \gamma)$ however depends on the off-diagonal entries of
$V_{u_R}$, $V_{u_L}$ and $V_{d_L}$ in Eq.\,(\ref{assump}) \cite{Kim:2008yta}.

In Fig.\,\ref{mu250}, we show the allowed region on ($m_{1/2}-A_0$) plane.
Here we fix 
$m_0=1$\,TeV, $m_{30}=200$\,GeV, $\tan\beta=10$, $\mu=250$\,GeV and $m_A=105$\,GeV.
The heavier Higgs boson in the MSSM is the SM like in this case,
and LEP II bound should be applied to $H$.
The red region of Fig.\,\ref{mu250} is excluded by the condition $m_H > 114\,$GeV.
The allowed region is widely extended 
to the small $A_0$ region compared with the normal case.
We choose Point B defined in Table\,\ref{point} as a representative parameter point.

\section{The characteristic signatures of  MUSM}
\label{charact}

\subsection{The number of $b$ jets:}

In MUSM scenario, $m_0$ is much larger than $m_{1/2}$.
Then, the gluino 2-body decay mode into the first two generation squarks, 
$\ti g \to \ti q q$, is closed.
On the other hand, in a wide parameter region
$\ti t_1$ and $\ti b_1$ are lighter than gluino due to the RG running and left-right mixing effects
even if $m_{30} \gsim m_{1/2}$.
Then, $\ti g \to \ti t_1 t$ and $\ti b_1 b$ modes 
entirely dominate the gluino decay.
Since a gluino is a flavor singlet,
the gluino decay chain contains at least 2 $b$ jets ($b$-$\bar b$ pairs).
Therefore $\ti g$-$\ti g$ and $\ti q$-$\ti g(\ti q)$ production events have 4 $b$ partons.
This characteristic feature can be observed at the LHC by counting the 
number of $b$ tagged jets.

To simulate the LHC signature at the model points selected in Section\,\ref{sec_nusm},
we calculate the low energy particle spectra
and the sparticle decay branching ratios by {\tt ISAJET}.
The SUSY events are generated by the parton shower Monte Carlo {\tt HERWIG}
\cite{Corcella:2000bw,Corcella:2002jc}, 
and the detector resolutions are simulated by {\tt AcerDET}
\cite{RichterWas:2002ch}.
We assume the collider center of mass energy is $\sqrt{s}=14$\,TeV. 
Throughout  this paper, we adopt the following SUSY cuts to reduce the SM background:
\begin{itemize}
\item $N^{\rm jets}_{50} \ge 4,~~~N_{100}^{\rm jets} \ge 1$,
\item $M_{\rm eff} \equiv \sum_{i=1}^4 |p_T^{(i)}| + \esla_T > 500$\,GeV,
\item $\esla_T > {\rm max}\{ 200\,{\rm GeV}, 0.2 M_{\rm eff} \}$,
\end{itemize}
where $p_T^{(i)}$ is the transverse momentum of $i$-th jet ($p_T^{(i)} > p_T^{(j)}$ for $i<j$) and 
$N^{\rm jets}_{50(100)}$ is the number of jets with $p_T>50(100)$\,GeV and $|\eta|<3$.

\begin{figure}[t!]
\begin{center}
\begin{tabular}{ccc}
\begin{minipage}{0.28\hsize}
\includegraphics[width=3.5cm,clip]{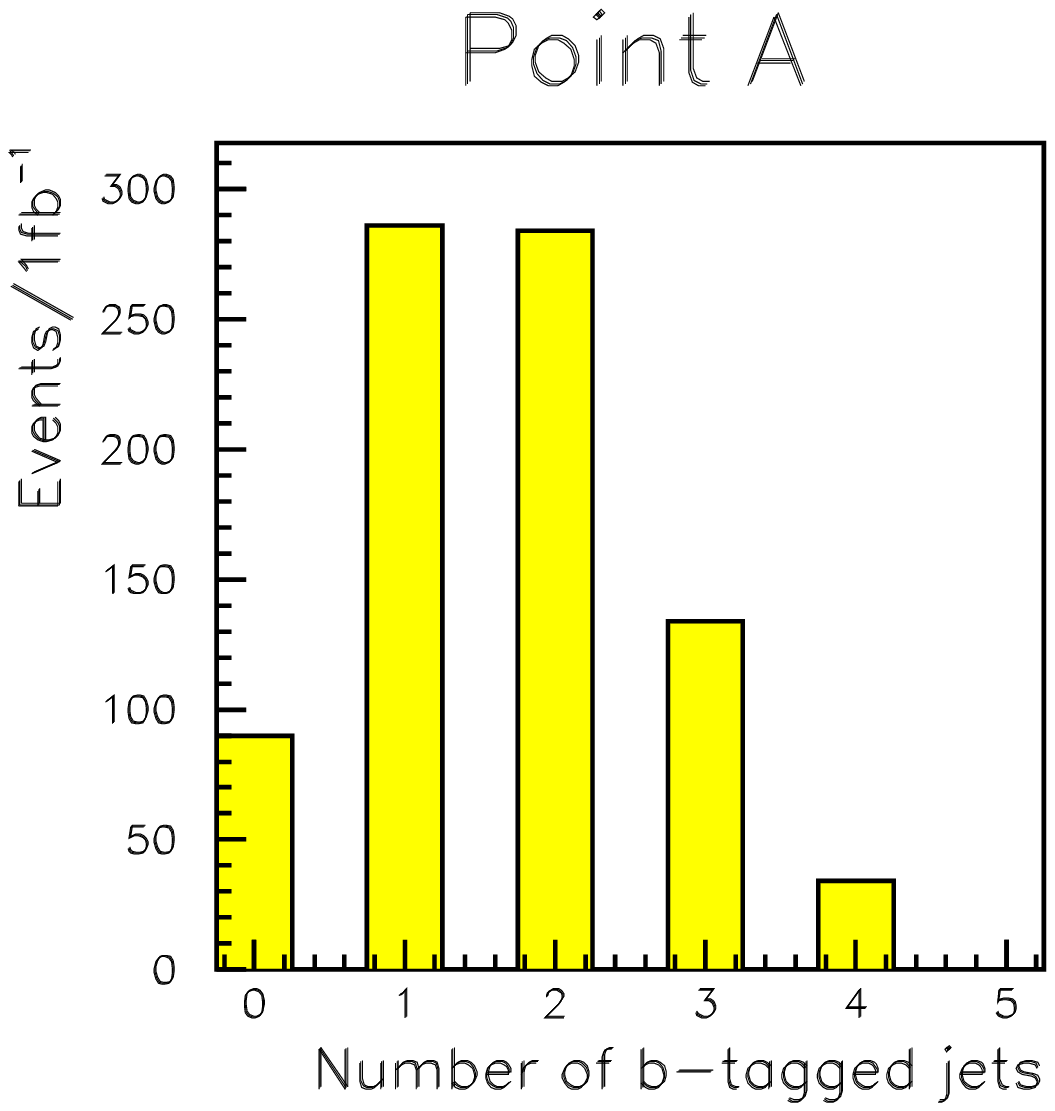}
\end{minipage}
\begin{minipage}{0.28\hsize}
\includegraphics[width=3.5cm,clip]{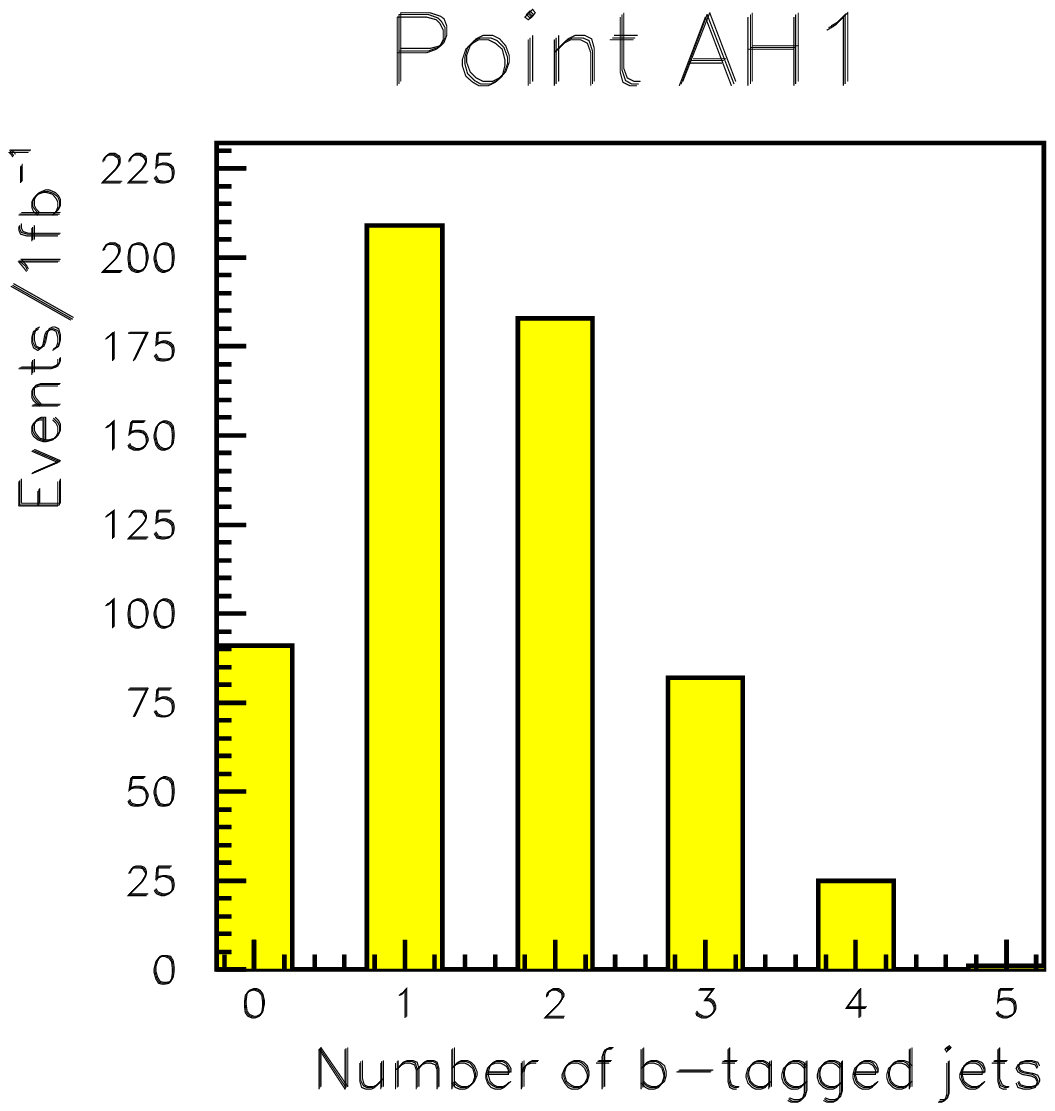}
\end{minipage}
\begin{minipage}{0.28\hsize}
\includegraphics[width=3.5cm,clip]{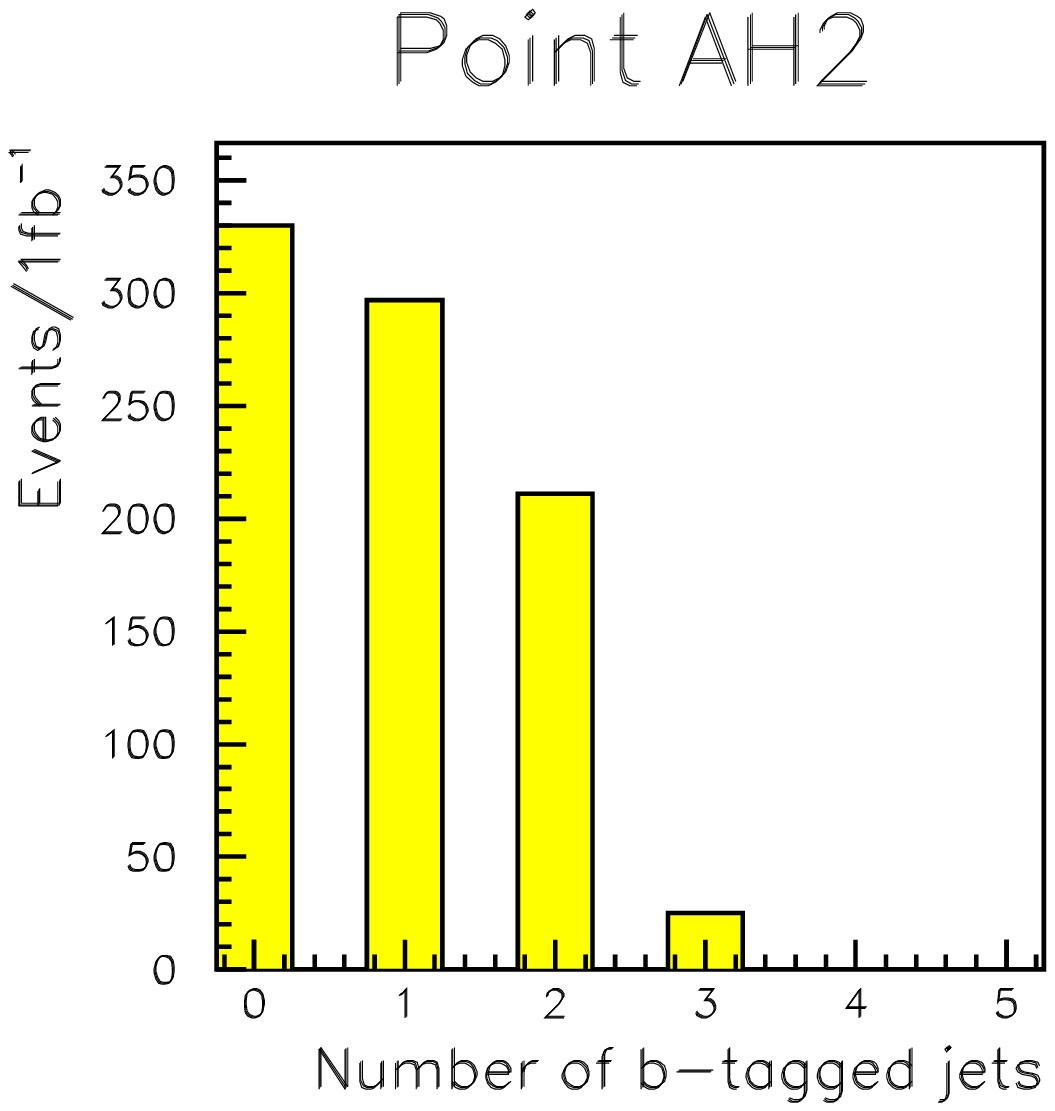}
\end{minipage}
\end{tabular}
\end{center}
%
%
\begin{center}
\begin{tabular}{ccc}
\begin{minipage}{0.28\hsize}
\includegraphics[width=3.5cm,clip]{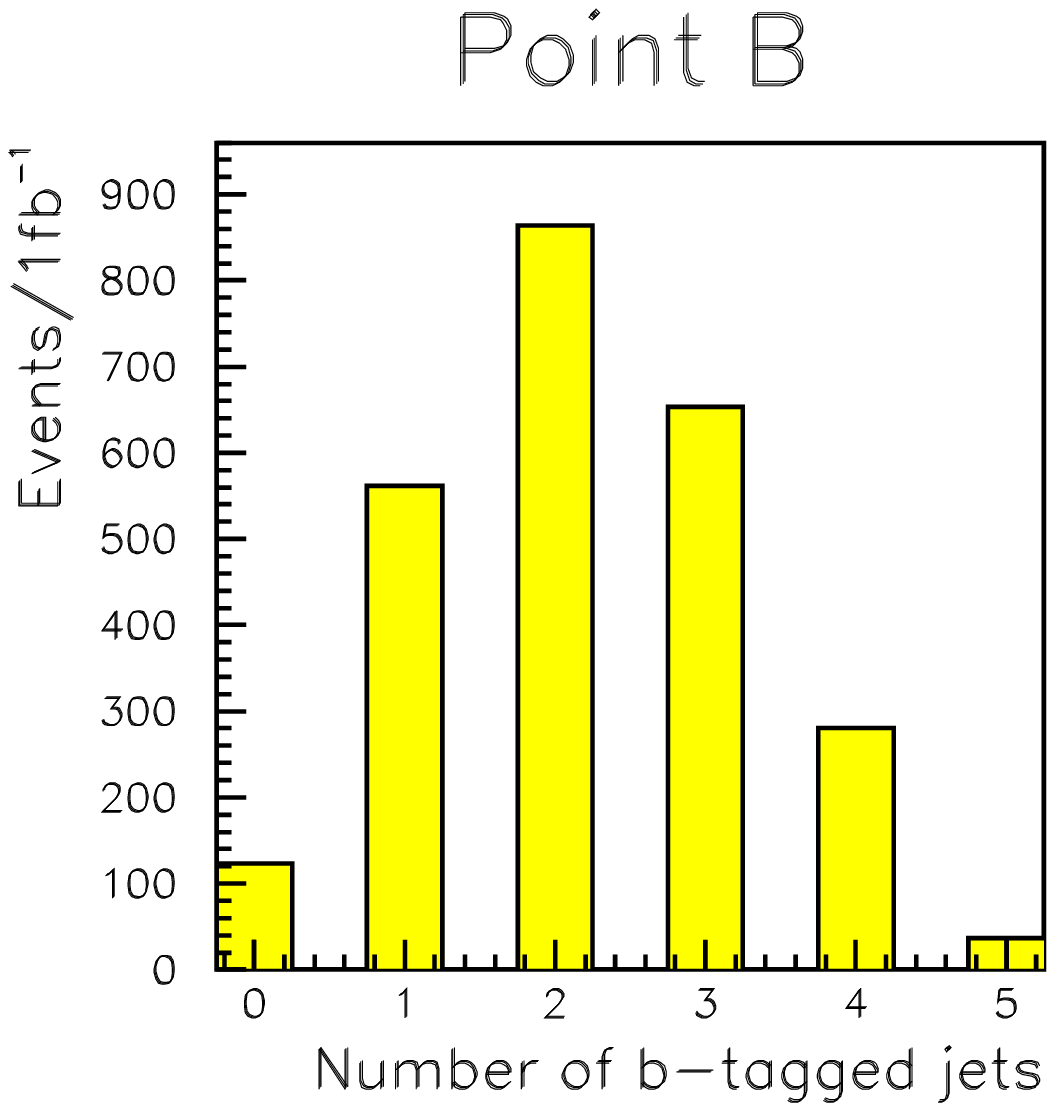}
\end{minipage}
\begin{minipage}{0.28\hsize}
\includegraphics[width=3.5cm,clip]{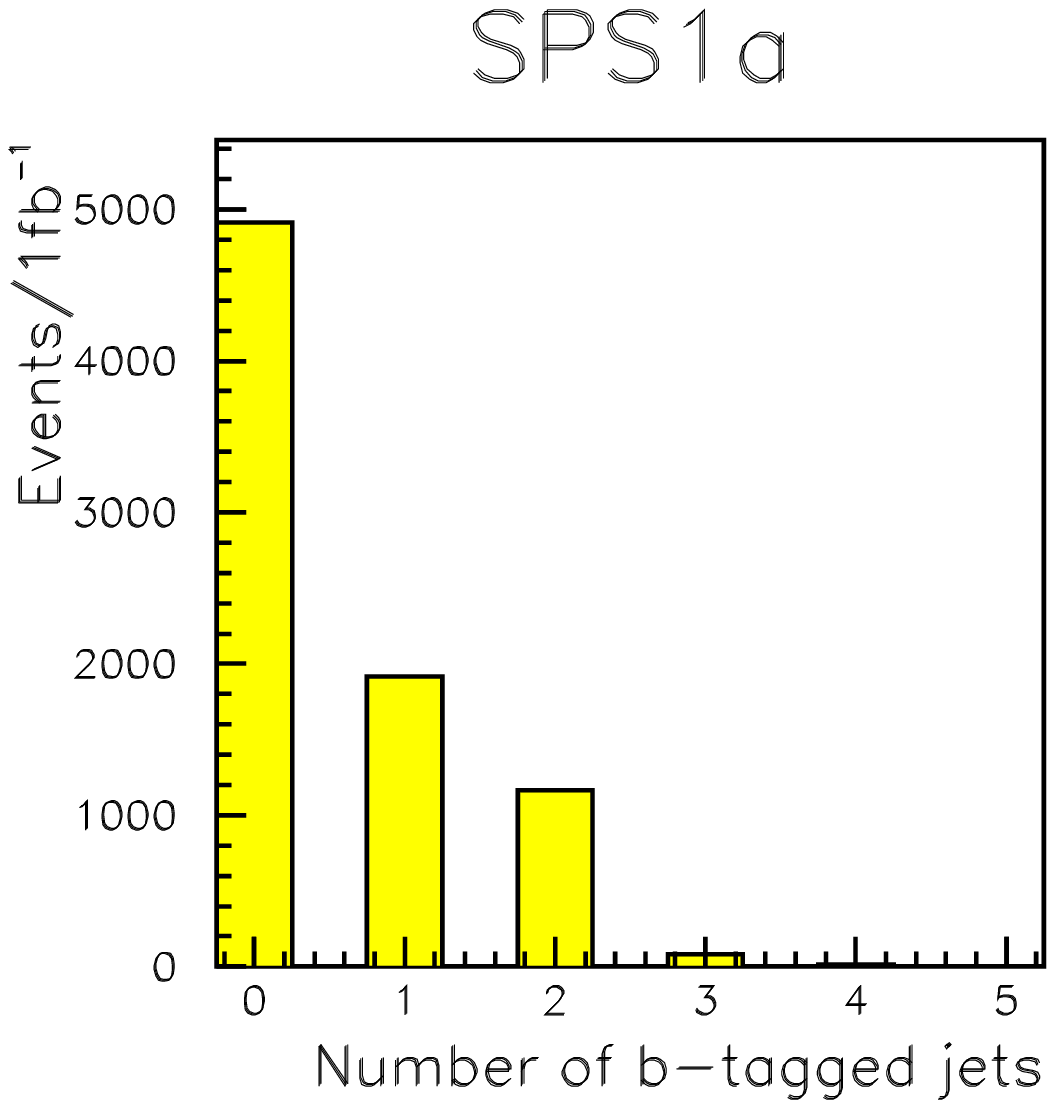}
\end{minipage}
\begin{minipage}{0.28\hsize}
\includegraphics[width=3.5cm,clip]{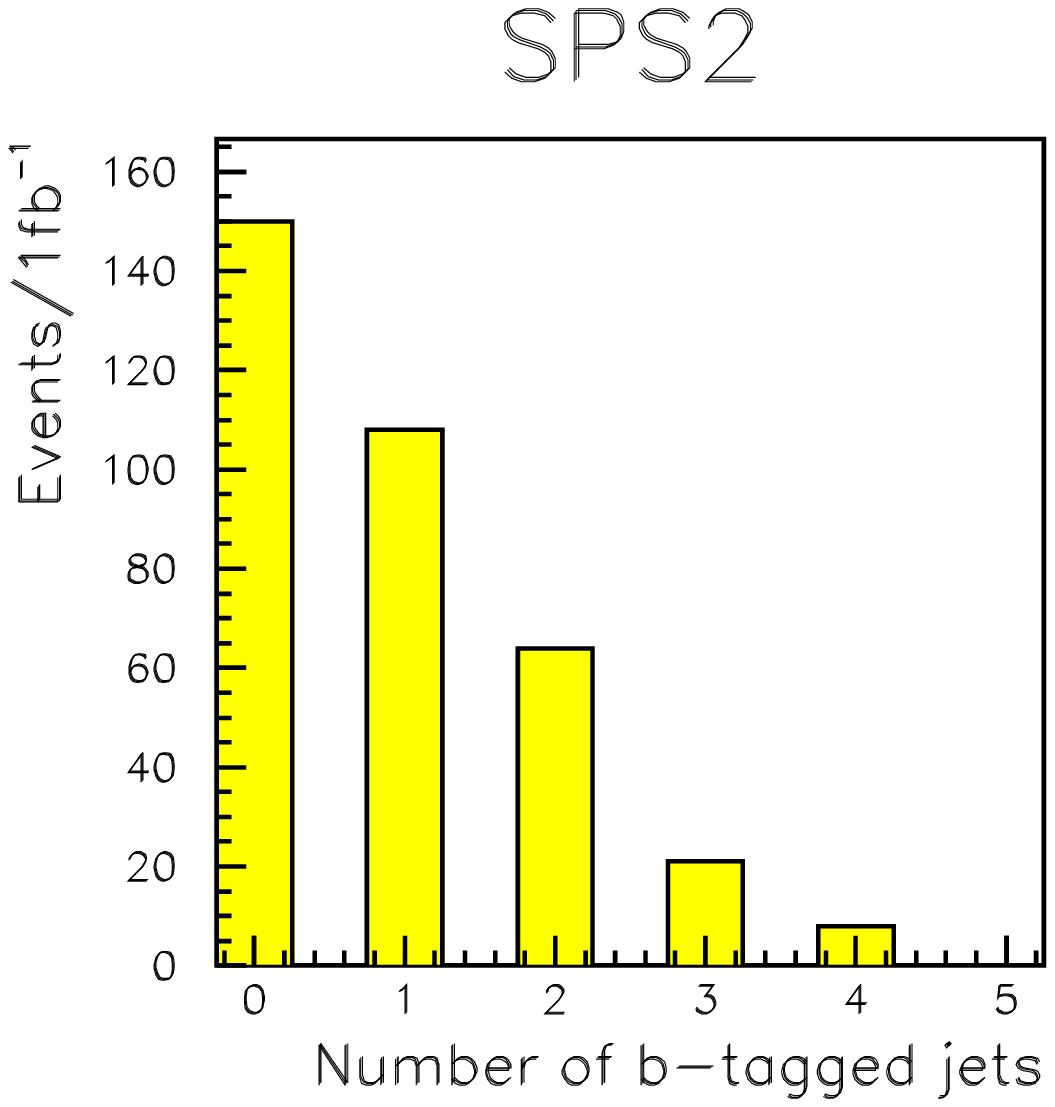}
\end{minipage}
\end{tabular}
\caption{\footnotesize{
The distribution of the number of $b$ tagged jets per $1 {\rm fb^{-1}}$ at each model point.
The $b$ tagging efficiency is assumed to be 60\,\%.
}}
\label{numb}
\end{center}
\end{figure}

The distribution of the number of $b$ jets 
at each model point is shown in Fig.\,\ref{numb}.
Here we count the $b$ tagged jets with $p_T(b)>50$\,GeV and $|\eta(b)|<2.5$, 
and require no isolated lepton in the event.
The number of SUSY events corresponds to 1\,${\rm fb^{-1}}$ of the integrated luminosity. 
We assume the $b$ tagging efficiency is 60\,\%.
To compare the distributions with those of the benchmark model points with universal sfermion masses,
the distributions for the SPS1a and SPS2 model points \cite{Allanach:2002nj} are also shown in Fig.\,\ref{numb}.
The peak of the distributions is at zero at SPS1a and SPS2, while
zero $b$ tagged events are suppressed at Point A, AH1 and B.

The suppression of no-$b$ jet SUSY events indicates the following mass relation:
\begin{equation}
m_{\ti q} > m_{\ti g} > m_{\ti t_1(\ti b_1)}+m_{t(b)}
~~~~~{\rm or}~~~~~
m_{\ti q} \gg m_{\ti t_1(\ti b_1)}+m_{t(b)} > m_{\ti g},
\label{mass_relation1}
\end{equation}
Indeed in the distributions of mSUGRA benchmark points SPS1a$-$SPS9,
this feature is not seen (See Appendix\,\ref{sps_appen}).
Even at SPS2, where ($m_{\ti g}$, $m_{\ti q}$, $m_{\ti t_1}$, $m_{\ti b_1})=($796, 1560, 963, 1301)\,GeV
and the gluino branching ratio into $2b+X$ is about 70\%, the peak is at zero.

An exceptional case is Point AH2.
At this point, the distribution peaks at zero, although the mass relation Eq.\,(\ref{mass_relation1})
is satisfied.
At this point the $\ti t_1$ is as light as top quark, $m_{\ti t_1}=187$\,GeV, 
due to the large $m_0$ value ($m_0=1700$\,GeV),
and the 2-body decay modes $\ti t_1 \to \chi_1^+ b$
and $\ti t_1 \to \none t$ are closed.
Then the flavor violating 2-body decay mode $\ti t_1 \to \none c$ dominates the $\ti t_1$ decay 
\cite{Hikasa:1987db}.
In such case, the distribution of the number of $b$ jets are not helpful for the model discriminations. 
It has been shown in \cite{Carena:2008mj} that
such a light stop may be detected at the LHC in $\gamma + \esla_T$ or $j + \esla_T$
channel.

\subsection{The highest $p_T$ jet:}

In MUSM scenario, there is large mass splitting between
$m_0$ and $m_{1/2}$, $m_{30}$,
but $m_0$ is bounded above by the CCB constraint.
The 1st and 2nd generation squarks, $\ti q$, may be produced enough to be seen at the LHC.
The number of SUSY events after the standard SUSY cuts at $\int {\cal L} dt = 1 {\rm fb^{-1}}$
are listed in Table\,\ref{production}.

\begin{table}[t!]
\begin{center}
\renewcommand{\arraystretch}{1.1}
\begin{tabular}{|r|c|c|c|c|c|}
\hline
 Production     & \multicolumn{5}{|c|}{ Production Ratios (\%)} \\
\cline{2-6}
 Processes      & A &  AH1 &  AH2 &  B & U  \\
\hline \hline
$\ti g \ti g$    & 32& 50 & 55 & 47 & 36 \\
$\ti q \ti g$    & 43& 22 & 9  & 38 & 45 \\
$\ti q \ti q$    & 7 &  2 &  1 &  3 &  9 \\
$\ti t_1 \ti t_1$ & 9 & 14 & 19 &  5 &  5 \\
$\ti b_1 \ti b_1$ & 2 &  5 &  9 &  3 &   0 \\
others           & 7 &  6 &  7 &  4 &   5 \\
\hline
\hline
Total Events & 1484  & 984  & 1096 & 3468  & 1677  \\
\hline
\end{tabular}
\caption{\footnotesize{
The number of SUSY events after standard SUSY cut at $\int {\cal L} dt = 1\,{\rm fb^{-1}}$.
Here we use the same character for a particle and the antiparticles.
}}
\label{production}
\end{center}
\end{table}
The SUSY production is dominated
by $\ti g$-$\ti g$ and $\ti q$-$\ti g$ production processes
at Points A, AH1 and B.
On the other hand, 
the fraction of $\ti q$-$\ti g$($\ti q$) production is relatively small (10\,\%) at Point AH2, 
due to the very large $m_0$ value ($m_0=1700$\,GeV).
The fractions of $\ti t$-$\ti t$ and $\ti b$-$\ti b$ productions are also small at all model points.
These events hardly survive the standard SUSY cut due to the small masses of $\ti t_1$ and $\ti b_1$.

Once a heavy squark is produced,
it decays mainly into $\ti g+q$.
The  quark jet is expected to have relatively large transverse momentum 
since the mass difference between $\ti g$ and $\ti q$ is large.
The  order of the transverse momentum is
around $m_{\ti q}/2$.
This quark jet tends to be the highest $p_T$ jet in the event.

\begin{figure}[t!]
\begin{center}
\begin{tabular}{ccc}
\begin{minipage}{0.3\hsize}
\begin{center}
\includegraphics[width=3.8cm,clip]{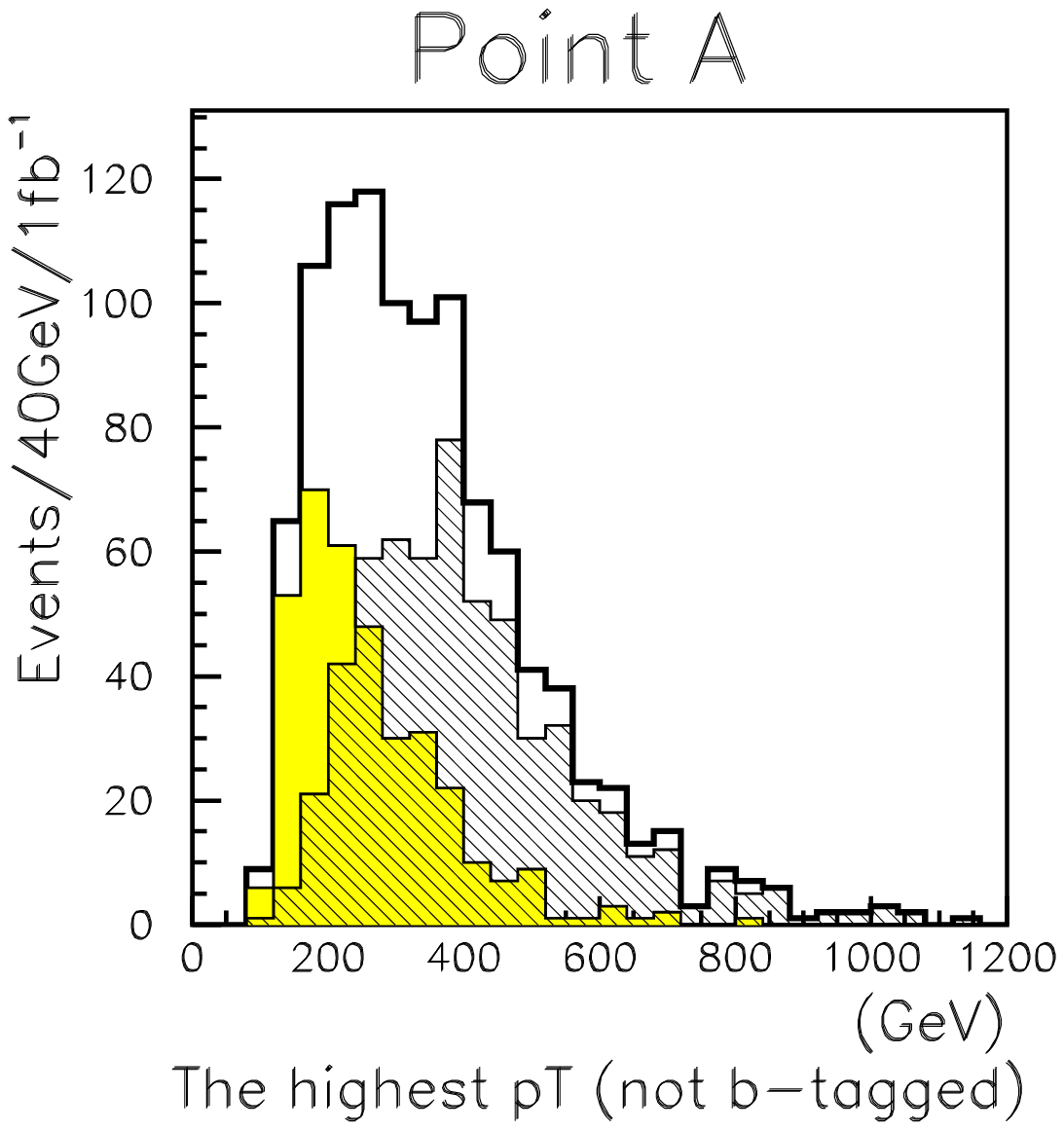}
\end{center}
\end{minipage}
\begin{minipage}{0.3\hsize}
\begin{center}
\includegraphics[width=3.8cm,clip]{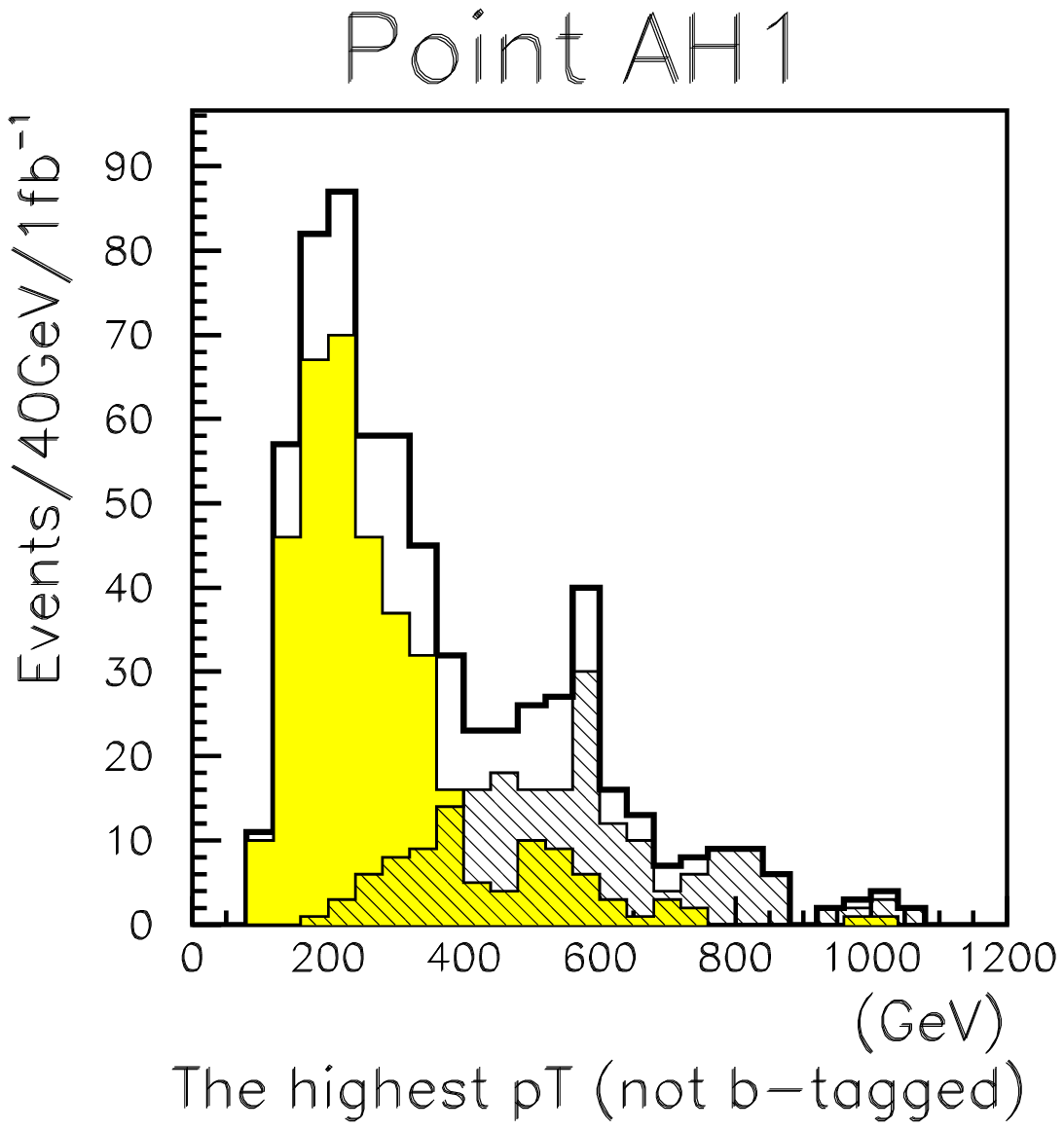}
\end{center}
\end{minipage}
\begin{minipage}{0.3\hsize}
\begin{center}
\includegraphics[width=3.8cm,clip]{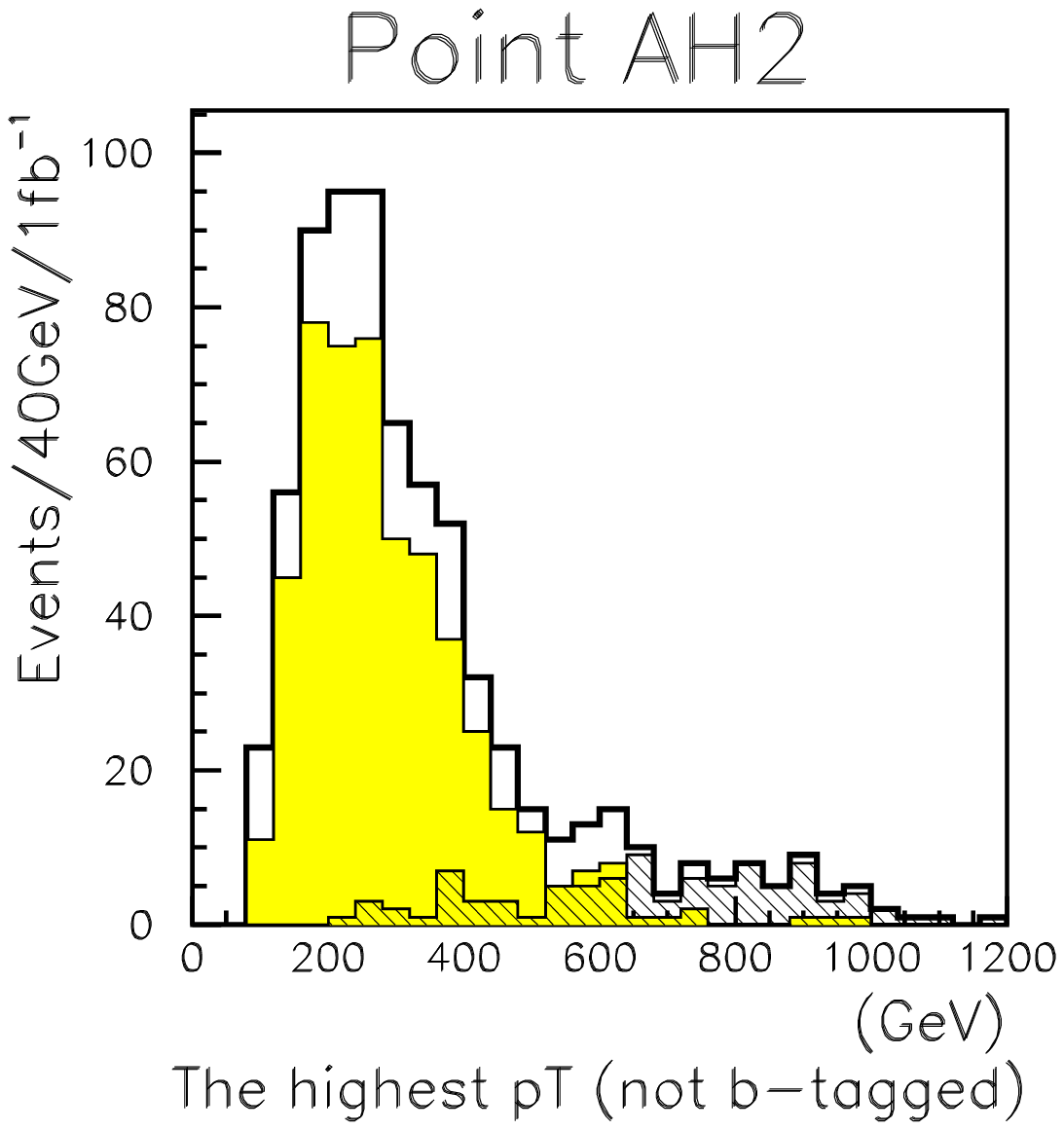}
\end{center}
\end{minipage}
\end{tabular}
~\\
\vspace{5mm}
\begin{tabular}{ccc}
\begin{minipage}{0.3\hsize}
\begin{center}
\includegraphics[width=3.8cm,clip]{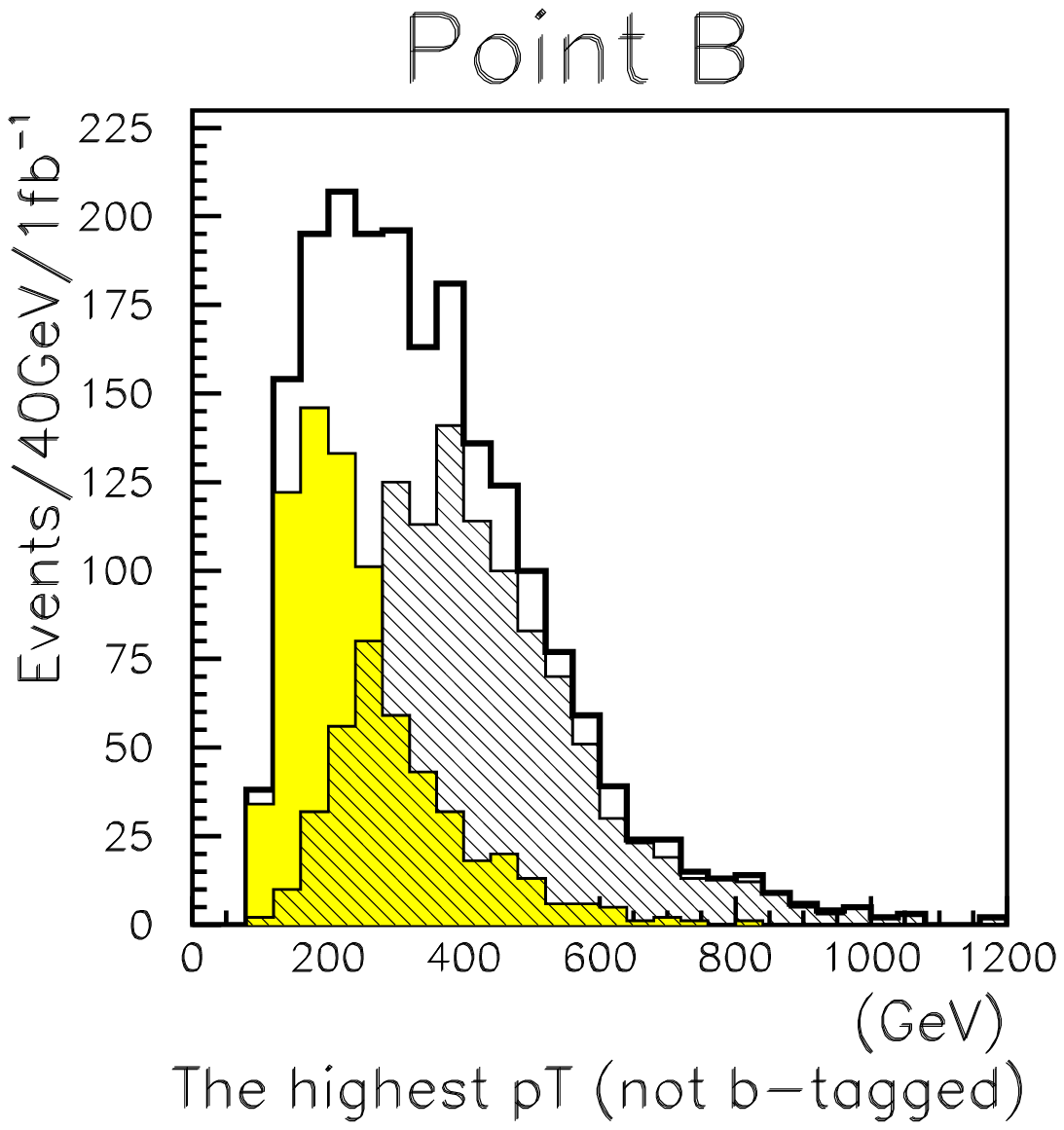}
\end{center}
\end{minipage}
\begin{minipage}{0.3\hsize}
\begin{center}
\includegraphics[width=3.8cm,clip]{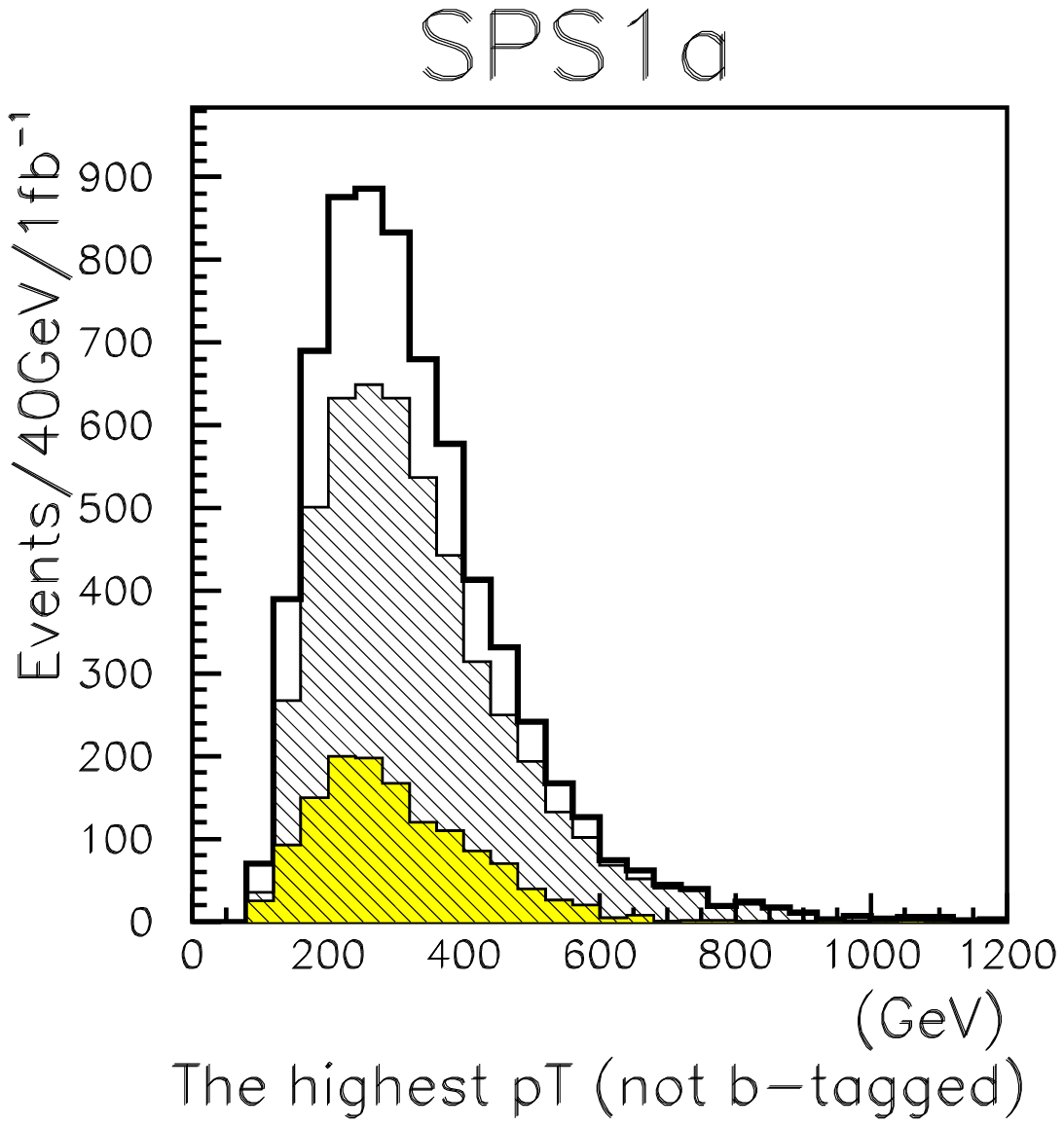}
\end{center}
\end{minipage}
\begin{minipage}{0.3\hsize}
\begin{center}
\includegraphics[width=3.8cm,clip]{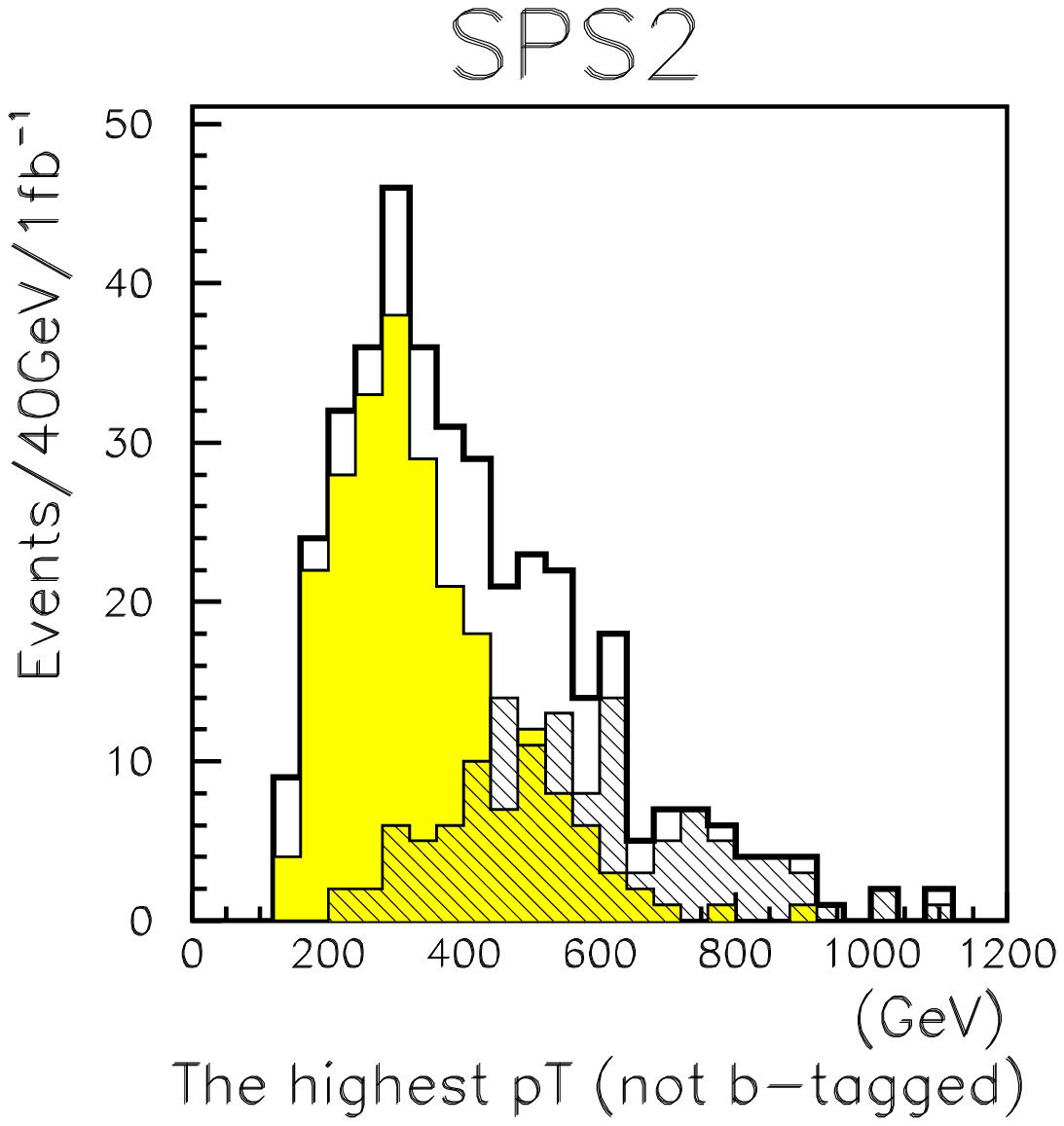}
\end{center}
\end{minipage}
\end{tabular}
\caption{\footnotesize{
The $p_T$ distribution of the non-b tagged highest $p_T$ jet in the event at each model points.
}}
\label{hstpt}
\end{center}
\end{figure}
\begin{figure}[t!]
\begin{center}
\begin{tabular}{ccc}
\begin{minipage}{0.3\hsize}
\begin{center}
\includegraphics[width=3.8cm,clip]{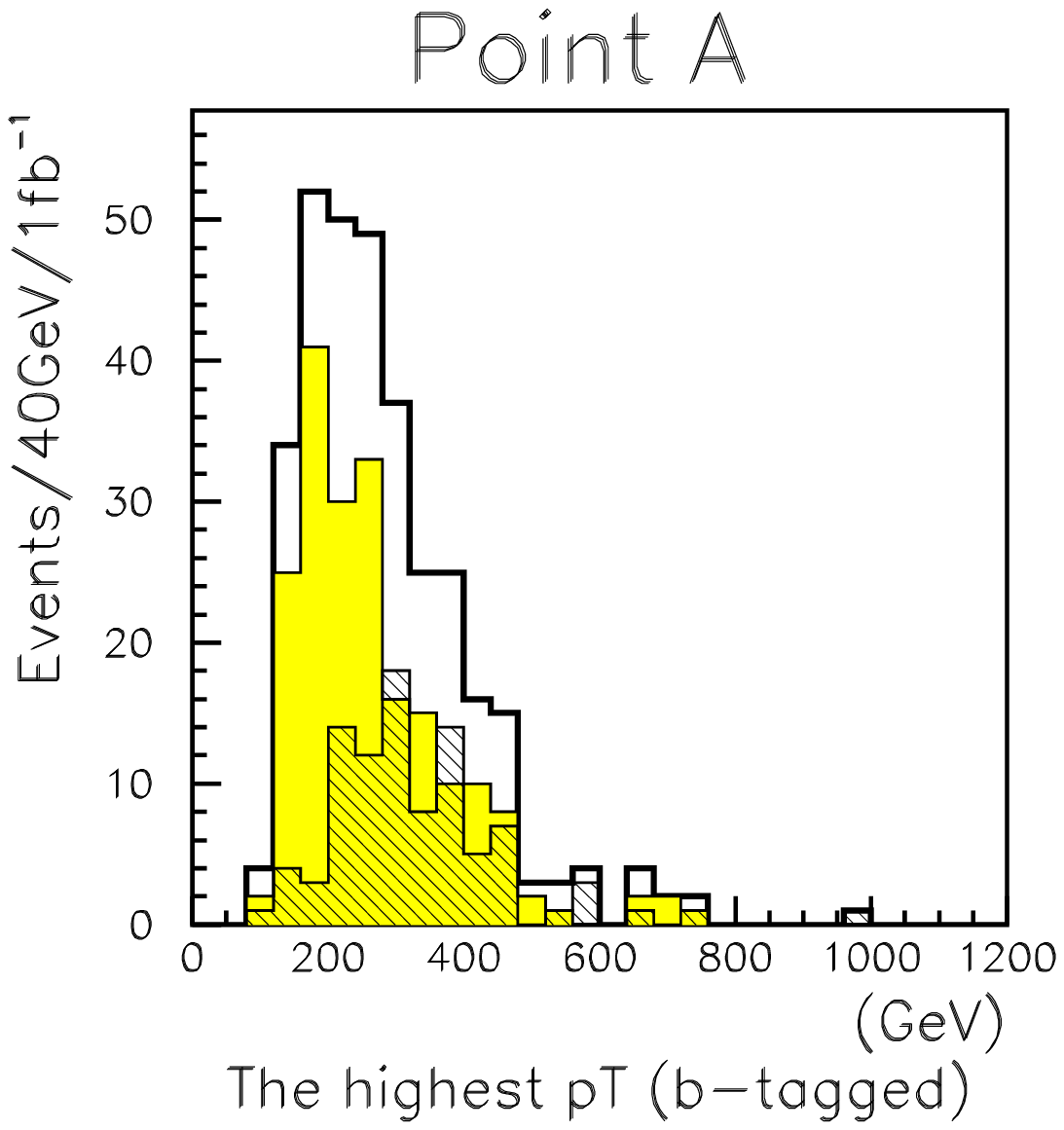}
\end{center}
\end{minipage}
\begin{minipage}{0.3\hsize}
\begin{center}
\includegraphics[width=3.8cm,clip]{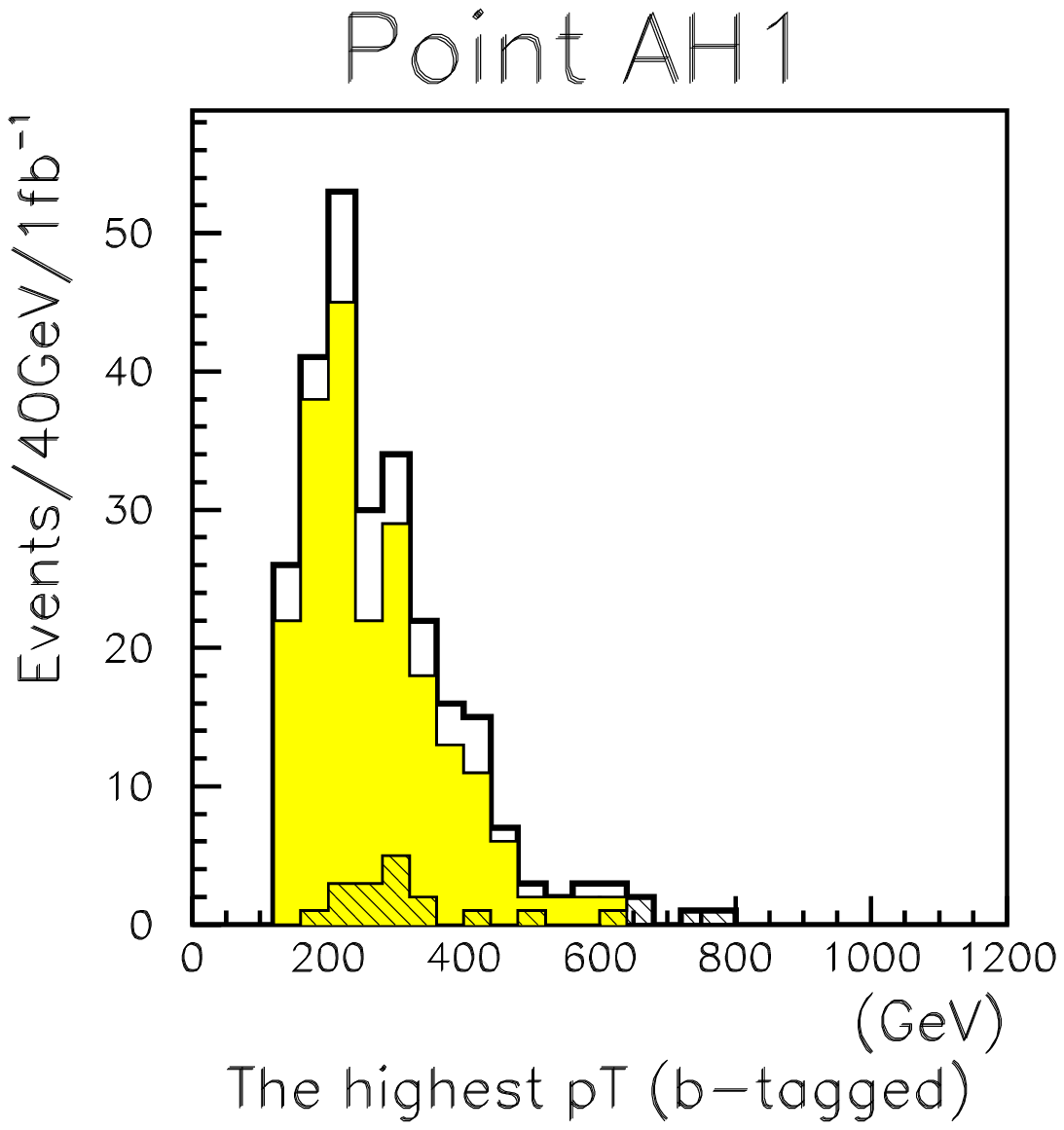}
\end{center}
\end{minipage}
\begin{minipage}{0.3\hsize}
\begin{center}
\includegraphics[width=3.8cm,clip]{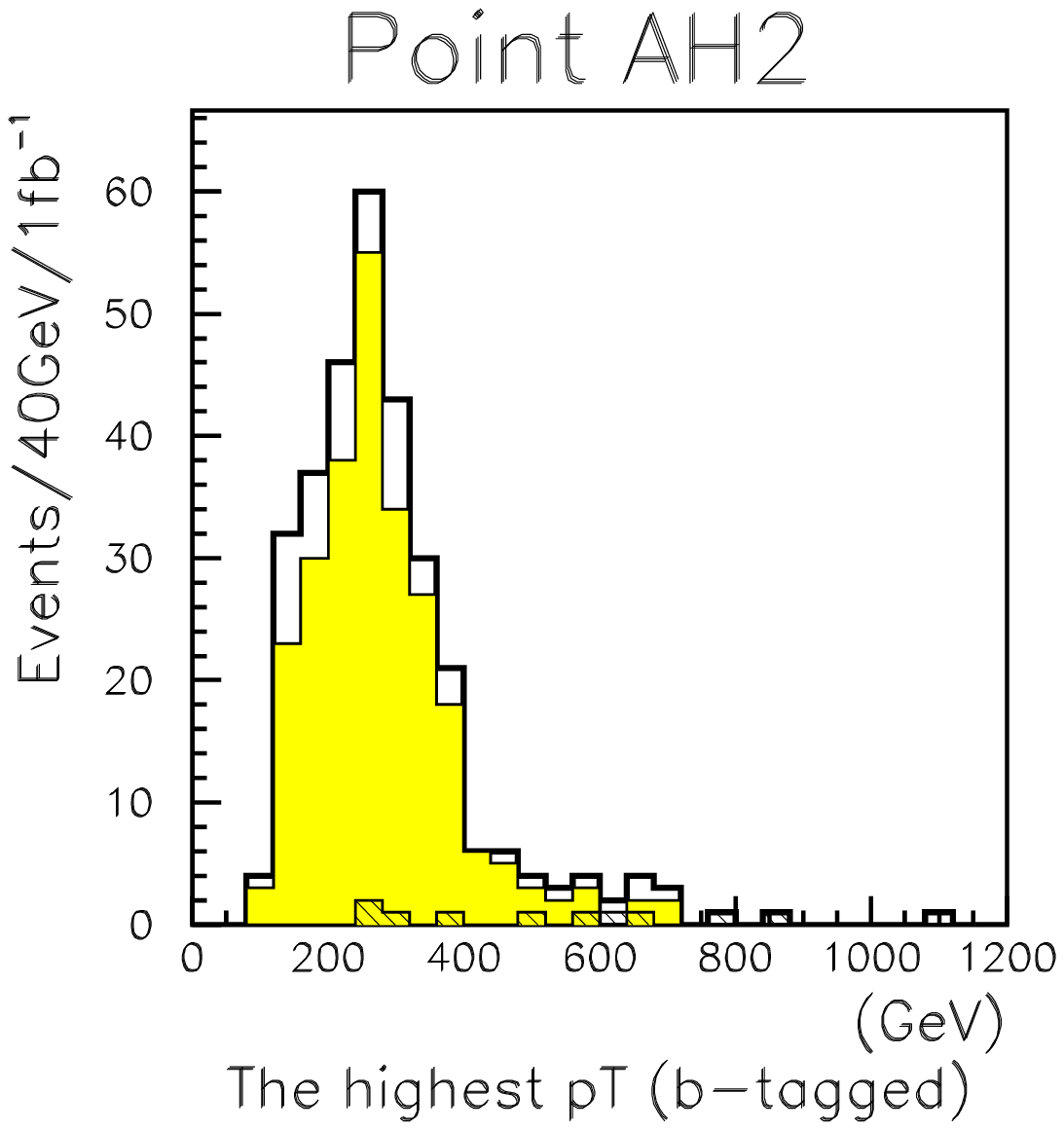}
\end{center}
\end{minipage}
\end{tabular}
~\\
\vspace{5mm}
\begin{tabular}{ccc}
\begin{minipage}{0.3\hsize}
\begin{center}
\includegraphics[width=3.8cm,clip]{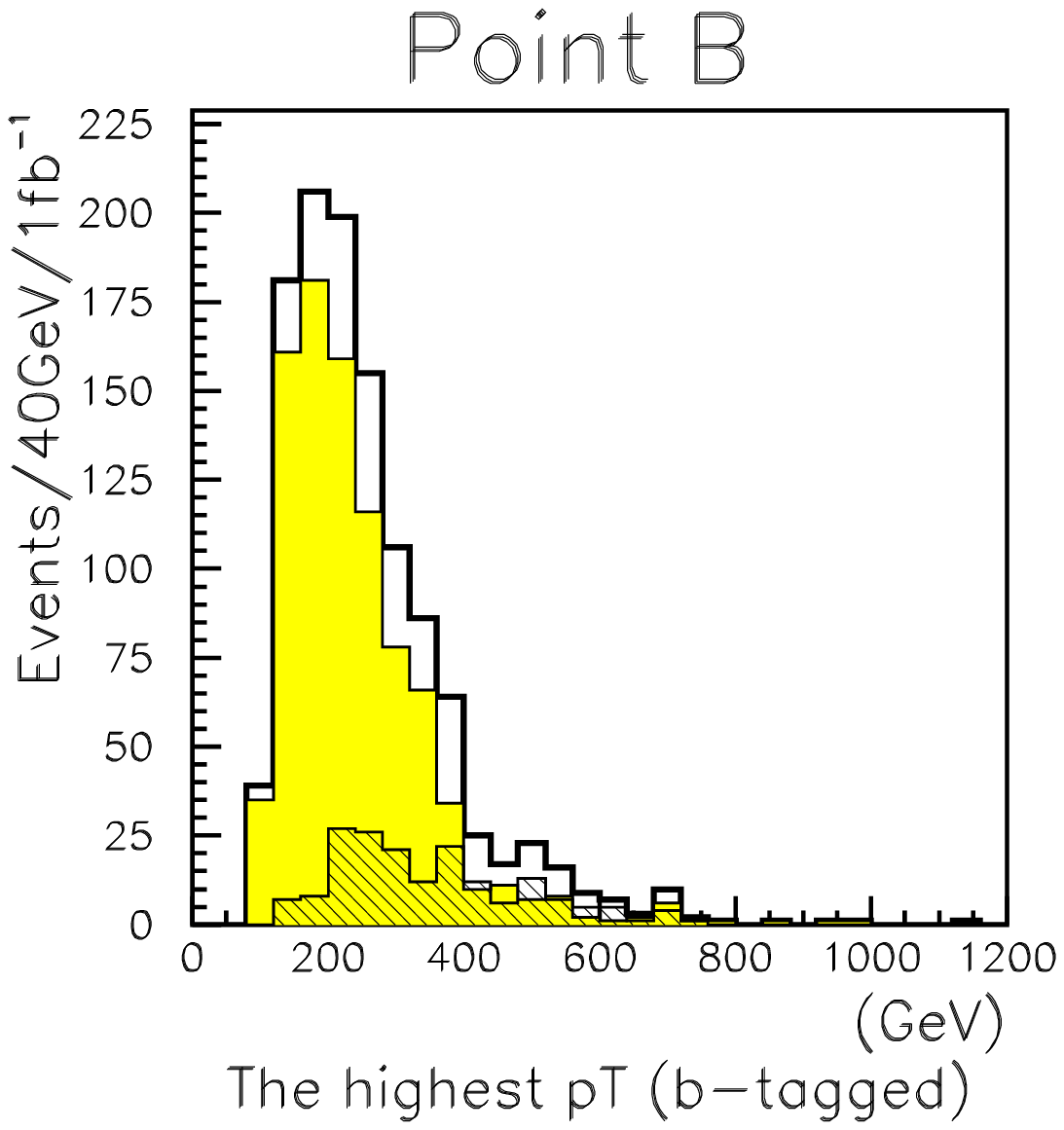}
\end{center}
\end{minipage}
\begin{minipage}{0.3\hsize}
\begin{center}
\includegraphics[width=3.8cm,clip]{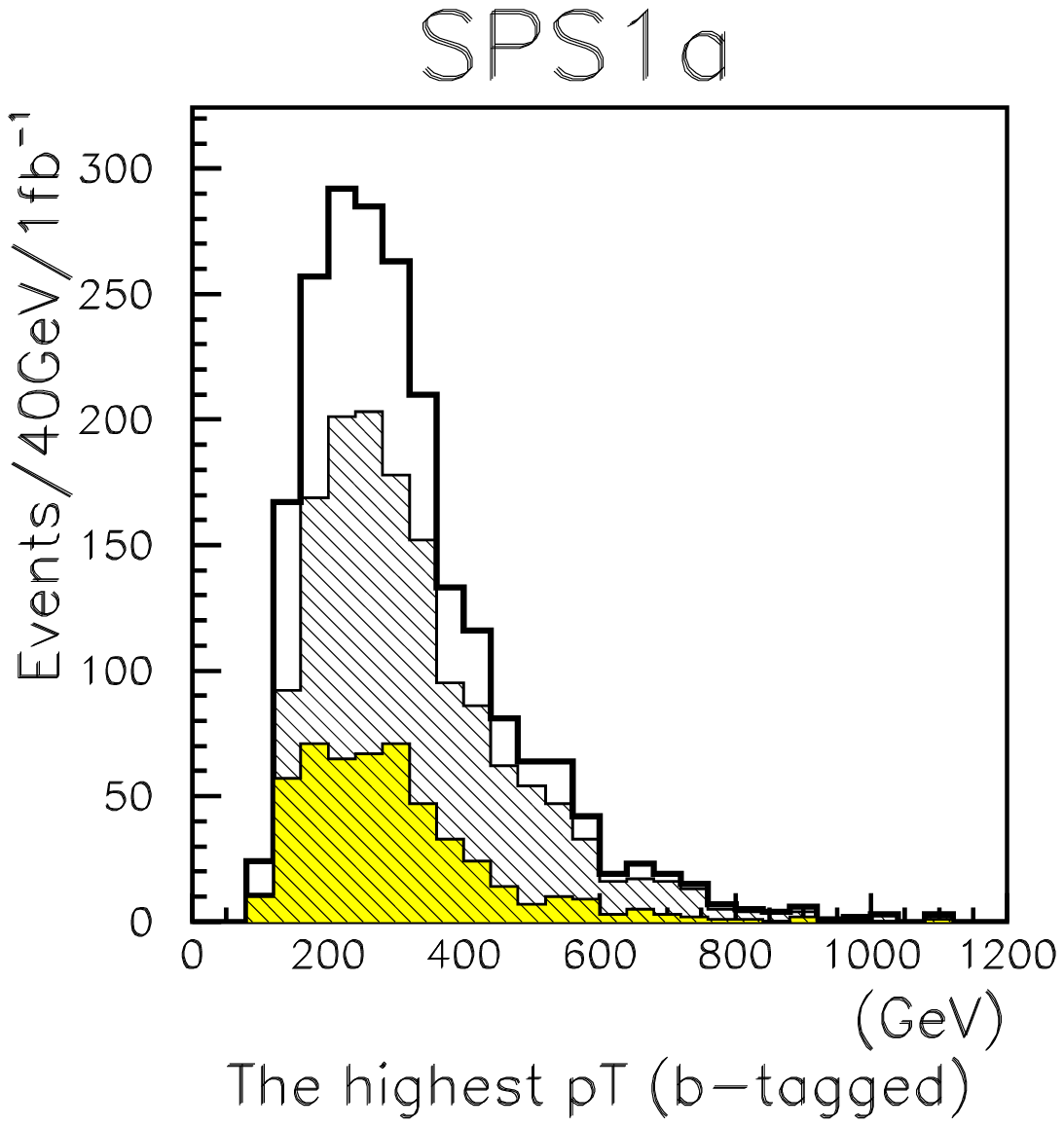}
\end{center}
\end{minipage}
\begin{minipage}{0.3\hsize}
\begin{center}
\includegraphics[width=3.8cm,clip]{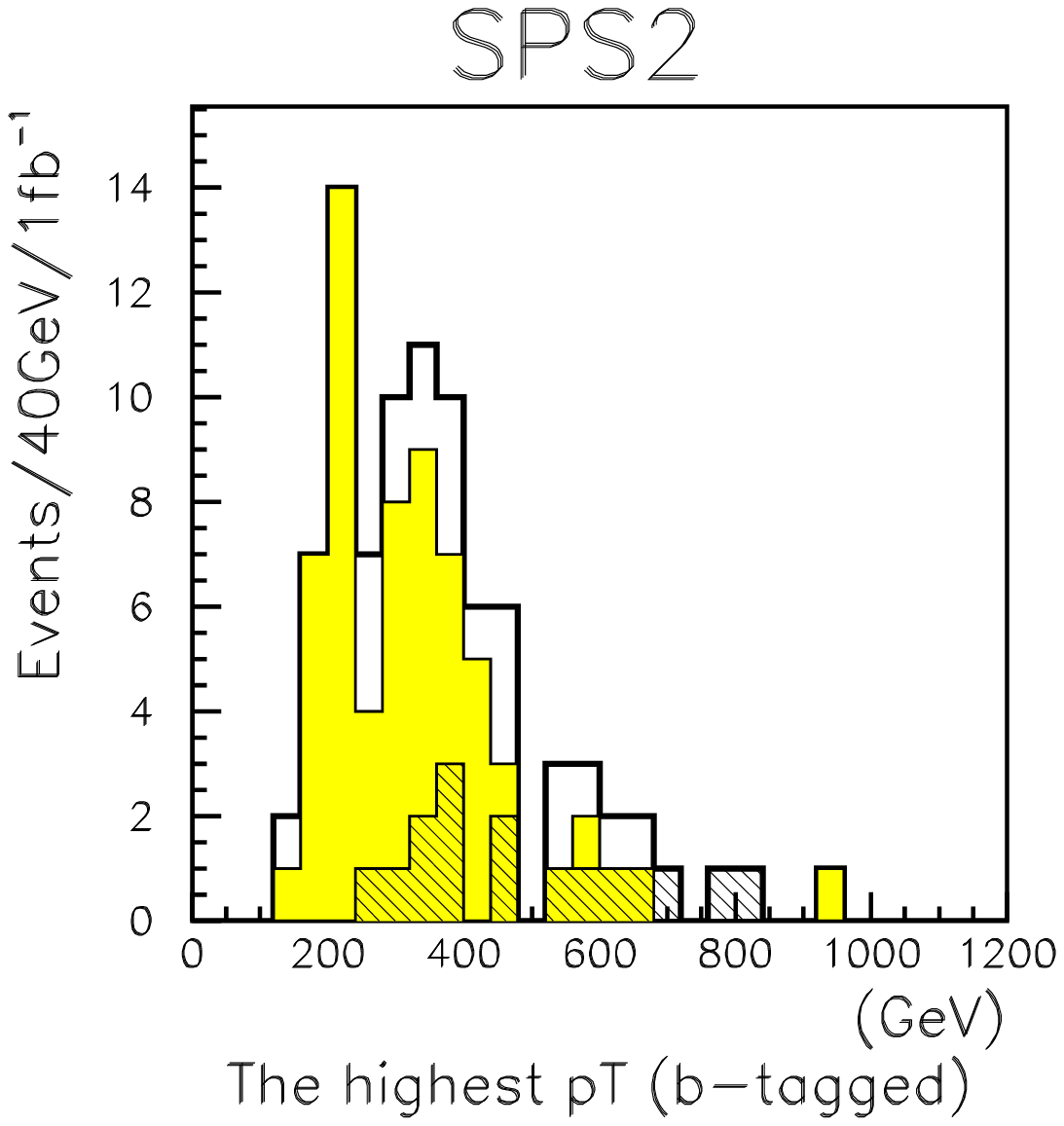}
\end{center}
\end{minipage}
\end{tabular}
\caption{\footnotesize{
The $p_T$ distribution of the $b$ tagged highest $p_T$ jet in the event at each model points.
}}
\label{hstptb}
\end{center}
\end{figure}

We show the $p_T$ distribution of the highest $p_T$ jet at each model point in Fig.\,\ref{hstpt}.
Here we require that the rapidity of the highest $p_T$ jet is less than 1.5 ($|\eta(j^{\rm 1st})|<1.5$)
and the jet is not $b$ tagged.   
The number of the generated events corresponds to $\int {\cal L} dt = 1\,{\rm fb^{-1}}$. 
In the figures the yellow histograms represent the contribution of  
$\ti g$-$\ti g$ and $\ti t$-$\ti t$ productions,
while the shaded histograms represent that of $\ti g$-$\ti q$ and 
$\ti q$-$\ti q$ productions.
The $p_T$ distribution of squark productions tends to be harder
than that of gluino pair production.
This is the contribution of the hard jet from the heavy squark decay.
Indeed, 
as $m_0$ increases ($m_0=1400$, 1700\,GeV at Points AH1 and AH2), 
the peak of the shaded distribution moves to the high energy side.
At the same time, the number of the events coming from squark production
decreases significantly.

The difference between the yellow and shaded distribution
indicates 
\begin{equation}
m_{\ti q} - m_{\ti g} \gg \max \{m_{\ti g} - m_{\ti t_1 (\ti b_1)} - m_{t(b)},\,m_{\ti t_1 (\ti b_1)} - m_{\none} \} ~,
\label{mass_relation2}
\end{equation} 
under the mass relation Eq.\,({\ref{mass_relation1}}).

In order to see the contributions from gluino pair production separately,
we can use the distribution of the $b$ tagged jets.
If the highest jet is a $b$ jet, the jet is not originated from the first two generation squark decay.
The contribution from the heavy squark decay can be removed 
by requiring that the highest $p_T$ jet is a $b$ jet.
In Fig.\,\ref{hstptb}, we show the $p_T$ distribution of the $b$ tagged highest $p_T$ jet.
The heavy squark contributions (the shaded histograms)  are significantly suppressed in the figures. 
The difference between the distributions of non-$b$ tagged jets and $b$ tagged jets
suggests the large mass splitting between $\ti g$ and $\ti q$.

For comparison,
we have done the same analysis for the benchmark 
model points SPS1a$-$SPS9 (See Appendix\,\ref{sps_appen}).
There is no clear difference between the ``$b$ tagged'' and ``non-$b$ tagged'' distributions except at SPS2.
At SPS2 the universal scalar mass, $m_0=1450$\,GeV, is much larger than 
the universal gaugino mass, $m_{1/2}=300$\,GeV.
So the decay of the 1st generation squarks produce the hard jet, 
while the gluino 3-body decays cannot produce hard $b$ jets.

\section{``Look alike" in CMSSM}
\label{alike_msugra}


The characteristic signatures of MUSM scenario 
in the number of $b$ tagged jets and the $p_T^{(1)}$ distributions
do not eliminate the possibility of the universal (CMSSM type) boundary condition 
of soft masses at the cutoff scale,
though they are good indication of MUSM scenario.
In this section we study a region of parameter space in CMSSM
whose signature is similar to that of MUSM scenario, 
and seek for the outstanding observable 
that is useful to distinguish MUSM scenario from the CMSSM parameter region.

If one takes $m_0 \gg m_{1/2}$ in CMSSM, 
the $p_T^{(1)}$ distributions of  $b$ tagged and non-$b$ tagged jets is similar to that of MUSM scenario
as we have seen at SPS2.
When one fixes $m_0$ and $m_{1/2}$,
CMSSM still has the other 2 free parameters $A_0$ and $\tan\beta$.
\begin{figure}[t!]
\begin{center}
\includegraphics[width=8cm,clip]{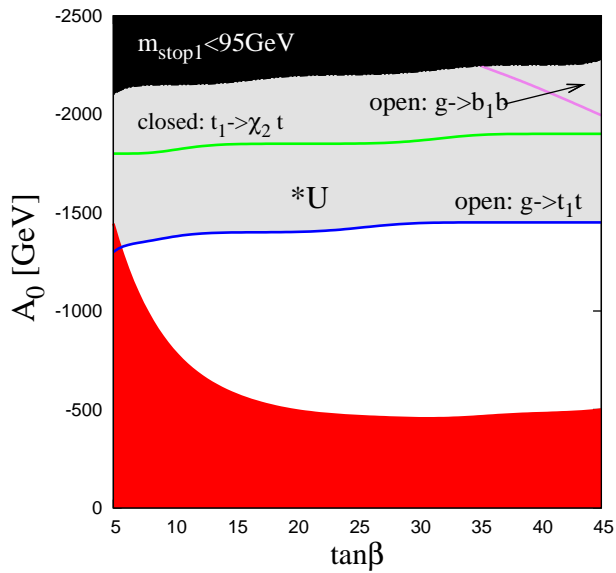}
\caption{\footnotesize{
The thresholds of the 2-body decays
on the ($\tan\beta-A_0$) parameter plane.
Other parameters are chosen as
$m_0=m_{H_u}(\Lambda)=m_{H_d}(\Lambda)=1000$\,GeV, $m_{1/2}=270$\,GeV, $\sgn(\mu)=+$.
}}
\label{a0_tanb}
\end{center}
\end{figure}
We scan ($\tan\beta-A_0$) parameter space 
with fixing $m_0=1$\,TeV, $m_{1/2}=270$\,GeV and sign$(\mu)=+$
in Fig.\,\ref{a0_tanb}.
In a gray region, $\ti g \to \ti t_1 t$ mode is open
and it entirely dominates the gluino decay.
Therefore there will be 4 $b$ partons in the  final state as in MUSM scenario. 
We choose such a model point U as a representative parameter point.  
The parameters are listed in Table\,\ref{point} and the sparticle masses are listed in Table\,\ref{mass}.

\begin{figure}[t!]
\begin{center}
\begin{tabular}{ccc}
\begin{minipage}{0.33\hsize}
\begin{center}
\includegraphics[width=4.0cm,clip]{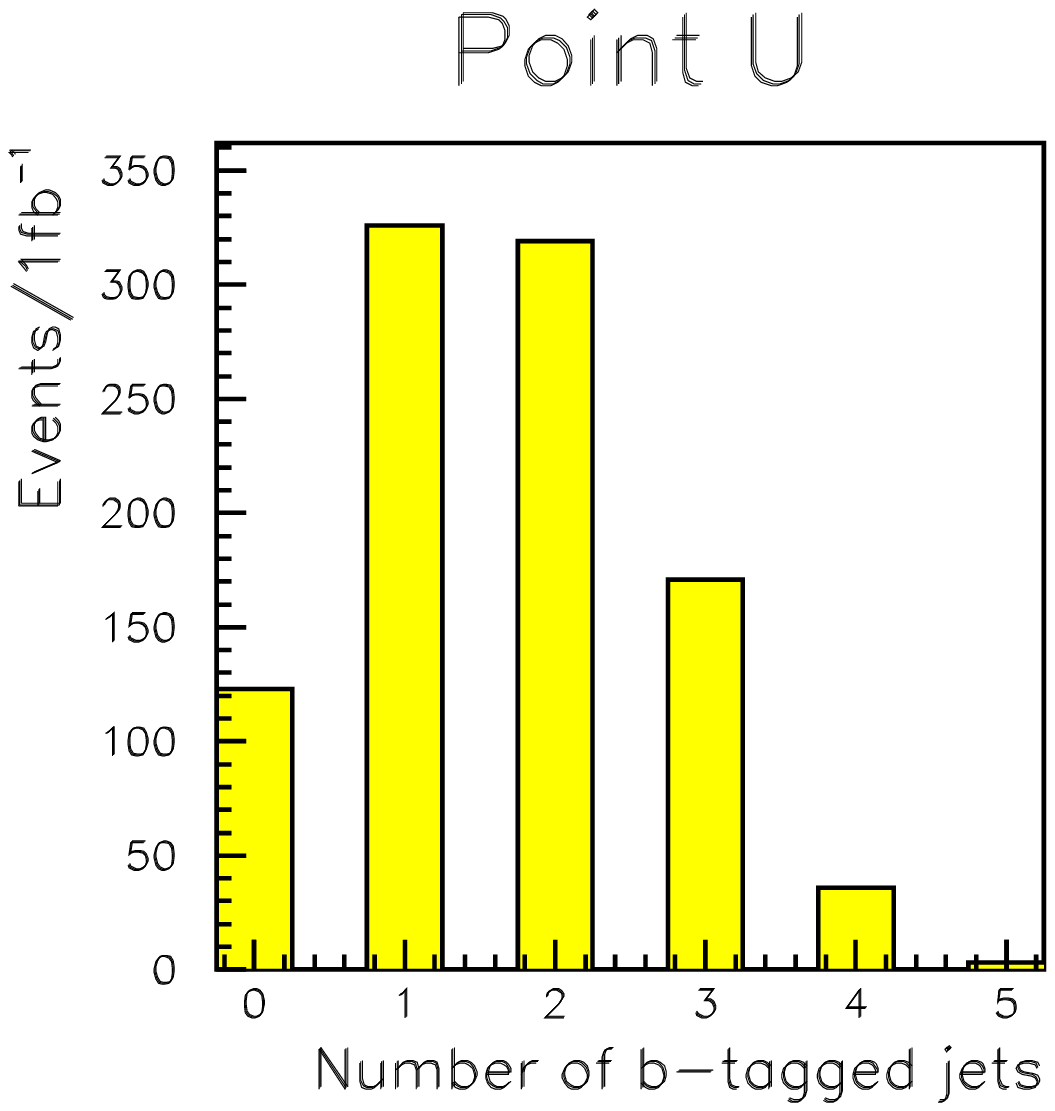}
\end{center}
\end{minipage}
\begin{minipage}{0.33\hsize}
\begin{center}
\includegraphics[width=4.0cm,clip]{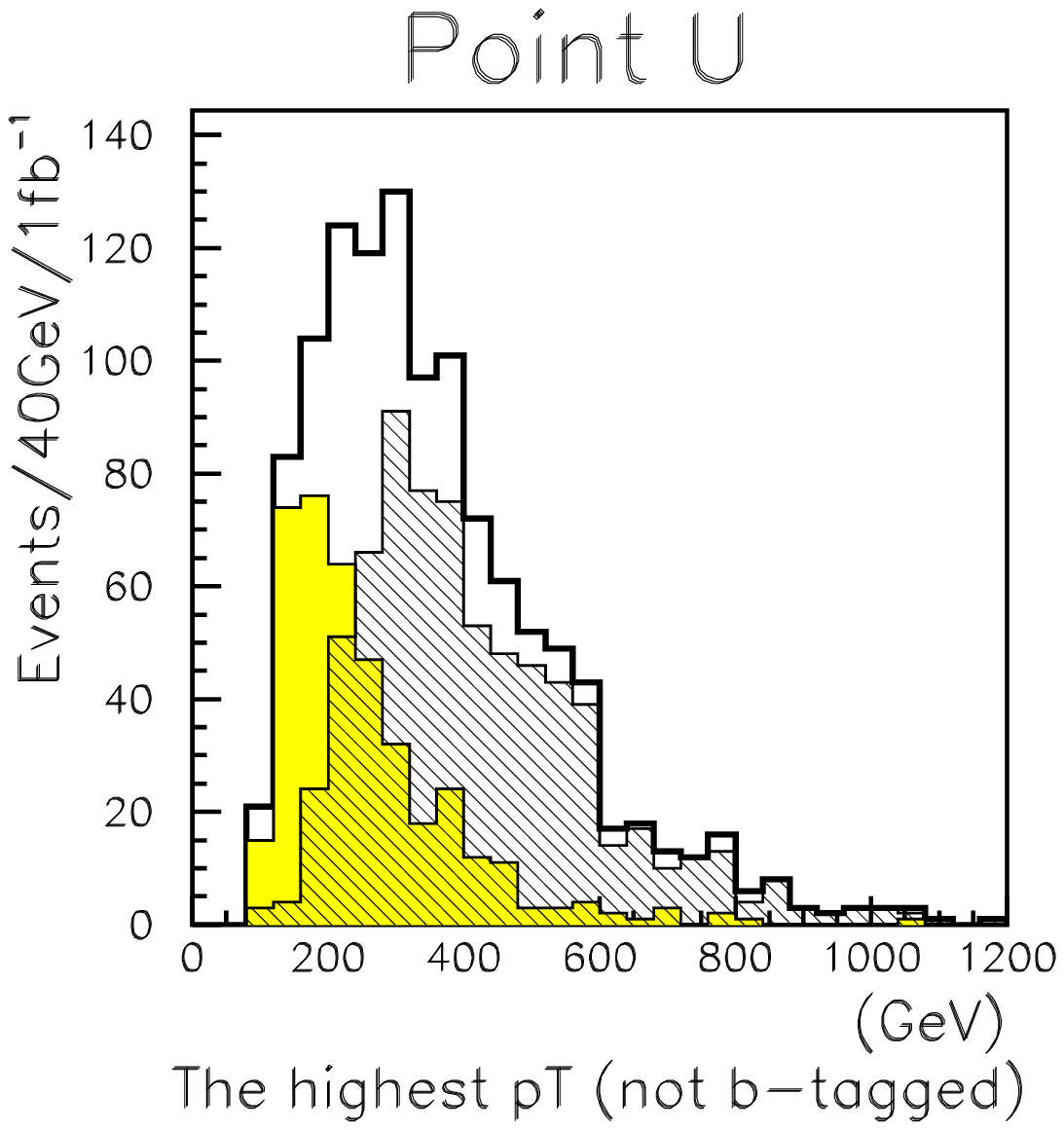}
\end{center}
\end{minipage}
\begin{minipage}{0.33\hsize}
\begin{center}
\includegraphics[width=4.0cm,clip]{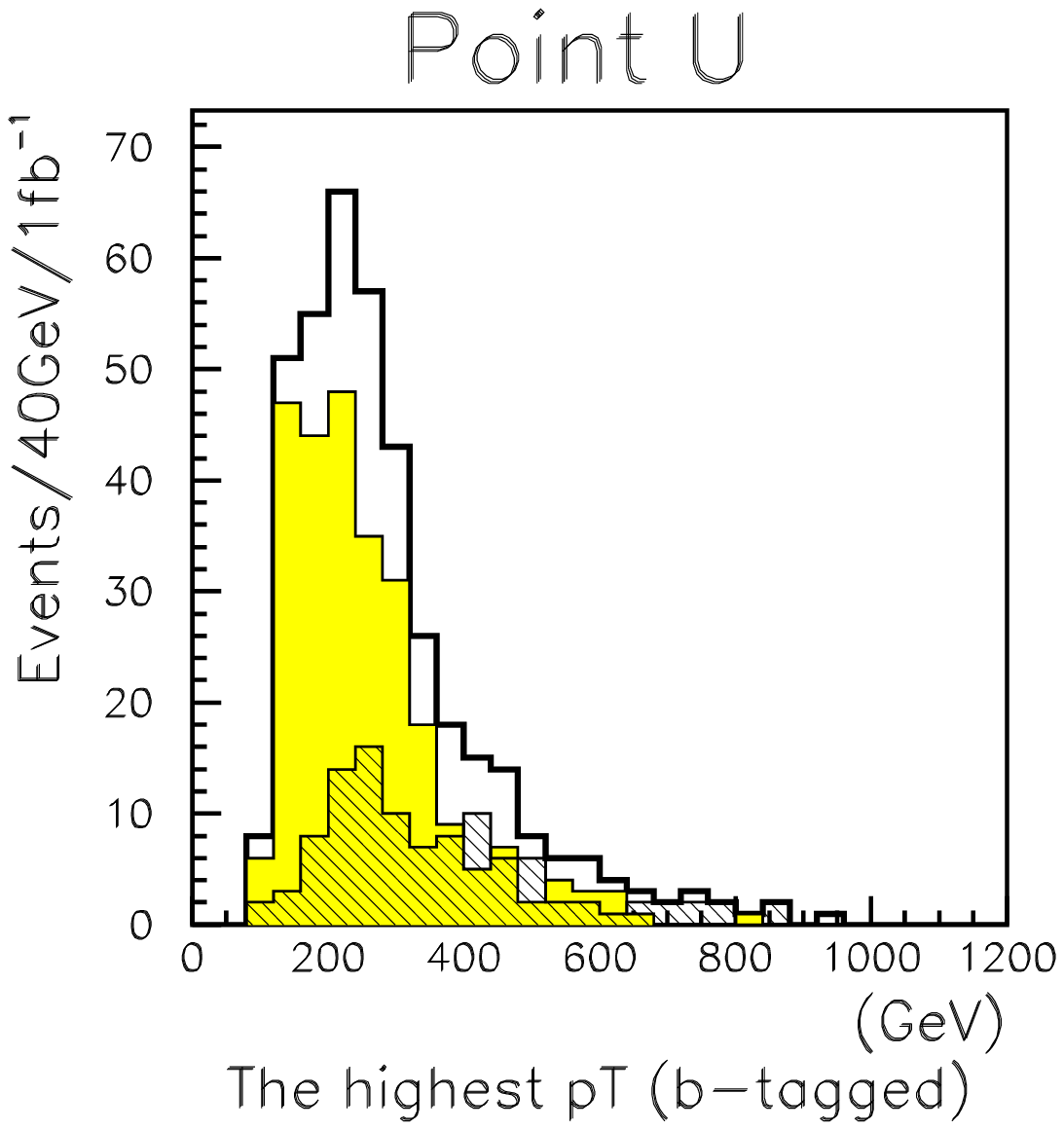}
\end{center}
\end{minipage}
\end{tabular}
\caption{\footnotesize{
The distributions of the number of $b$ jets  (Left) and 
the $p_T$ of the highest $p_T$ jet for non-b tagged(Center) and
$b$ tagged(Right) jets at point U. 
}}
\label{u_incl}
\end{center}
\end{figure}

We show the number of $b$ tagged jets and the $p_T^{(1)}$ distributions at Point U
in Fig.\,\ref{u_incl}.
The distributions are similar to those of MUSM scenario.
This means the analysis presented in the previous section
cannot discriminate MUSM scenario from the gray region in CMSSM,
though the soft masses are universal at the cutoff scale.
Therefore more information is required to distinguish these two scenarios. 

There are some differences in gluino decay branching ratios.
In CMSSM case, the gluino decay can be dominated only by $\ti g \to \ti t_1 t$ mode.
If one takes a large $\tan\beta$ along with a large $|A_0|$,
$\ti g \to \ti b_1 b$ mode is open as shown in Fig.\,\ref{a0_tanb}.
However $\ti t_1$ is much lighter than $\ti b_1$ in this region, 
and $Br(\ti g \to \ti b_1 b)$ cannot be significant.
On the other hand,
$Br(\ti g \to \ti b_1 b)$ can be as large as $Br(\ti g \to \ti t_1 t)$ in MUSM scenario, 
unless $\ti t_1$ is extremely lighter than $\ti b_1$ due to the large $|A_0|$ effect. 
Therefore we can regard a sizable  $Br(\ti g \to \ti b_1 b)$
as an indication of MUSM scenario.

Distinction of  $\ti g \to \ti b_1 b$ decay from $\ti g \to \ti t_1 t$ decay is not so easy at these model points.
A difficulty  comes from similarity of final states of these two decay modes.
The main decay chains of $\ti t_1$ and $\ti b_1$ are as follows.
\vspace{-5mm}
\begin{figure}[h!]
\begin{flushleft}
\begin{tabular}{rl}
\begin{minipage}{0.5\hsize}
\begin{eqnarray}
\ti t_1 &\to& \chi^{\pm}_1 b \to \none W^{(*)} b ~~\cdots (a) \nonumber\\
&\to& \none t \nonumber  ~~~~~~\,~~~~~~~~~\cdots (b) \\
&\to& \ntwo t \to \none Z^{(*)} t ~~\,~~\cdots (c) \nonumber 
\end{eqnarray}
\end{minipage}
\hspace{-5mm}
\begin{minipage}{0.5\hsize}
\begin{eqnarray}
\ti b_1 &\to& \ti t_1 W^{(*)} ~~~~~~~~~~~~~\cdots (d) \nonumber \\ 
&\to& \chi^{\pm}_1 t \to \none W^{(*)} t ~~\cdots(e)\nonumber \\
&\to& \ntwo b \to \none Z^{(*)} b ~~~\cdots (f)~~~~~~~
\label{tbdecay}
\end{eqnarray}
\end{minipage}
\vspace{-5mm}
\end{tabular}
\end{flushleft}
\end{figure}\\
%
Here we ignore $\ti b_1 \to \none b$ mode because the branching ratio is tiny 
due to the hypercharge and $U(1)_Y$ gauge coupling suppression if
$\ti b_1 \sim \ti b_L$ and the $\none$ is bino-like. 
In MUSM scenario or in the gray region of CMSSM,
$\ti \chi^{\pm}_1 \to \ti l \nu (\ti \nu l)$ and $\ntwo \to \ti ll (\ti \nu \nu)$ modes are not open 
due to a large $m_0$.
Therefore  we assume $\chi^{\pm}_1 \to \none W$ and $\ntwo \to \none Z$ modes are open 
and dominate $\chi^{\pm}_1$ and $\ntwo$ decay.
Except for the decay modes $(c)$ and $(f)$,
$\ti g \to \ti t t$ and $\ti g \to \ti b_1 b$ have the same final state $2W + 2b +\none$.

The decay modes $(c)$ and $(f)$ are useful although
their branching ratios are less than $10\%$ in a wide parameter region.
The final states of the gluino decay via the decay modes $(c)$ and $(f)$
are $2W+2b+Z+\none$ and $2b+Z+\none$, respectively.
If one of the $W$ bosons decays into jets,
the number of associate jets is differ by 2 between these modes.

\begin{figure}[t!]
\begin{center}
\includegraphics[width=5.0cm,clip]{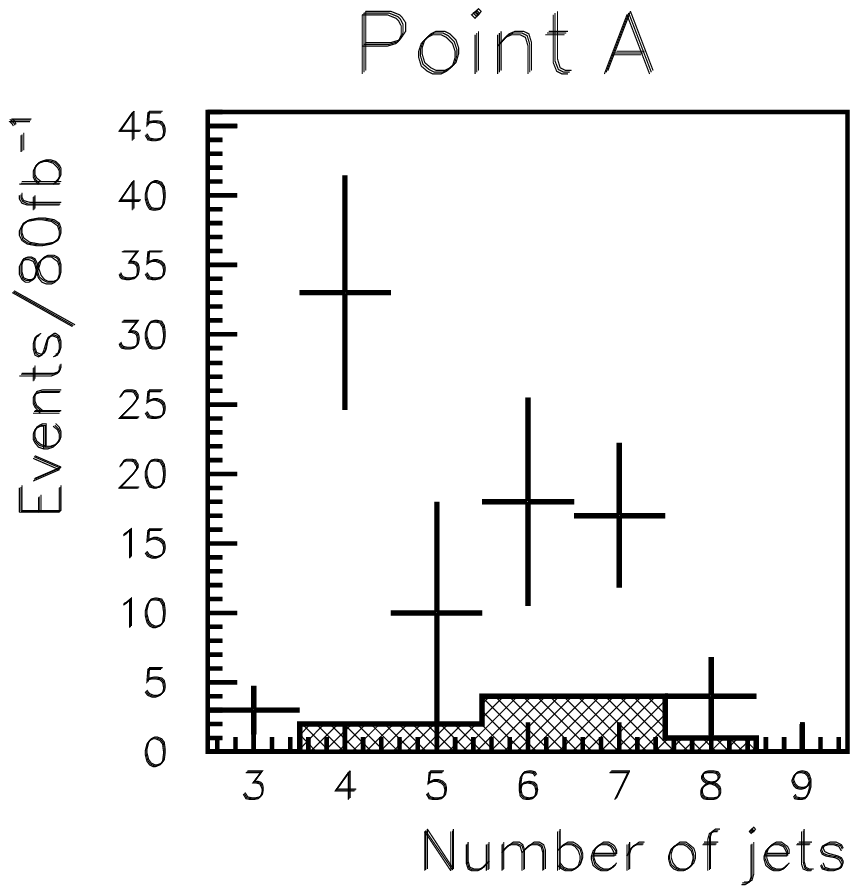}
\hspace{10mm}
\includegraphics[width=5.0cm,clip]{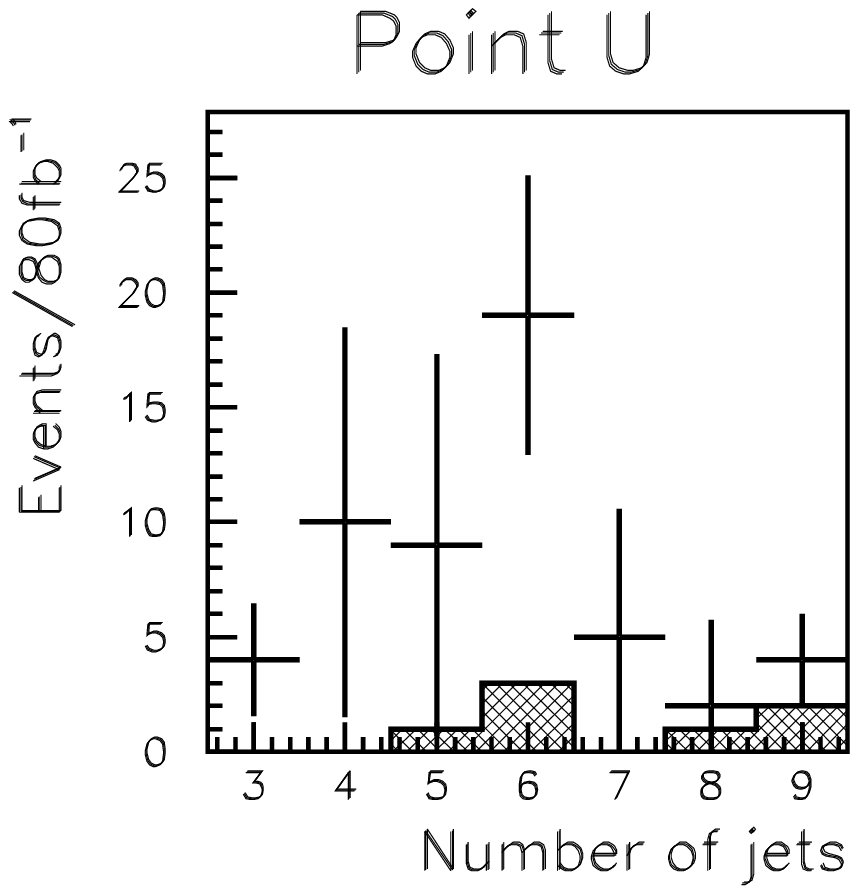}
\caption{\footnotesize{
The distributions of the number of jets in the event with $Z \to l_i^+ l_i^-$ at Point A and U.
}}
\label{z_njets}
\end{center}
\end{figure}

We show the distributions of the number of the jets with $p_T>50$\,GeV 
and $|\eta|<2.5$,  $(N^{\rm jets})$ 
in events with a $Z \to l^+_i l^-_i$ candidate at Point A and U in Fig\,\ref{z_njets}.
We require there are opposite sign same flavor (OSSF) lepton pairs 
with $p_T>15$\,GeV and $|\eta|<3$ 
whose invariant mass satisfy $|m_{ll}-m_Z|<5$\,GeV.\footnote{
If $\ntwo \to \none Z$ mode is not open, the distribution of two lepton invariant mass 
would show kinematical edge induced by $\ntwo \to \none l^+_i l^-_i$ mode 
as shown in Section 5.1.2.
In this case, we should require $m_{ll}^{\rm edge}-m_{ll}<10$\,GeV 
instead of $|m_{ll}-m_Z|<5$\,GeV.
}
To reduce an additional $Z$ boson source from $\ti q \to \ntwo q \to \none Z q$,
we adopt $\ti g$-$\ti g$ selection cuts: 
\begin{itemize}
\item The highest $p_T$ jet is $b$ tagged,
\item $p_T^{(1)}<300$\,GeV.
\end{itemize}
The shaded histograms in these figures
represent the contribution from the $\ti q$-$\ti g$ and $\ti q$-$\ti q$ production events.
The contamination from the squark decay $\ti q \to \ntwo q \to \none Z q$
is negligible after the $\ti g$-$\ti g$ selection cuts.

At Point A the peak is at $N^{\rm jets} = 4$,
while the peak is at $N^{\rm jets} = 6$ at Point U.
This suggests that the $Z$ boson comes from $\ti b_1$ decay at Point A, 
while it comes from $\ti t_1$ decay at Point U.
By this analysis, we can discriminate MUSM scenario from the CMSSM parameter region,
though a large integrated luminosity may be required.\footnote{ 
Note in the region above a green line in Fig.\,\ref{a0_tanb},
the decay mode ($c$) in Eq.\,(\ref{tbdecay}) is closed.
In this region, $Z$ boson does not come from $\ti g$ decay in CMSSM, 
so the discrimination from this region is easy.}

\section{The sparticle mass measurement}


\subsection{Exclusive analyses}

\subsubsection{Conditions to have  OSSF lepton pair signature in MUSM scanario}
SUSY particle masses may be determined by measuring the kinematical endpoints of 
sparticle decay products.
Especially, the distribution of the events with OSSF leptons
coming from $\ntwo \to \ti l^{\pm}_i l_i^{\mp} \to \none l^{\pm}_i l_i^{\mp}$ are useful
\cite{Hinchliffe:1996iu,Hinchliffe:1999zc,Bachacou:1999zb,Allanach:2000kt,Abdullin:1998pm,Atlas}.
However, there are several reasons for this channel may not be available in MUSM scenario.

In this scenario gluino entirely decays into the 3rd generation squarks,  
$\ti t_1$ and $\ti b_1$.
From a phase space consideration
$Br(\ti t_1 \to \cpl_1 b) \gg 2 Br(\ti t_1 \to \ntwo t)$ and
$Br(\ti t_1 \to \none t) \gg Br(\ti t_1 \to \ntwo t)$,
where the factor 2 in the first relation comes from the Dirac nature of the chargino.
Thus, $\ti t_1 \to \cpl_1 b$ and $\ti t_1 \to \none t$ dominate the stop decay.
If the gluino decay is dominated by $\ti g \to \ti t_1 t$ mode,
$\ntwo$  does not appear  with sufficiently high rate 
in the gluino cascade decays.

In this scenario the $\ti b_1$ branching ratio into $\ntwo$ may also be small, 
though gluino can decay into $\ti b_1 b$.   
Assuming $Y_b=0$, $\ntwo = \ti W_3$ and $\ti b_1 = \ti b_L$,
the tree level  formulae of the $\ti b_1 \to \ntwo b$ and $\ti b_1 \to \ti t_1 W$ decay widths 
are given by \cite{Bartl:1994bu}
\begin{eqnarray}
\Gamma(\ti b_1 \to \ntwo b) &\simeq& \frac{g^2 m_{\ti b_1}}{32 \pi} \Big( 1- \frac{m_{\ntwo}^2}{m_{\ti b_1}^2} \Big)^2,  \\
\Gamma(\ti b_1 \to \ti t_1 W) &\simeq& \frac{g^2 m_{\ti b_1}}{32 \pi} \over \lambda^{3/2} (x_{\ti t_1}, x_W )  
\Big( \frac{a_t^2 m^2_{\ti b_1}}{(m^2_{\ti t_L}-m^2_{\ti t_1})^2+ a_t^2 m_t^2} \Big) \Big( \frac{m^2_t}{m^2_W} \Big),
\label{decayrate}
\end{eqnarray}
where $x_{\ti t_1} = m^2_{\ti t_1}/m^2_{\ti b_1}$, $x_W = m^2_W/m^2_{\ti b_1}$ and
$\over \lambda(x,y)=1+x^2+y^2-2(x+y+xy)$ and $a_t = (A_t -\mu \cot\beta)$.
In Eq.\,(\ref{decayrate})  $\Gamma(\ti b_1 \to \ti t_1 W)$ has a enhancement factor $(m^2_t/m^2_W)$.
Thus, $\ti b_1$ decays dominantly into $\ti t_1 W$ unless it is kinematically suppressed.

In MUSM scenario, the first two generation sfermions are much heavier than the second
lightest neutralino.
Therefore the 2-body decay mode of $\ntwo$ into the first two generation sleptons 
$\ntwo \to \ti l^{\pm}_i l_i^{\mp}$ 
is kinematically forbidden. 
The OSSF leptons may arise from 
the 3-body decay mode $\ntwo \to \none l^{\pm}_i l^{\mp}_i$ from the off-shell $Z$ boson exchange.
If the 2-body decay mode $\ntwo \to \none Z$ is open, it dominates the $\ntwo$ decay.

In summary,  the sparticle mass measurement 
using the exclusive analysis with OSSF leptons may  work  if  following conditions are satisfied:
\begin{equation}
\arraycolsep=1.6pt
\def\arraystretch{1.5}
\begin{array}{rll}
\text{\bf (i) } & & \ti g \to \ti t_1 t \text{~mode does not dominate the gluino decay.  } \\
\text{\bf (ii) } & & \ti b_1 \to \ti t_1 W \text{~mode is kinematically suppressed.}\\ 
\text{\bf (iii) }& & \ntwo \to \none Z \text{~mode is kinematically forbidden.}
\end{array}
\label{3cond}
\end{equation}
The three conditions can be satisfied when the mass differences
$m_{\ti b_1} - m_{\ti t_1}$ and $m_{\ntwo} -m_{\none}$ are small,
and it can be realized for small $m_{1/2}$ and $|A_0|$.
If all conditions are satisfied, a gluino cascade chain,
\begin{equation}
\ti g \to \ti b_1 b^{(1)} \to \ntwo b^{(1)} b^{(2)} \to \none b^{(1)} b^{(2)} l^+_i l^-_i,
\label{sbdc}
\end{equation}
can have enough branching fraction.
Here subscripts of the $b$ quarks in Eq.\,(\ref{sbdc}) 
are used to distinguish two $b$ quarks appear subsequently.

The low energy mass spectra and 
the decay branching ratios of various sparticles at Points A and B are shown in 
 Table\,\ref{mass_spectrum} in Appendix\,\ref{masses}.
At Point A, 
the three conditions in Eq.\,(\ref{3cond}) are not satisfied. 
On the  other hand, at point B all of the three conditions are satisfied 
due to the very low SUSY breaking scales,
$m_{1/2}=m_{30}=200$\,GeV and $A_0=0$\,GeV. 

\subsubsection{The exclusive analysis using $\ntwo \to \none l_i^{+} l_i^{-}$ channel at Point B}

\begin{figure}[t!]
\begin{center}
\includegraphics[width=5.0cm,clip]{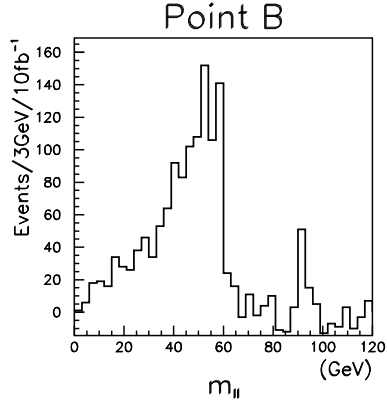}
\caption{\footnotesize{
The invariant mass distributions of $l^+_i l^-_i$.
The background distributions estimated from OSOF leptons are subtracted.
The number of events corresponds to $\int {\cal L} dt =$10\,${\rm fb^{-1}}$.
 }}
\label{end:ll}
\end{center}
\end{figure}

At Point B, SUSY events contain the cascade decay chain (\ref{sbdc}).
The OSSF lepton pair from the decay chain is relatively clean signal 
\cite{Hinchliffe:1996iu,Hinchliffe:1999zc,Bachacou:1999zb,Allanach:2000kt,Abdullin:1998pm,Atlas}. 
The background distributions of fake OSSF leptons can be estimated by distributions of OSOF leptons.
The kinematical maximum of the two lepton invariant mass coming from $\ntwo \to \none l^+_i l^-_i$ at Point B 
is given by
\begin{equation}
m_{l_i^+ l^-_i}^{\rm max} = m_{\ntwo} - m_{\none} = 61\,{\rm GeV}.~
\label{end0}
\end{equation}
The distribution is shown in Fig\,\ref{end:ll}.
Throughout this section, leptons are required to satisfy $p_T>15$\,{\rm GeV} and $|\eta|<3$.

For the decay chain (\ref{sbdc}), the invariant mass distributions
$m_{bb}$, $m_{l_i^+ l_i^- b^{(1)}}$, $m_{l_i^+ l_i^- b^{(2)}}$ and $m_{bbl_i^+ l^-_i}$
may determine sparticle masses.
Their kinematical maxima at Point B are given as
\begin{equation}
m^{\rm max}_{b b} =
m_{\gluino} \sqrt{\Big(1- \frac{m^2_{\ti b_1}}{m^2_{\gluino}} \Big)
\Big(1- \frac{m^2_{\ntwo}}{m^2_{\ti b_1}} \Big) }
= 336\,{\rm GeV},
\label{end1}
\end{equation}
\begin{eqnarray}
m^{\rm max}_{l^+_i l^-_i b^{(1)}} &=&
m_{\gluino} \sqrt{\Big(1- \frac{m^2_{\ti b_1}}{m^2_{\gluino}} \Big)
\Big(1- \frac{m^2_{\none}}{m^2_{\ntwo}} \Big) }
=296\,{\rm GeV}, 
\label{end2} \\
m^{\rm max}_{l^+_i l^-_i b^{(2)}} &=& m_{\ti b_1} - m_{\none} 
= 322\,{\rm GeV}, 
\label{end3}
\end{eqnarray}
\begin{eqnarray}
m^{\rm max}_{l^+_i l^-_i bb} = m_{\ti g} - m_{\none} 
= 460\,{\rm GeV}.
\label{end4}
\end{eqnarray}

Since there are typically 4 $b$ partons in the SUSY events,
the invariant mass distributions suffer from the combinatorial background. 
To reduce the background, we adopt hemisphere method
\cite{hemi,Matsumoto:2006ws}.
The method divides jets and leptons from cascade decay chains into two groups called hemispheres
whose entries are much likely to originate from the same mother particle.
The groups are defined by hemisphere momenta,
\begin{equation}
p^{(1)}_{\rm hemi} = \sum_{i} p_i^{(1)},~~~p^{(2)}_{\rm hemi} = \sum_{i} p_i^{(2)},
\label{p_hemi}
\end{equation}
where  $p_i^{(1)}$  and $p_i^{(2)}$ are jet or lepton momenta that satisfy
\begin{equation}
d(p^{(1)}_{\rm hemi},p_i^{(1)}) < d(p^{(2)}_{\rm hemi},p_i^{(1)}),~~~~~~
d(p^{(2)}_{\rm hemi},p_i^{(2)}) < d(p^{(1)}_{\rm hemi},p_i^{(2)}).
\label{select_near}
\end{equation}
Here the function $d$ is defined by
\begin{eqnarray}
d(p_{\rm hemi}^{(i)},p_k) = (E_{\rm hemi}^{(i)}-|p_{\rm hemi}^{(i)}|\cos\theta_{ik})\frac{E^{(i)}_{\rm hemi}}{(E^{(i)}_{\rm hemi}+E_k)^2},
\label{distance}
\end{eqnarray}
where $\theta_{ik}$ is an angle between $p^{(i)}_{\rm hemi}$ and $p_k$.
We require that jets involved in hemisphere satisfy $p_T > 50$\,GeV and
$|\eta|<2.5$ to reduce the contamination of soft jets.
To find the two groups of jets and leptons that satisfy Eq.\,(\ref{p_hemi}),  we first choose the highest $p_T$ jet and
the jet with the largest $p_T \Delta R$ as $p_1{\rm (seed)}$ and $p_2{\rm (seed)}$,
where $\Delta R=\sqrt{\Delta \phi^2 + \Delta \eta^2}$ is the angle difference between the jet and the highest jet.
Next we group jets into two groups 
under the condition $d(p^{(1)}({\rm seed }) ,p_i^{(1)}) < d(p^{(2)}{(\rm seed)},p_i^{(1)})$ etc.. 
Then axis momenta are defined as  
$p^{(1)}{\rm (ax)}=  \sum p_i^{(1)}$, $p^{(2)}{\rm (ax)}=  \sum p_i^{(2)}$.
For the axis momenta, new groups are defined so that 
$d(p^{(1)}({\rm ax }) ,p_i^{(1)}) < d(p^{(2)}{(\rm ax)},p_i^{(1)})$ etc.. 
The procedure is iterated several times so that assignment converges.

Although the probability that hemisphere correctly reconstructs original cascade chains is not so high,
this method has an advantage for endpoint analyses.
In the algorithm, two objects whose momentum directions are roughly 
the same each other tend to be in the same hemisphere 
due to Eqs.\,(\ref{select_near}) and (\ref{distance}). 
Because of this property, 
any invariant mass distribution of jets and leptons in the same hemisphere tends to be lower than
that in the different hemispheres. 
Therefore wrong combinations whose invariant mass exceed the signal endpoint
are removed with high probability if we take jet pairs in the same hemisphere.

\begin{figure}[t!]
\begin{center}
\begin{tabular}{lcr}
\begin{minipage}{0.3\hsize}
\includegraphics[width=4.7cm,clip]{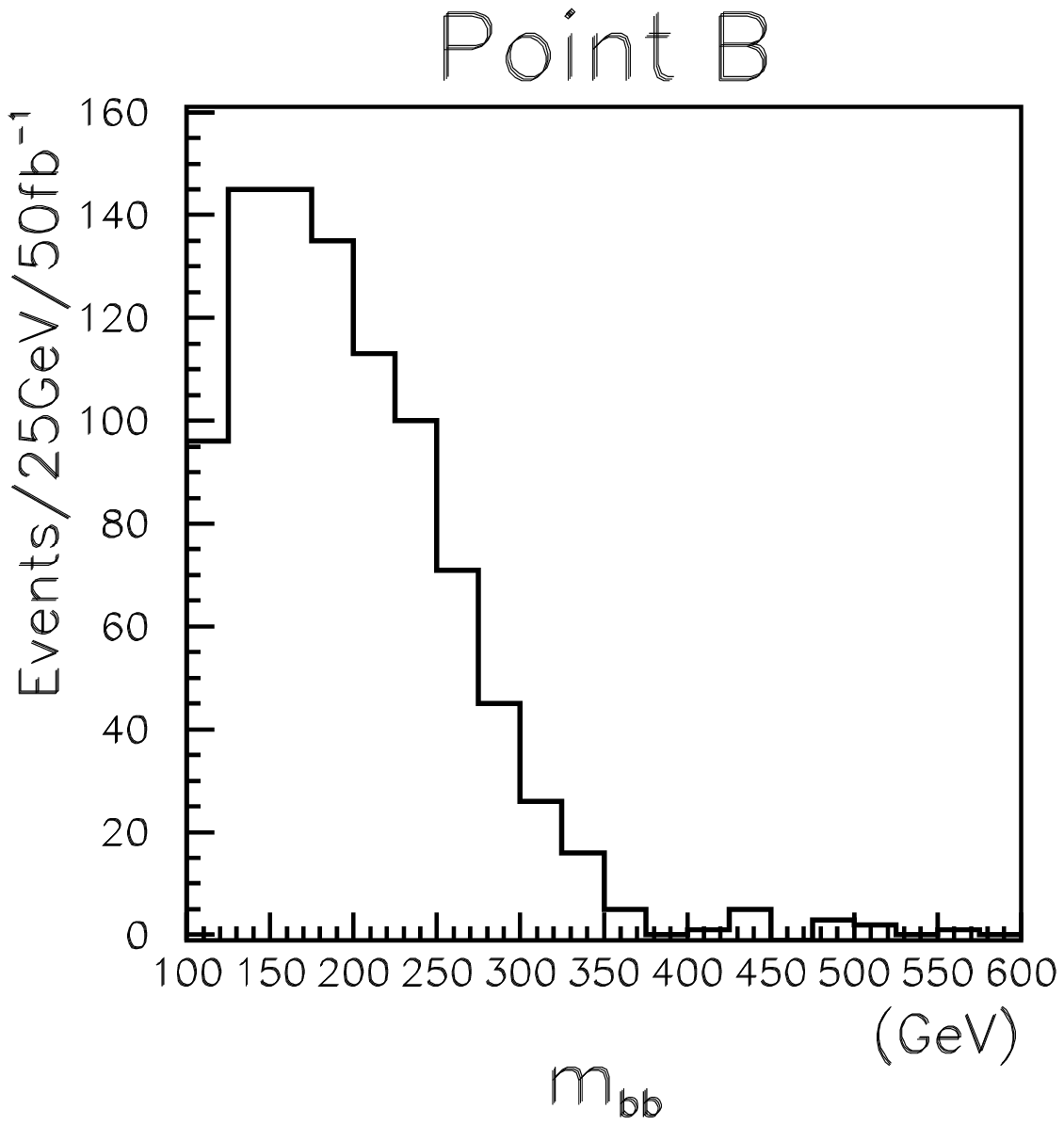}
\end{minipage}
\hspace{1mm}
\begin{minipage}{0.3\hsize}
\includegraphics[width=4.7cm,clip]{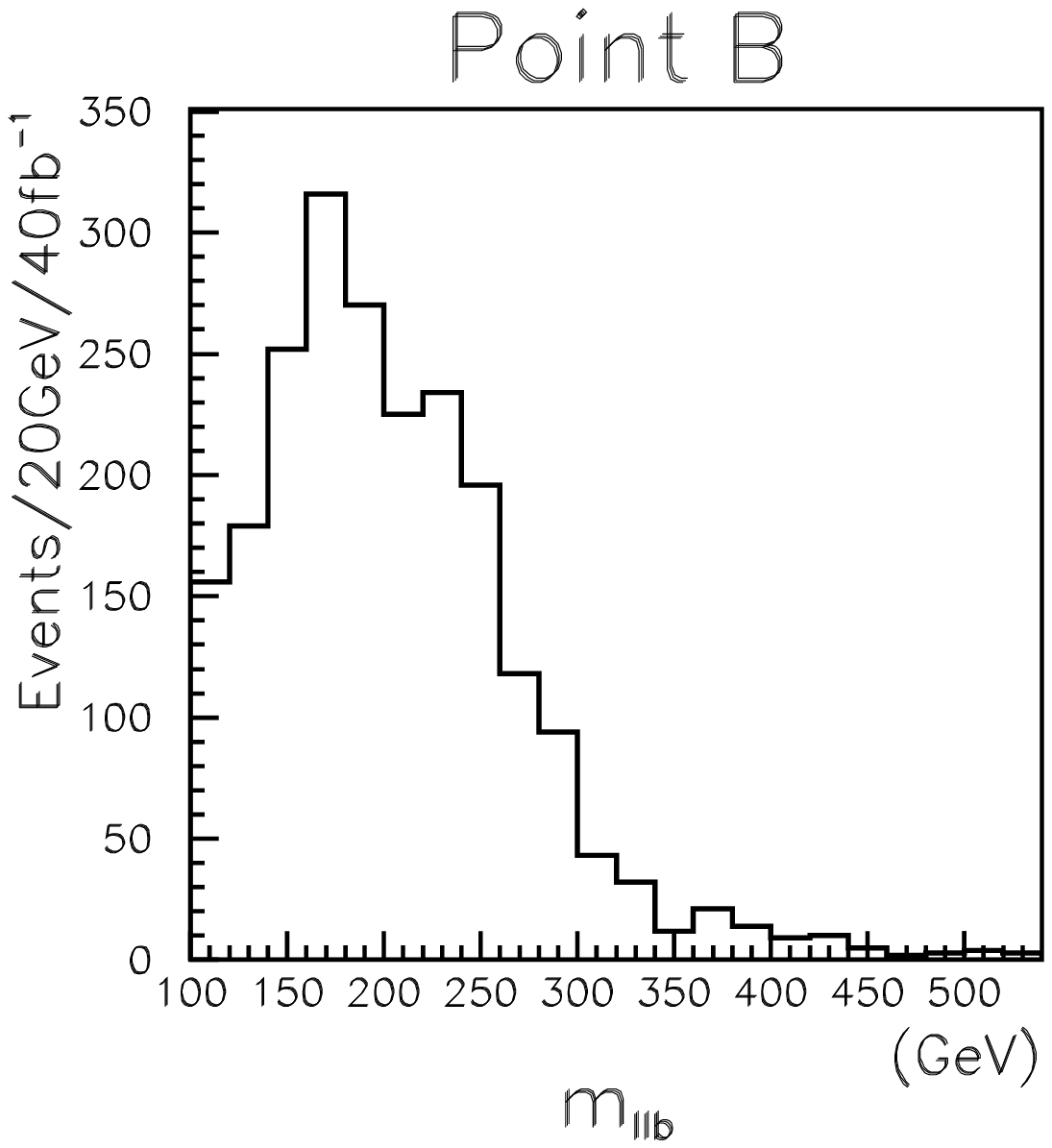}
\end{minipage}
\hspace{1mm}
\begin{minipage}{0.3\hsize}
\includegraphics[width=4.7cm,clip]{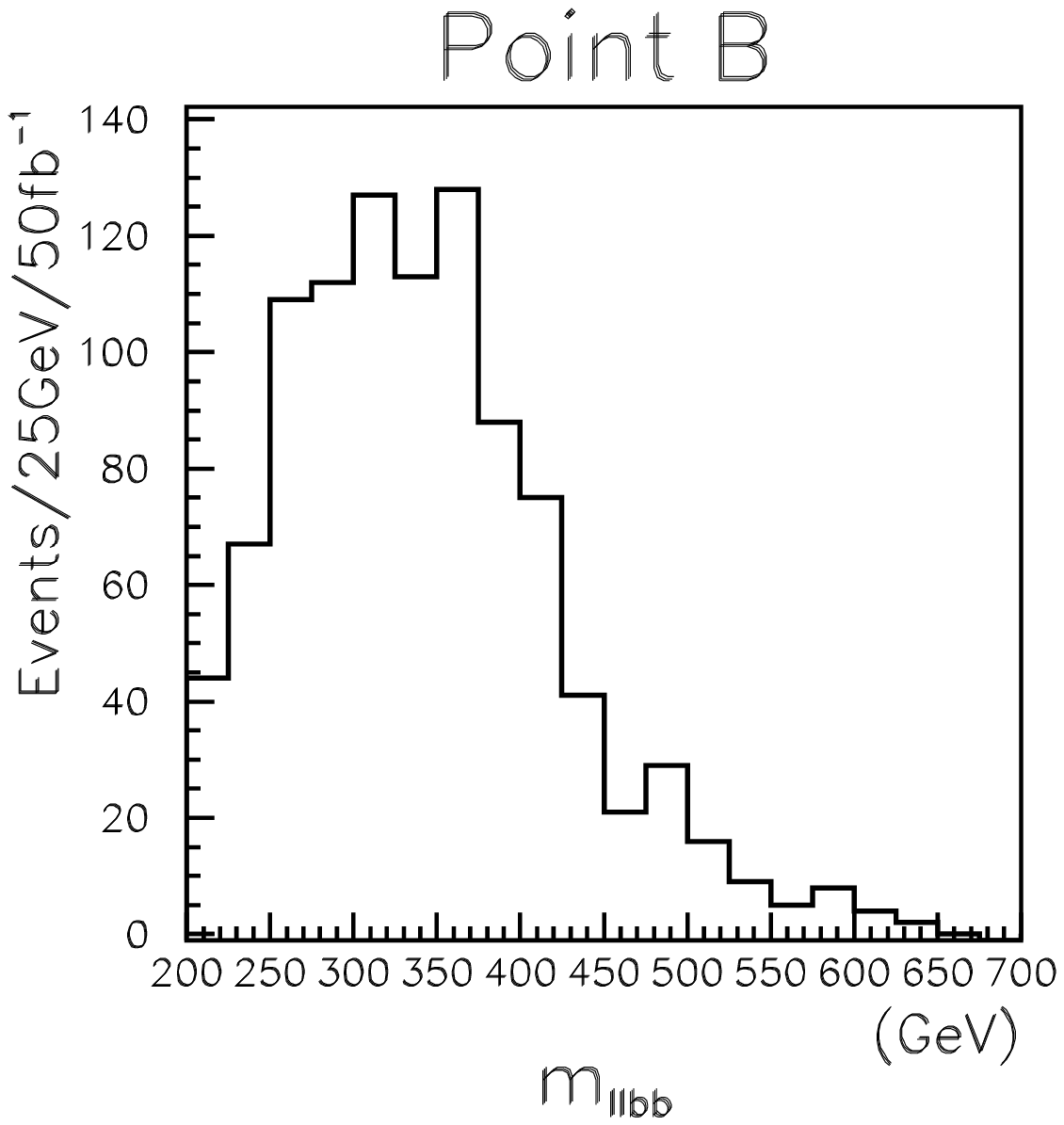}
\end{minipage}
\end{tabular}
\caption{\footnotesize{
The invariant mass distributions of $bb$ (Left), $b l^+_i l^-_i$ (Center) and $l_i^+ l_i^- bb$ (Right).
The background distributions estimated from OSOF leptons are subtracted.
The corresponding luminosities are $\int {\cal L} dt =$50, 40 and 50\,${\rm fb^{-1}}$ 
for the left, center and right distributions.
 }}
\label{end:ll_llb}
\end{center}
\end{figure}

We show the invariant mass distributions of $bb$, $l^+_i l^-_i b$ and $l^+_i l^-_i bb$ in Fig.\,\ref{end:ll_llb}.
Here all $b$ jets and leptons are required to be in the same hemisphere.
If there are more than one or two $b$ jet candidates, 
we take high $p_T$ $b$ jets.
We also require $m_{l^+_i l^-_i} \le m^{\rm max}_{l^+_i l^-_i}$
to reduce background OSSF leptons from $Z$ boson decay.
The distributions have endpoints near the theoretical expected values shown in Eq.\,(\ref{end1}) to (\ref{end4}).  
However $bb$ distribution does not show expected sharp edge structure due to the hemisphere selection.

At this stage, we have 4 measured values $m_{l^+_i l^-_i}^{\rm max}$, $m_{bb}^{\rm max}$, 
$m_{l^+_i l^-_i b}^{\rm max}$ and $m_{l^+_i l^-_i bb}^{\rm max}$ 
 for 4 unknown sparticle masses
$m_{\none}$, $m_{\ntwo}$, $m_{\ti b_1}$ and $m_{\ti g}$.  
By solving 4 equations, (\ref{end0}), (\ref{end1}), (\ref{end3}) and (\ref{end4}),
we can get  
all sparticle masses appeared in the decay chain (\ref{sbdc})
in the stage around $50$\,${\rm fb^{-1}}$ at Point B.

\begin{figure}[t!]
\begin{center}

\begin{tabular}{cc}

\begin{minipage}{0.4\hsize}
\begin{center}
\includegraphics[width=5.0cm,clip]{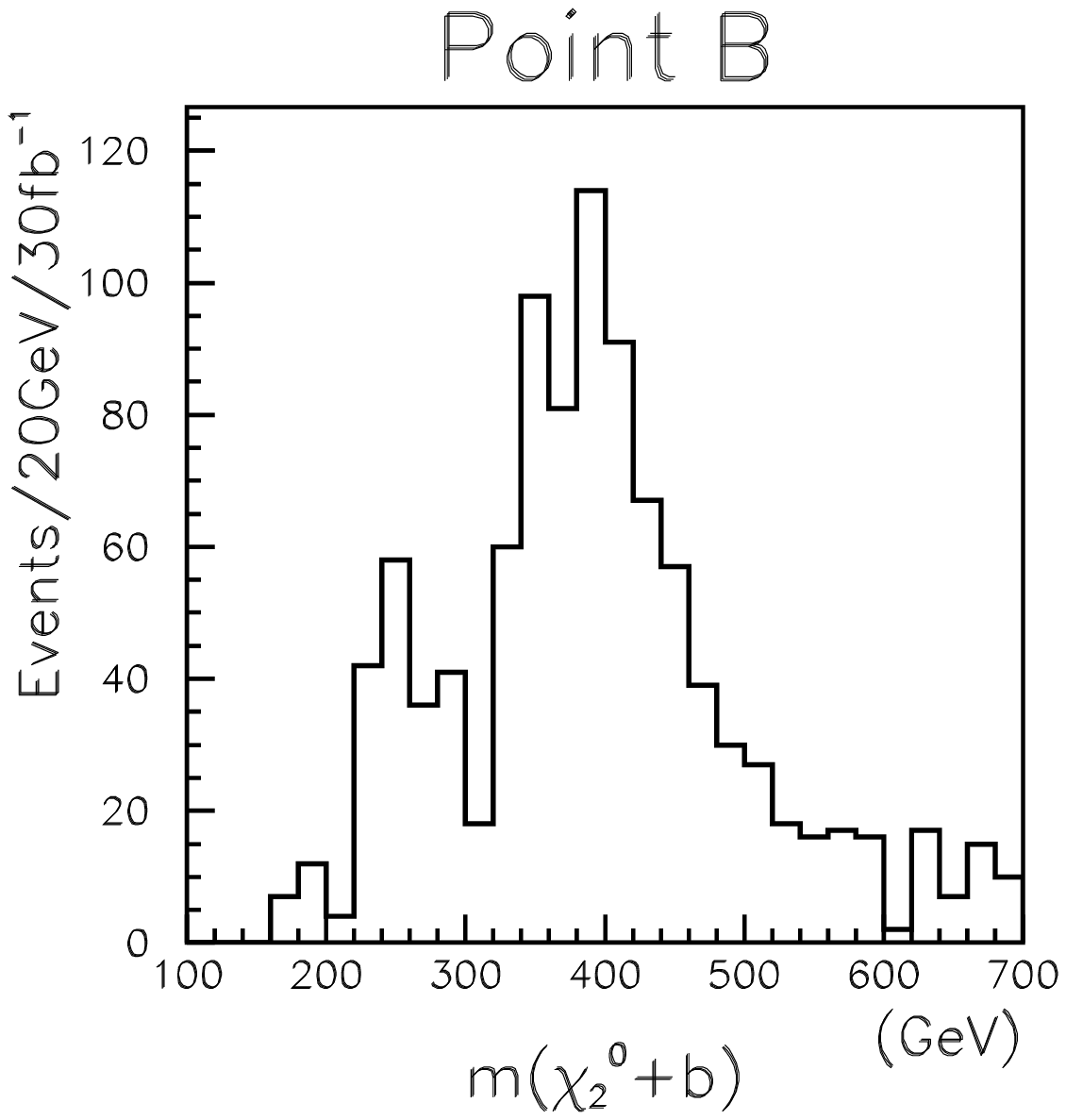}
\end{center}
\end{minipage}

\hspace{0mm}

\begin{minipage}{0.4\hsize}
\begin{center}
\includegraphics[width=5.0cm,clip]{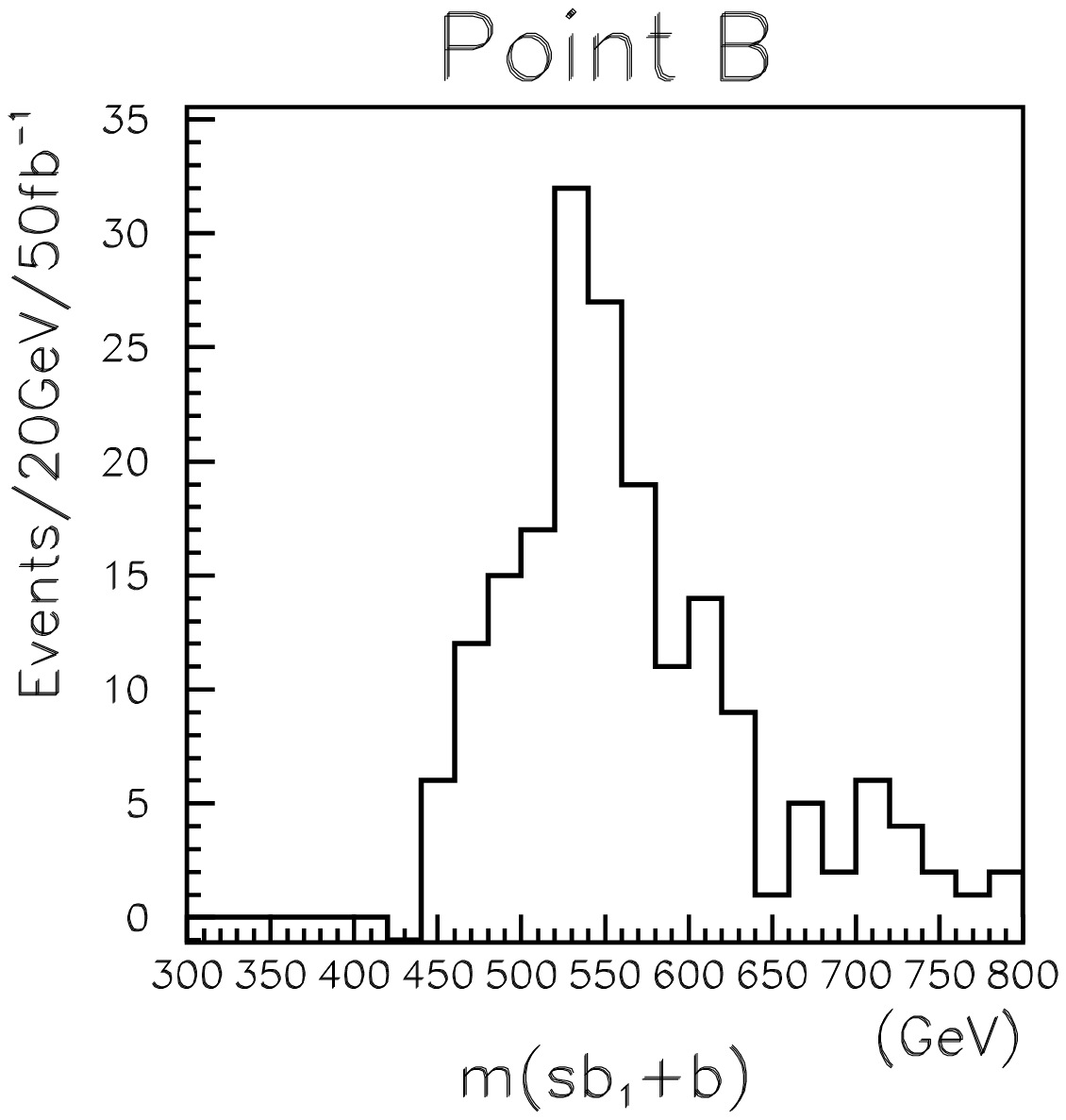}
\end{center}
\end{minipage}

\end{tabular}

\caption{\footnotesize{
The invariant mass distributions of $\ntwo b$ (Left) and $\ti b_1 b$ (Right).
In the left figure 4 momenta of the $\ntwo$ are estimated from the velocity of the lepton pair system whose invariant mass
satisfies $0\,{\rm GeV} \le m^{\rm max}_{l^+_i l^-_i}-m_{l^+_i l^-_i} \le 10$\,GeV
by assuming correct $\ntwo$ mass. 
In the right figure, momentum of $\ntwo b$ system whose invariant mass satisfies 
$|m_{\ntwo b}^{\rm peak} - m_{\ntwo b}| \le 15$\,GeV is regarded as $\ti b_1$ momentum.
The corresponding luminosities are $\int {\cal L} dt =$30\,${\rm fb^{-1}}$ (Left) and 50\,${\rm fb^{-1}}$ (Right).
}}
\label{peak:msb1_mg}
\end{center}
\end{figure}

In addition, we can check our results by selecting the events near $m_{ll}$ endpoint.
The two lepton system from the 3-body decay $\ntwo \to \none l^+_i l^-_i$
must be at rest in the $\ntwo$ rest frame 
when $m_{l^+_i l^-_i}=m_{l^+_i l^-_i}^{\rm max}$.
The $\ntwo$ momentum can be estimated from a velocity of the two lepton system
for the events with $m_{l^+_i l^-_i} \lsim m_{l^+_i l^-_i}^{\rm max}$, if $m_{\ntwo}$ is known
\cite{Hinchliffe:1996iu,Atlas}. 
Using the observed $\ntwo$ mass from 4 endpoint measurements, 
(\ref{end0}), (\ref{end1}), (\ref{end3}) and (\ref{end4}),
we can calculate the invariant mass distribution of $\ntwo + b$ from estimated $\ntwo$ momentum. 
For the decay chain (\ref{sbdc}), 
the peak of the $m_{\ntwo b^{(2)}}$ distribution
gives the $\ti b_1$ mass. 
We show the $m_{\ntwo b}$ distribution in Fig.\,\ref{peak:msb1_mg} (Left).
Here we use events that satisfy $0\,{\rm GeV} \le m^{\rm max}_{l^+_i l^-_i}-m_{l^+_i l^-_i} \le 10$\,GeV.
Here we use all $b$ tagged jets and leptons as the candidates 
because only a few events survive under the requirement
that all $b$ jets and leptons are in the same hemisphere.
The distribution has a peak near the correct $\ti b_1$ mass, $m_{\ti b_1} = 400$\,GeV.

By using events around the $\ti b_1$ peak, we can subsequently estimate the $\ti b_1$ momentum,
and calculate the invariant mass of $\ti b_1 b$.
For the decay chain $(\ref{sbdc})$, the $\ti g$ mass can be measured from
the peak of the $m_{\ti b_1 b^{(1)}}$ distribution.
We show the $m_{\ti b_1 b}$ distribution in the right figure in Fig.\,\ref{peak:msb1_mg}.  
Here we use events that satisfy $|m^{\rm peak}_{\ntwo b}-m_{\ntwo b}| \le 15$\,GeV.
The distribution has a peak near the correct $\ti g$ mass, $m_{\ti g}=537$\,GeV.

\subsubsection{Top reconstruction and the $tb$ endpoint at Point A}

The sparticle mass measurement is challenging for the class of the points A, AH1, AH2.
The gluino entirely decays into $\ti t_1$ and $\ti b_1$,
and decay modes of the $\ti t_1$ and $\ti b_1$ are dominated by
$(a)$, $(b)$ and $(d)$, $(e)$ in Eq.\,(\ref{tbdecay}), respectively.
All these cascade chains have the same decay products $2b+2W+\none$, 
and a gluino pair system leads to 
$4b+4W+\esla_T$.\footnote{
This event topology is essentially the same as in the focus point like region \cite{Chattopadhyay:2000qa}.
However in our scenario 
the branching ratio to the $4b+4W+\esla_T$ final states
is very large because gluino can decay to the 3rd generation squarks in 2-body way.}
Therefore, any exclusive analysis using jets suffers from a large combinatorial background.
The gluino decay products contain a top quark with high probability.
If the top quark is  detected with a significant rate in SUSY events, 
it indicates the  existence of  the light 3rd generation squarks.

The lighter stop mass is relatively light at Point A, $m_{\ti t_1}=321$\,GeV, and the 
phase space of the decay is not small, 
$m_{\ti g} - (m_{\ti t_1} + m_t) \simeq 200$\,GeV. 
The top quark from the gluino decay, $\ti g \to \ti t_1 t$, is boosted,
and the top decay products, $b + 2j$, tend to go in the same direction.
The jets that goes in the same direction are efficiently  picked up by the hemisphere method.
In the analysis therefore we require all $b+2j$ are in the same hemisphere.

Since there are a large number of jets 
in SUSY events,
the combinatorial background   for $W$ reconstruction still remains 
after the restriction. 
In order to estimate the background distribution
from fake jet pairs whose invariant mass is $\sim m_W$,
we apply sideband subtractions \cite{Atlas,Hisano:2002xq}.
\begin{figure}[t!]
\begin{tabular}{lr}
\begin{minipage}{0.33\hsize}
\begin{center}
\includegraphics[width=4.2cm,clip]{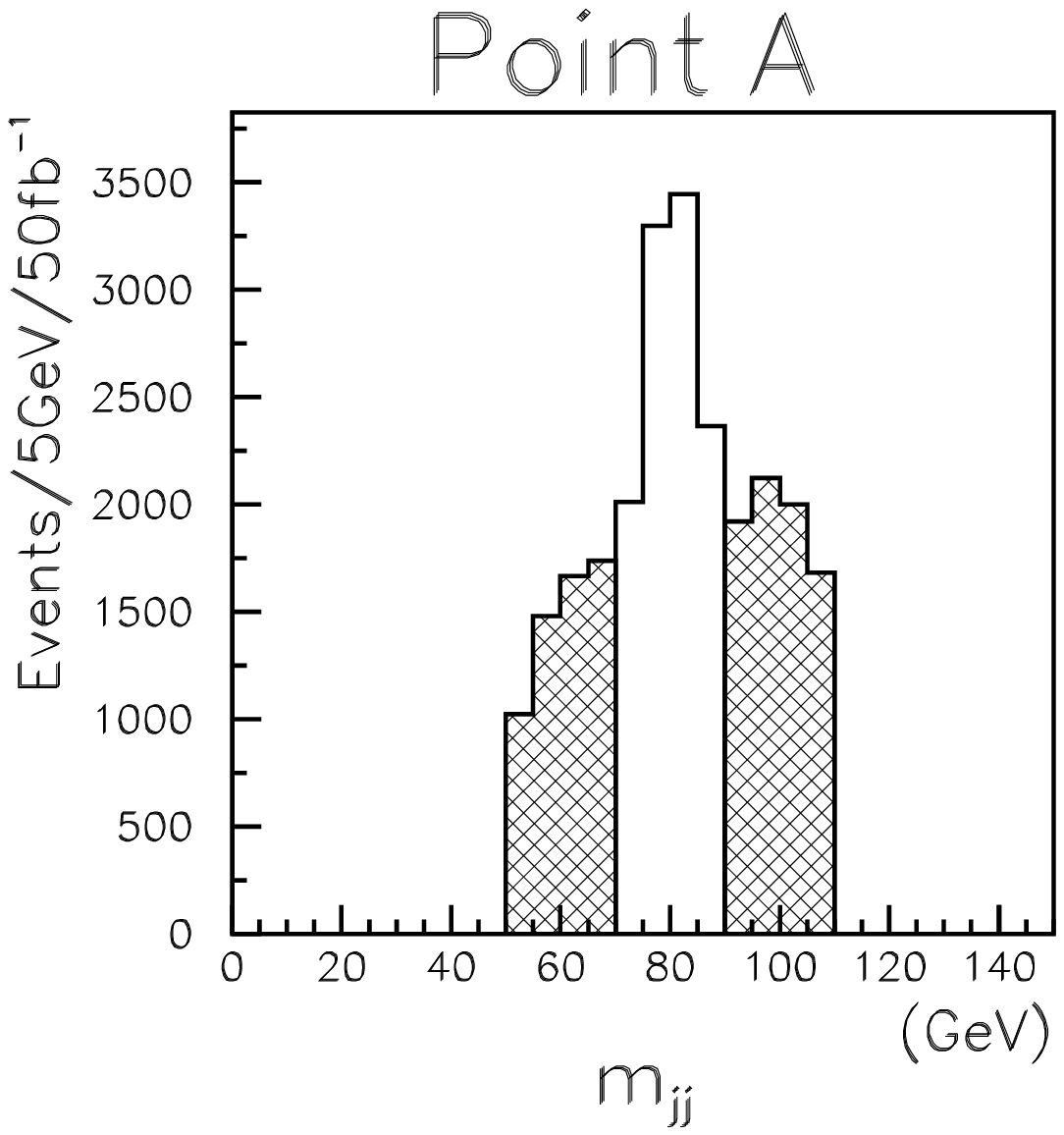}
\end{center}
\end{minipage}
\begin{minipage}{0.33\hsize}
\begin{center}
\includegraphics[width=4.2cm,clip]{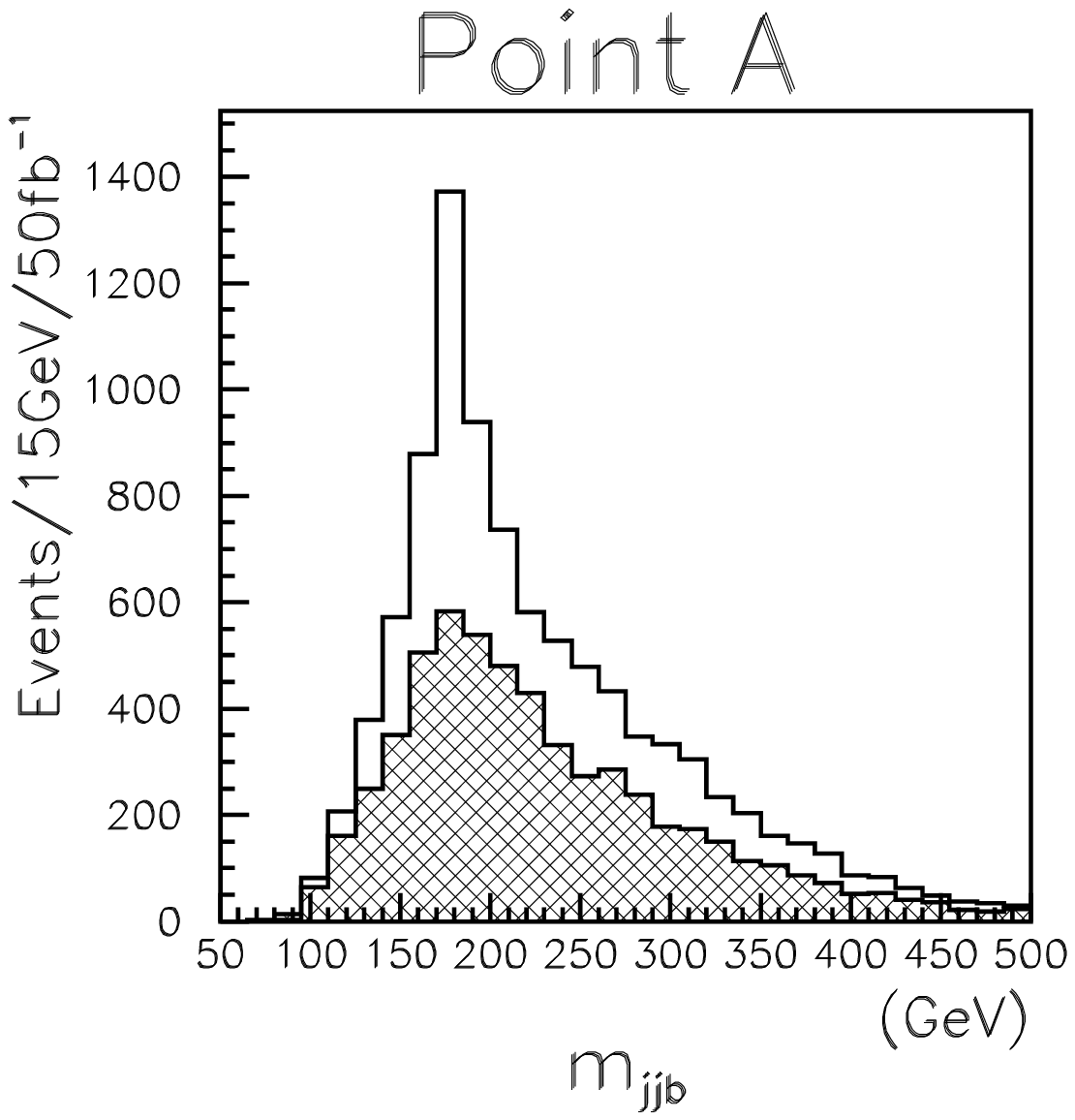}
\end{center}
\end{minipage}
\begin{minipage}{0.33\hsize}
\begin{center}
\includegraphics[width=4.2cm,clip]{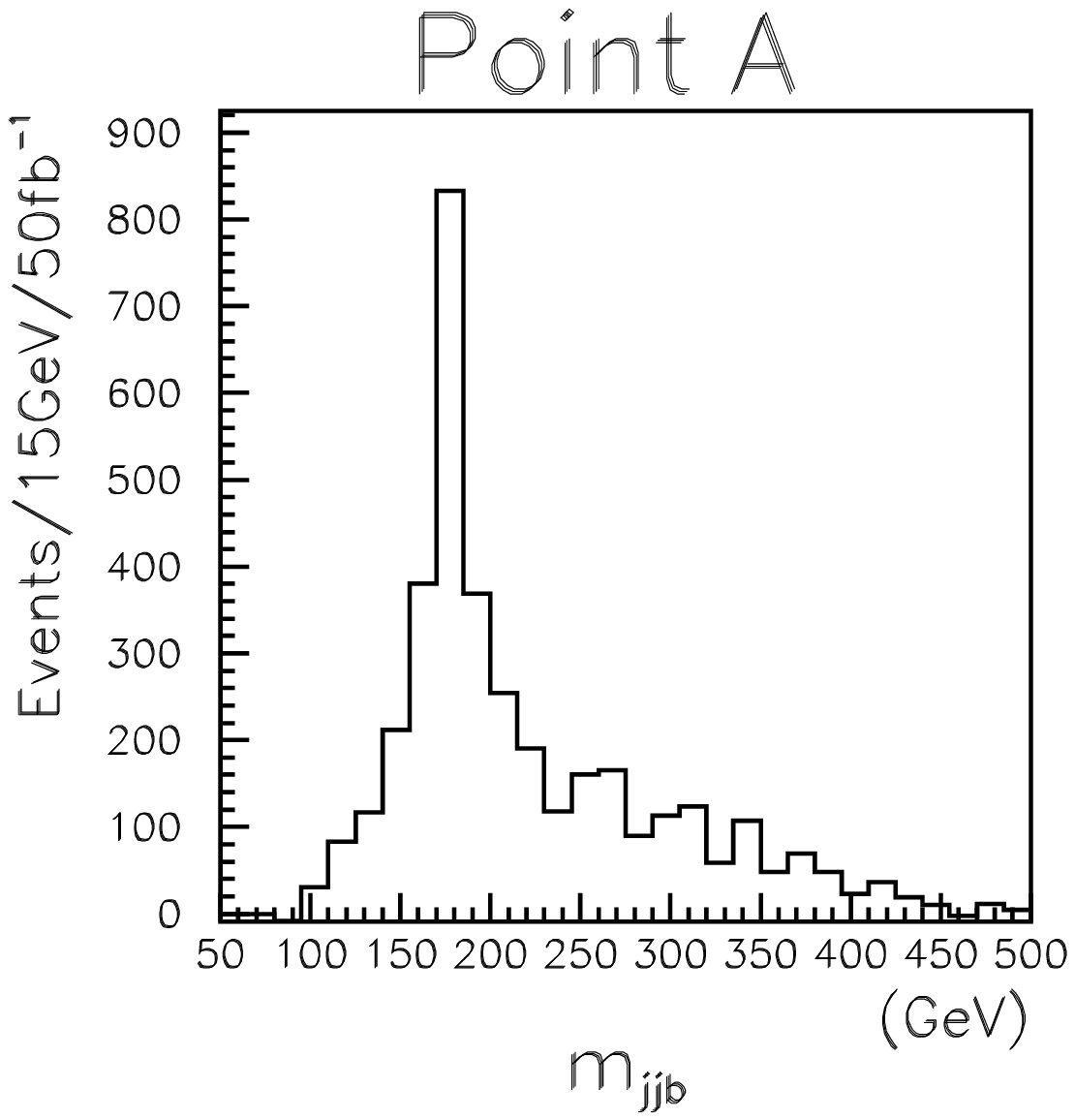}
\end{center}
\end{minipage}
\end{tabular}
\caption{\footnotesize{
{\bf Left;} The invariant mass distribution of the jet pairs that give the closest value to 
$m_{jj}=m_W-20$\,GeV for sideband region I,  $m_{jj}=m_W$ for $W$ mass region, $m_{jj}=m_W+20$\,GeV for sideband region II.
{\bf Center;} The invariant mass distributions of $jjb$, where the jet pair is in the $W$ mass region. 
The shaded region represents the estimated background distribution from the fake $W$ jet pairs. 
The background is estimated from events in the sideband region I and II by rescaling the momenta and normalizing the 
number of events events. 
{\bf Right;} The $jjb$ invariant mass distribution after the background subtraction.
}}
\label{mjjb}
\end{figure}
We first define the W mass region and sideband regions as follows, 
\begin{eqnarray}
|m_{jj}-m_W| \le 10 \, {\rm GeV} &\cdots& (W~{\rm mass~region}), \nonumber \\
|m_{jj}-(m_W-20 \, {\rm GeV})|<10\,{\rm GeV} &\cdots& (W~{\rm sideband~region~I}), \nonumber \\
|m_{jj}-(m_W+20 \, {\rm GeV})|<10 \,{\rm GeV} &\cdots& (W~{\rm sideband~region~II}). 
\end{eqnarray}
If the event contains several jet pairs in the same region, we choose the 
jet pair whose invariant mass is the closest to 
$m_W$ for $W$ mass region 
and to $m_W \mp 20$\,GeV for sideband region I (II), respectively.\footnote{The events may be double counted  in the different regions.}  
In  Fig.\,\ref{mjjb} (left)   we show the  selected two jets invariant mass distributions. 
The number of events in $W$ mass region is 
clearly bigger than those in the sideband regions. 

The central figure shows the $jjb$ invariant mass distributions, 
where the two jets are the closest jet pair to $m_W$.
Here the open histogram shows the distribution of the events in $W$ mass region.
On the other hand, the shaded distribution shows the  background estimated from the 
sideband events. 
Namely,  the jet pair momentum 
in region I (II) is rescaled  by a factor ($m_W/(m_W \mp 20\,{\rm GeV})$) before calculating $m_{jjb}$, and  
the two distributions are averaged. 
The right figure shows the $W+b$ invariant mass distribution
after subtracting the sideband distribution. 
It has a clear peak at the top mass, 
which indicates that the SUSY events contain top quarks with a significant rate.

Because we find very prominent top quarks, it is natural to
think $tb$ distribution also show the clear kinematical structure.
At  pont A, the dominant gluino cascade decay chain  is
$\ti g \to \ti t_1 t \to \chi^{\pm}_1 b t$ with $47\%$ of the branching ratio.
The $m_{tb}$ distribution should have an edge. 
The kinematical maximum of the $tb$ invariant mass is given as
\begin{eqnarray}
m_{tb}^{\rm max} &=& \Bigg(
m_t^2 + \frac{m^2_{\tilde{t}_1}-m^2_{\tilde{\chi}_1^{\pm}}}{2 m^2_{\tilde{t}_1}} 
\Big[
(m^2_{\gluino}-m^2_{\tilde{t}_1}-m_t^2)+\sqrt{(m^2_{\gluino}-m^2_{\tilde{t}_1}-m_t^2)^2-4m^2_{\tilde{t}_1} m_t^2}
\Big]
\Bigg)^{1/2}
\nonumber \\ 
&=&475\,{\rm GeV} ~.
\end{eqnarray}

However searching the endpoint in a tagged $tb$ distribution is not successful.
This is because there are typically 4 $b$ partons  in the final state. 
When an event has 2 tagged $b$ jets in the final state and one of the $jjb$ system 
has a mass consistent with top mass, the probability
that  the other  $b$ jets is  coming from the  same  gluino cascade decay chain is only 1/3. 
In addition, the probability that both the two $b$ tagged jets are satisfy the hemisphere cut
$p_T>50$\,GeV and $|\eta|<2.5$ is not large.  
One would also loose events by requiring $m(bW) \sim m_t$.

Instead, we study the $Wbj$ distirbution. 
The $W$ is selected by the sideband method, while one of the other two jets  in the same hemisphere 
is  tagged as $b$. 
The other jet is the highest jet in the hemisphere except for the used jets. 
When there are more than one $b$ tagged jets in a hemisphere,
we take the highest $p_T$ $b$ jet. 

\begin{figure}[t!]
\begin{center}
\begin{tabular}{cc}
\begin{minipage}{0.4\hsize}
\begin{center}
\includegraphics[width=6.5cm,clip]{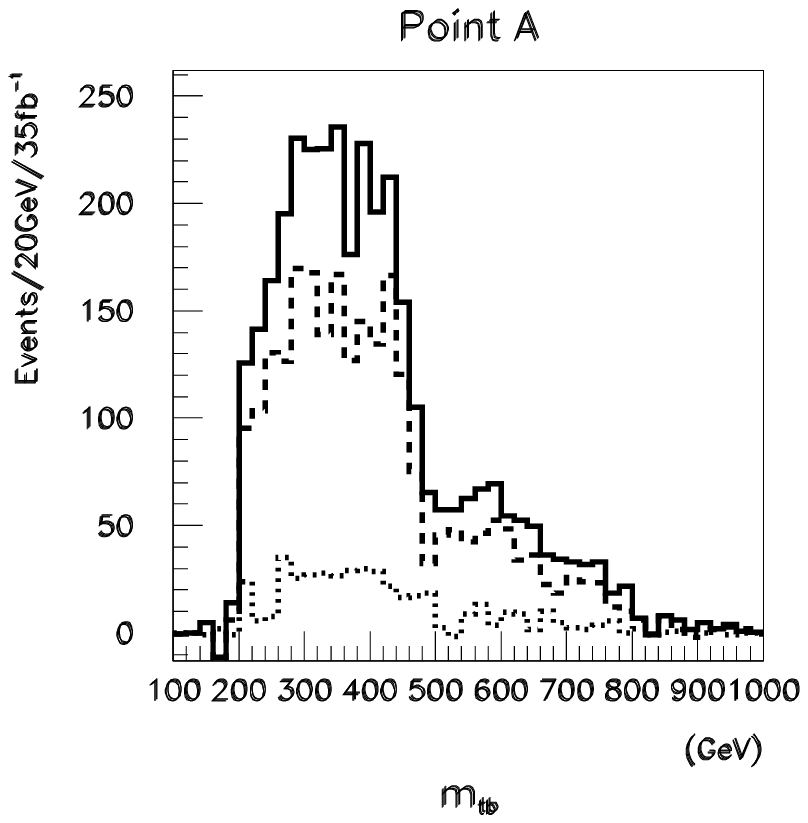}
\end{center}
\end{minipage}
\begin{minipage}{0.4\hsize}
\begin{center}
\includegraphics[width=6.5cm,clip]{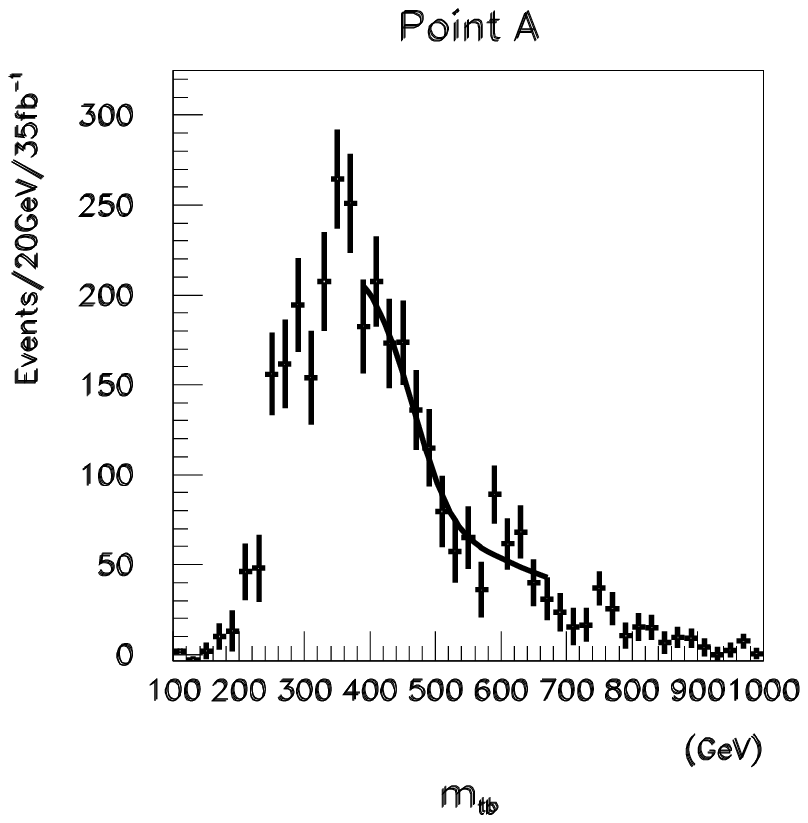}
\end{center}
\end{minipage}
\end{tabular}
\caption{\footnotesize{
The $Wbj$ invariant mass distribution after the $W$ sideband subtraction.
The left (right) figure is a result of parton (jet) level analysis. 
}}
\label{mtb}
\end{center}
\end{figure}

We show the $Wbj$ invariant mass distributions after the $W$ sideband subtraction in Fig.\,\ref{mtb}.
The left figure shows the result of a  parton level analysis.
Here the parton momenta from cascade decays are taken from HERWIG event record and 
they are processed by the same hemishpere algorithm. 
We assume 60\% tagging efficiency for $b$ quark.
The total distribution has a clear endpoint of the cascade decay, $\ti g \to \ti t_1 t \to \chi^{\pm}_1 b t$, 
expected at 475\,GeV.   
The contribution of the events where at least one gluino decays though
the decay chain ($a$) is shown by a dashed line in the figure. 
A dotted line represents the contribution from the events that contains subdominant decay chain ($d$), 
which does not show such structure at $\sim$ 475\,GeV.  
The other gluino decay modes do not give statistically significant contributions.

The jet level distribution is shown in Fig.\,\ref{mtb} (right).  
The distribution is smeared due to the jet energy resolution.
Note that this is a distribution of 4 jet system, where two of the jets are from $b$ parton.  
Nevertheless, the endpoint structure is still visible.
We fit the fitting function used in Ref.\,\cite{Hisano:2002xq} to the distribution.
The fitted  endpoint is at $468\pm 7 $\,GeV, which is consistent with the theoretically 
expected value 475\,GeV. The significant edge proves the existence of 
the decay $\ti g \to \ti t_1 t \to \ti \chi_1^{\pm} tb$.

\subsection{Inclusive $M_{T2}$ and $M^{\rm min}_{T2}$ distributions}

Recently, it is realized that $M_{T2}$ variable is useful to
determine the squark and gluino masses at the LHC \cite{Nojiri:2008hy, Nojiri:2008vq}.
The $M_{T2}$ variable is defined for a system of sparticle pair production and decay, 
namely the system with two visible objects ${\bf p}_{T1}^{\rm vis}$ and ${\bf p}_{T2}^{\rm vis}$ 
and missing momentum arising from 2 LSPs $\bfpsla_T = {\bf p}_{T}^{\rm LSP1}$+${\bf p}_{T2}^{\rm LSP2}$, 
as a function of an arbitrary test LSP mass $m_{\chi}$, as follows:
\begin{equation}
M_{T2}(m_{\chi}) = \min_{{\bf p}_{T1}^{\rm miss}+{\bf p}_{T2}^{\rm miss}=\bfpsla_T}
\big[
\max \{
m_T^{(1)}({\bf p}_{T1}^{\rm vis},{\bf p}_{T1}^{\rm miss}),m_T^{(2)}({\bf p}_{T2}^{\rm vis},{\bf p}_{T2}^{\rm miss})
\} \big],
\label{mT2_def}
\end{equation}
The minimization is taken for the test   LSP momenta, 
${\bf p}_{T1}^{\rm miss}$ and ${\bf p}_{T2}^{\rm miss}$, under the constraint 
${\bf p}_{T1}^{\rm miss}+{\bf p}_{T2}^{\rm miss}=\bfpsla_T$.
The transverse mass, $m_T^{(i)}$, is defined as
\begin{equation}
[m_T^{(i)}({\bf p}_{Ti}^{\rm vis},{\bf p}_{Ti}^{\rm miss})]^2 = (m_i^{\rm vis})^2 + m^2_{\chi}
+2 (E_{Ti}^{\rm vis} E_{Ti}^{\rm miss} - {\bf p}_{Ti}^{\rm vis} \cdot {\bf p}_{Ti}^{\rm miss}),
\end{equation} 
where  $m_i^{\rm vis}$ is 
the invariant mass of the ``visible object'',  $(m_i^{\rm vis})^2=
(p_i^{\rm vis})^2$  and $E_{Ti}=\sqrt{{\bf p}_{Ti}^2+m_{\chi}^2}$\,.
The kinematical upper bound of the transverse mass $m_T^{(i)}$ is given by the mother particle mass
if ${\bf p}_{Ti}^{\rm miss}={\bf p}_{T}^{{\rm LSP}i}$ and $m_{\chi}=m_{\none}$.
Because of this property the kinematical upper bound of the $M_{T2}$ variable is given by
\begin{equation}
M_{T2}(m_{\none}) \le \max\{m_1, m_2\},
\label{mt2max}
\end{equation} 
where $m_1$ and $m_2$ are masses of the initially produced sparticles.
Thus, the endpoint of the $M_{T2}$ distribution relates to the heavy squark mass in our scenario as:
\begin{equation}
M^{\rm end}_{T2}(m_{\none}) \sim m_{\ti q}.
\end{equation}

Squark and gluino production events often produce
${\cal O}(10)$ jets in the final state. 
The central question for the application of $M_{T2}$ analysis
is how to define the ${\bf p}_{Ti}^{\rm vis}$  in such cases.
The ``inclusive $M_{T2}$'' is defined so that
${\bf p}_{Ti}^{\rm vis}$ is taken as a hemisphere momentum in Eq.\,(\ref{p_hemi}).
In addition, a sub-sytem $M_{T2}$ is introduced in Refs.\,\cite{Nojiri:2008vq,Burns:2008va}
which is defined as an inclusive $M_{T2}$ but the
highest $p_T$ jet is removed before hemisphere reconstruction. 
As discussed in the previous section, 
a quark jet from the decay $\ti q \to q \ti g$ or $\ti q \to q \ti \chi_i$ 
 tends to be the highest $p_T$ jet if $m_{\ti q} \gg m_{\ti g}$, 
which is also the case for our scenario (See Eq.\,({\ref{mass_relation2}})). 
The sub-system after removing the highest jet tends to be $\ti g$-$\ti g$ or $\ti g$-$\ti \chi_i$ system
for $\ti g$-$\ti q$ production events.
Thus the endpoint of the subsystem $M_{T2}$ should be the gluino mass.

More systematical approach  to observe the gluino mass is using the minimum $M_{T2}$.
The minimum $M_{T2}$ ($M_{T2}^{\rm min}$) is defined by
\begin{equation}
M_{T2}^{\rm min} = \min_{i=1,...,5} \big[M^{\rm sub}_{T2}(i) \big],
\end{equation}
where $M^{\rm sub}_{T2}(i)$ is a generalized sub-system $M_{T2}$ defined
by removing the $i$-th high $p_T$ jet before hemisphere reconstruction.
The $M^{\rm min }_{T2}$  has been defined in Ref.\,\cite{Alwall:2009zu} for the leading 5 jets 
of the events  $pp \to \ti g \ti g \to 4 j + 2 \none$ 
to reduce an effect of the initial state radiation (ISR) to a $\ti g$-$\ti g$ system.
Namely, if a gluino pair is produced via a $gq \to \ti g \ti g j$ process, 
this jet may have a large $p_T$ compared with some of jets from a sparticle cascade decay.  
By using $M_{T2}^{\rm min}$ one can effectively reduces the contamination from the ISR jet 
in $\ti g$-$\ti g$ production events. 

In this paper, we use $M_{T2}^{\rm min}$ to reduce 
the additional jet from $\ti q \to q \ti g$ in $\ti g$-$\ti q$ events
as well as the effect of the ISR jet in $\ti g$-$\ti g$ events.
Note that for our model points, a gluino decay may lead 6 jets in the final state, 
therefore the chance that those jets are soft compared with 
the ISR is not small. 
The endpoint of the $M_{T2}^{\rm min}$ is given by the gluino mass:
\begin{equation}
(M_{T2}^{\rm min}(m_{\none}))^{\rm end} \sim m_{\ti g}.
\end{equation}

\begin{figure}[t!]
\hspace{-11mm}
\begin{tabular}{lr}
\begin{minipage}{0.26\hsize}
\includegraphics[width=5.0cm,clip]{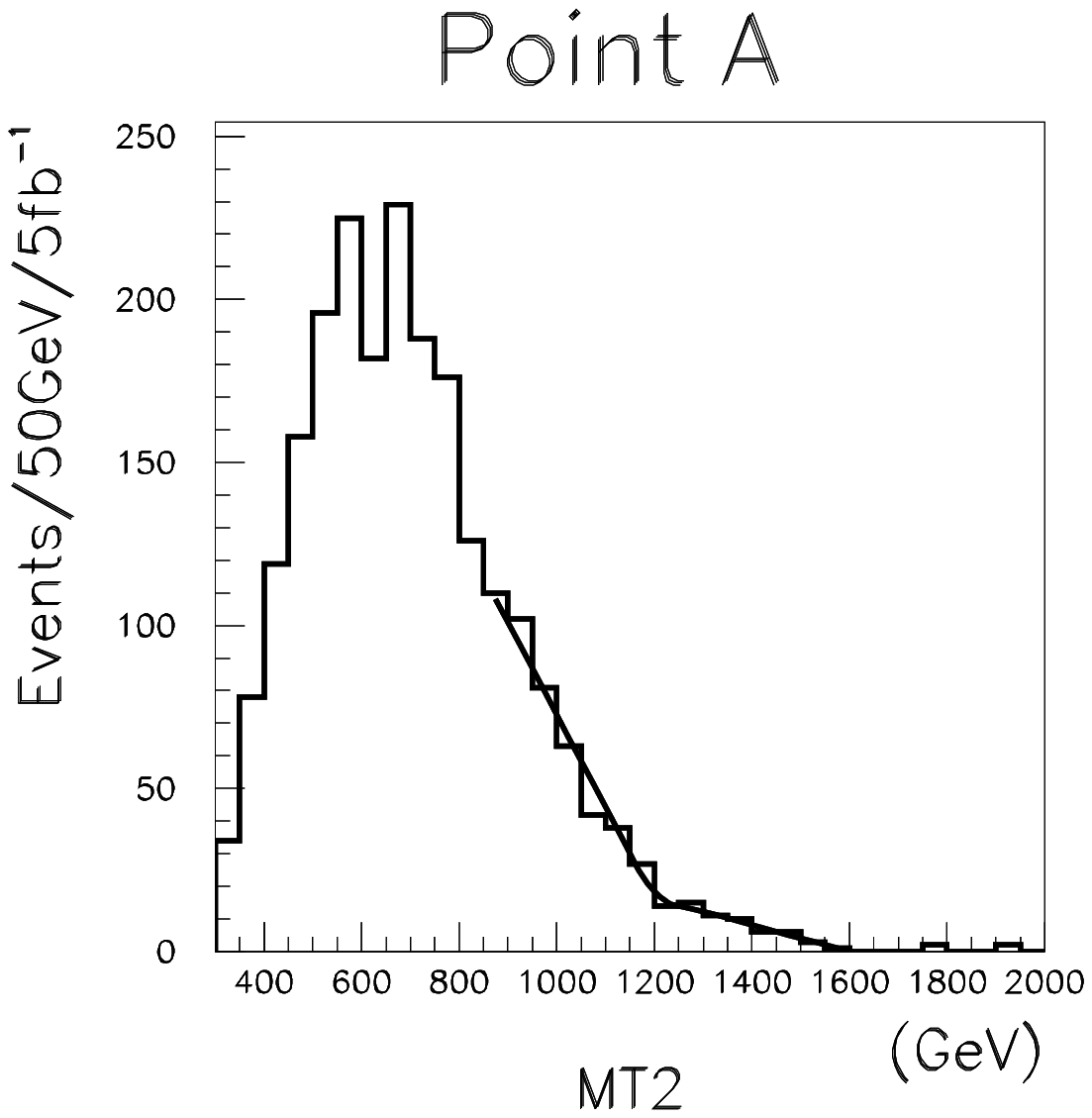}
\end{minipage}
\begin{minipage}{0.26\hsize}
\includegraphics[width=5.0cm,clip]{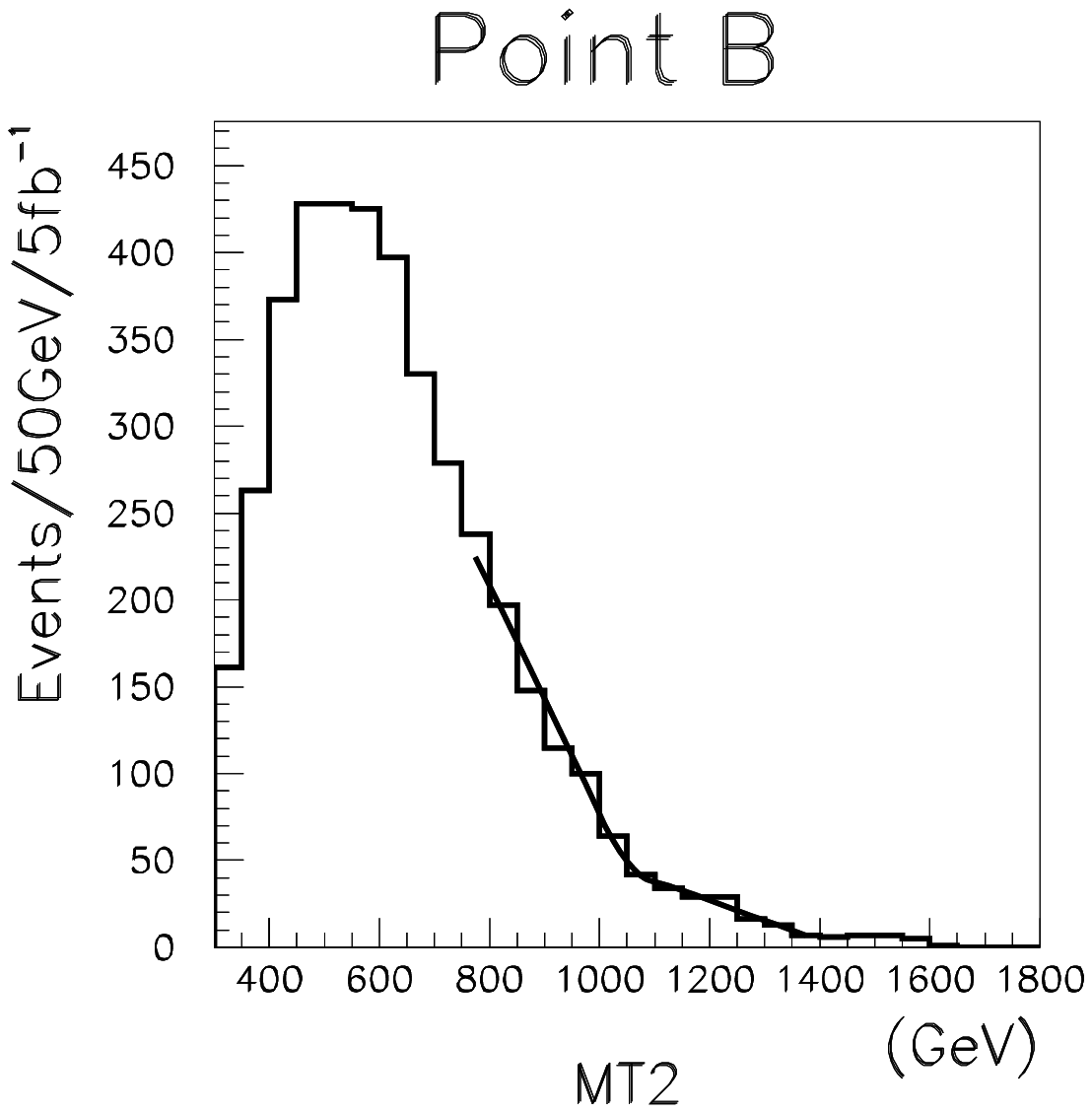}
\end{minipage}
\begin{minipage}{0.26\hsize}
\includegraphics[width=5.0cm,clip]{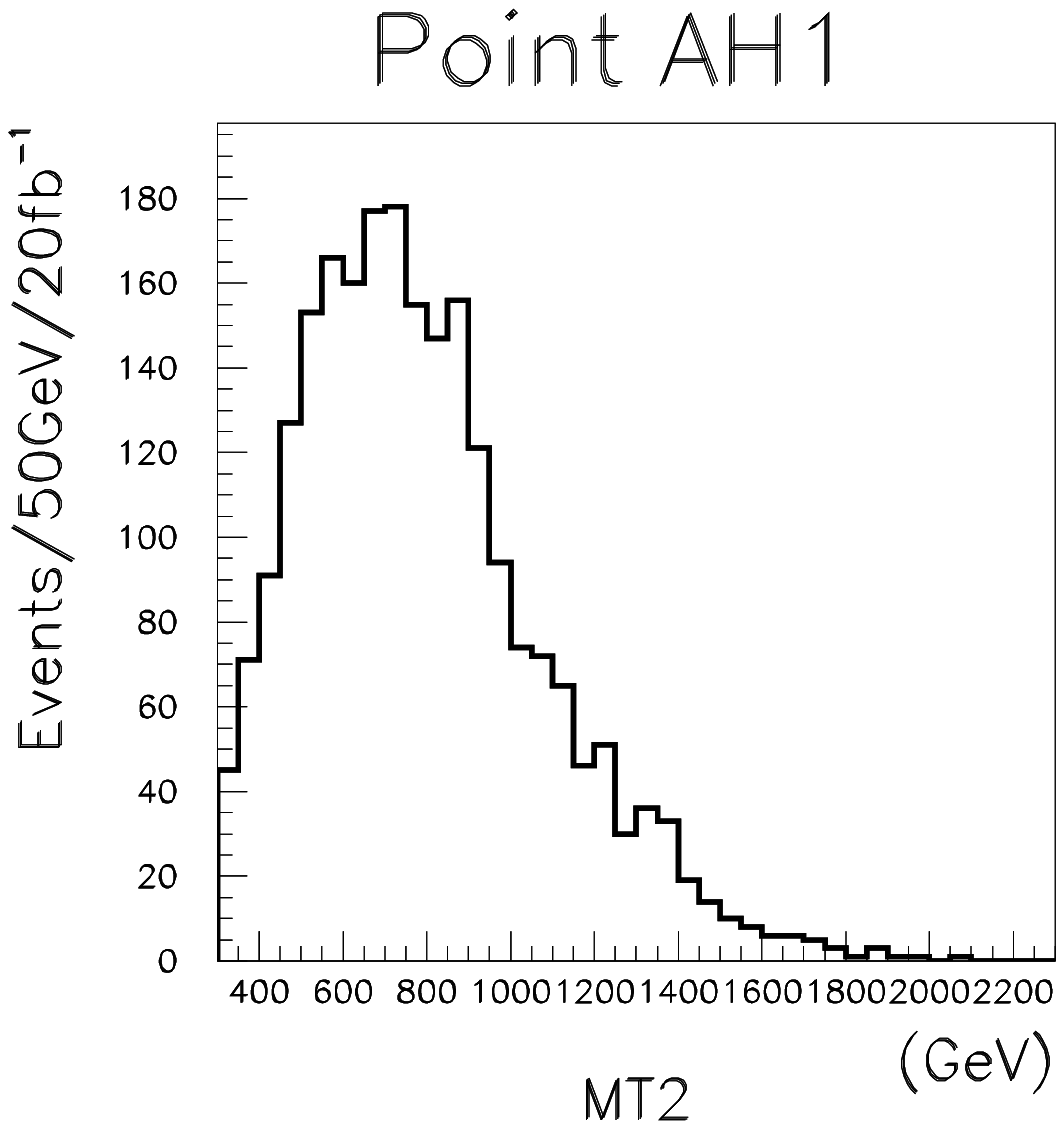}
\end{minipage}
\begin{minipage}{0.26\hsize}
\includegraphics[width=5.0cm,clip]{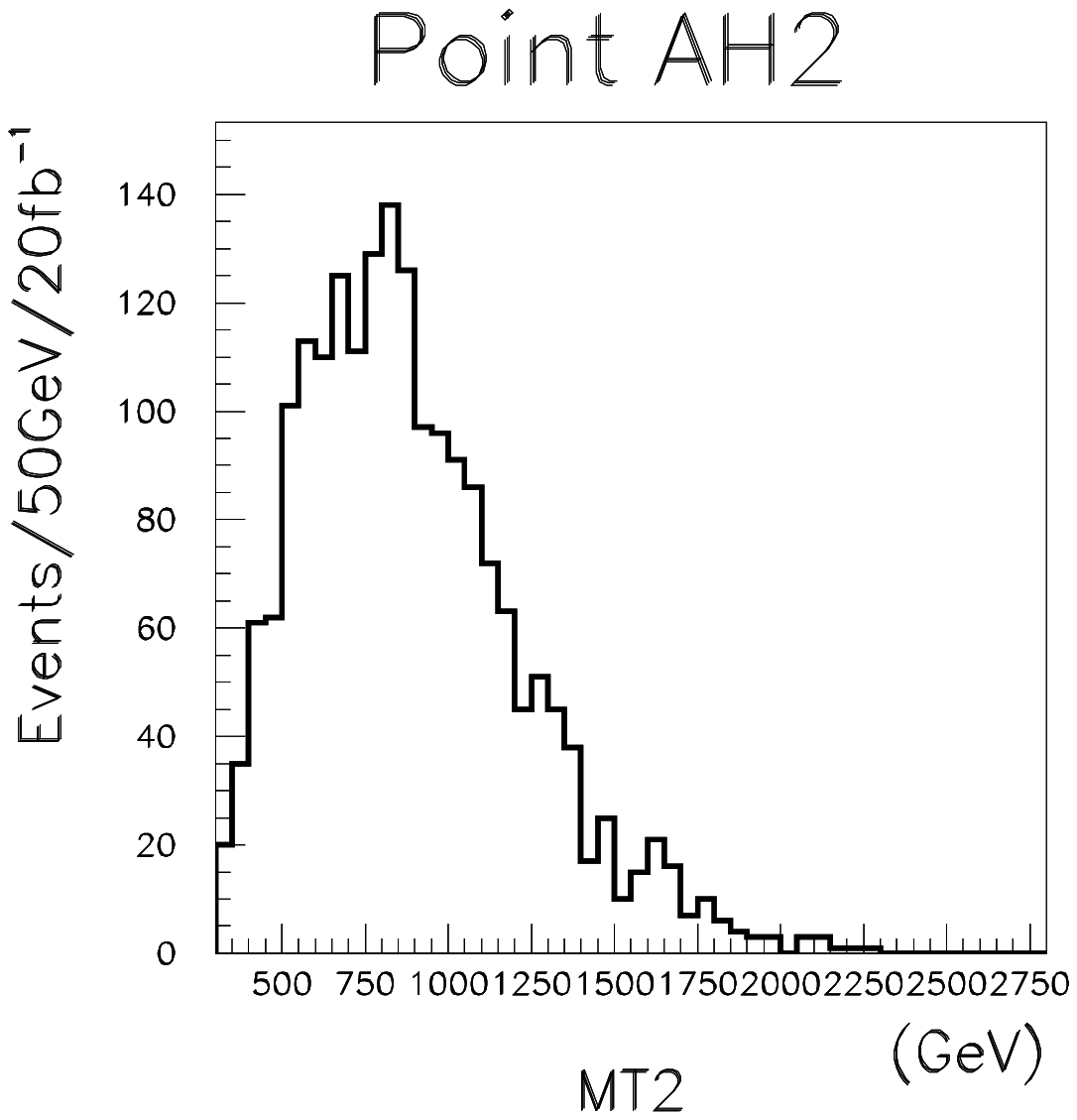}
\end{minipage}
\end{tabular}
\caption{\footnotesize{
The inclusive $M_{T2}$ distributions at Points A, B, AH1 and  AH2.
The result of the endpoint fit are $M^{\rm end}_{T2} = 1199\pm23$ and $1051\pm16$\,GeV 
for Point A and B, respectively.
The first two generation squark masses are $m_{\ti q} \simeq 1150$ and 1080\,GeV 
for Points A and B, respectively.
}}
\label{mt2}
\end{figure}

We show the  inclusive $M_{T2}$ distributions of our sample model points in Fig.\,\ref{mt2}.
Here we choose $m_{\chi}=100\,$GeV, and $p^{(1)}_{\rm vis}= p^{(1)}_{\rm hemi}$ and 
$p^{(2)}_{\rm vis}= p^{(2)}_{\rm hemi}$ for the $M_{T2}$ calculation.\footnote{
We require $p^1({\rm seed})$ and $p^2({\rm seed} )$ remain in the 
different hemispheres when we calculate $M_{T2}$ and $M^{\rm min}_{T2}$. 
The condition seems important to keep the 
events near the endpoint of $M^{\rm min}_{T2}$. 
}
In addition to the standard SUSY cut, 
we require no isolated lepton, $p_T^{\rm jet}>50$\,GeV 
and $|\eta^{\rm jet}|<2.5$ for the jet involved in a hemisphere.
We also require $N^{\rm jets}_{300} \ge 1$ for Points A and B, 
and $N^{\rm jets}_{400(600)} \ge 1$ for Point AH1 (AH2) to select $\ti q$-$\ti g$($\ti q$) production events.
We can optimize this cut 
from the difference of the $p_T^{(1)}$ distributions for $b$ tagged and non-$b$ tagged jets
(See Figs.\,\ref{hstpt} and \ref{hstptb}).
The cross sections of squark productions are small for Points AH1 and AH2
(See Table\,\ref{production}).
We use events correspond to 20\,${\rm fb^{-1}}$ for Points AH1 and AH2, 
while 5\,${\rm fb^{-1}}$ for Points A and B.

The $M_{T2}$ distributions at Point A and B have endpoints near the squark mass:
$m_{\ti q} \simeq 1150$ and 1080 for Points A and B, respectively. 
We fit the distributions by a simple fitting function
\begin{equation}
f(m)=\Theta(m-M^{\rm end}_{T2})[a_1(m-M_{T2}^{\rm end})+b]+
\Theta(M^{\rm end}_{T2}-m)[a_2(M_{T2}^{\rm end}-m)+b],
\label{fit}
\end{equation}
to see if the endpoints are recovered correctly.
We obtain $M_{T2}^{\rm end}= 1199\pm23$ and $1051\pm16$\,GeV 
for Points A and B, respectively.
They are roughly consistent with the input squark masses.

For Points AH1 and AH2 (especially for AH2),
there are a few events in the regions where $M_{T2} \lsim m_{\ti q}$. 
At these model points, a half of the heavy squark mass is larger than the gluino mass, 
$m_{\ti q}/2 \gsim m_{\ti g}$.
It is therefore expected that a gluino from the  squark decay $\ti q \to \ti g q$
is boosted and goes in the opposite direction from the direction of the quark jet $q$.
In that case, decay products of the $\ti g$ and the quark jet $q$ 
are not likely to be in the same hemisphere.

To  reconstruct  the heavy squark mass,
we should  separate the final states of $\ti g$-$\ti q \to \ti g$-$\ti g$-$q$ events into three parts,
two groups of decay products of the gluinos and the quark jet $q$. 
We adopt  the same technique as that used in the sub-system $M_{T2}$. 
Namely,  we remove the highest $p_T$ jet in the event, and 
the rest of jets and leptons are grouped into the each hemisphere. 
Next we assign  the highest $p_T$  jet into one of the hemispheres and calculate $M_{T2}$.
This gives two $M_{T2}$ values, $M_{T2}^{(1)}$ and $M_{T2}^{(2)}$,
depending on which hemisphere the  quark jet is assigned.
Finally, we choose the smaller $M_{T2}^{(i)}$, $M_{T2}=\min\{ M_{T2}^{(1)}, M_{T2}^{(2)} \}$.

\begin{figure}[t!]
\begin{center}
\begin{tabular}{cc}
\begin{minipage}{0.36\hsize}
\includegraphics[width=5.2cm,clip]{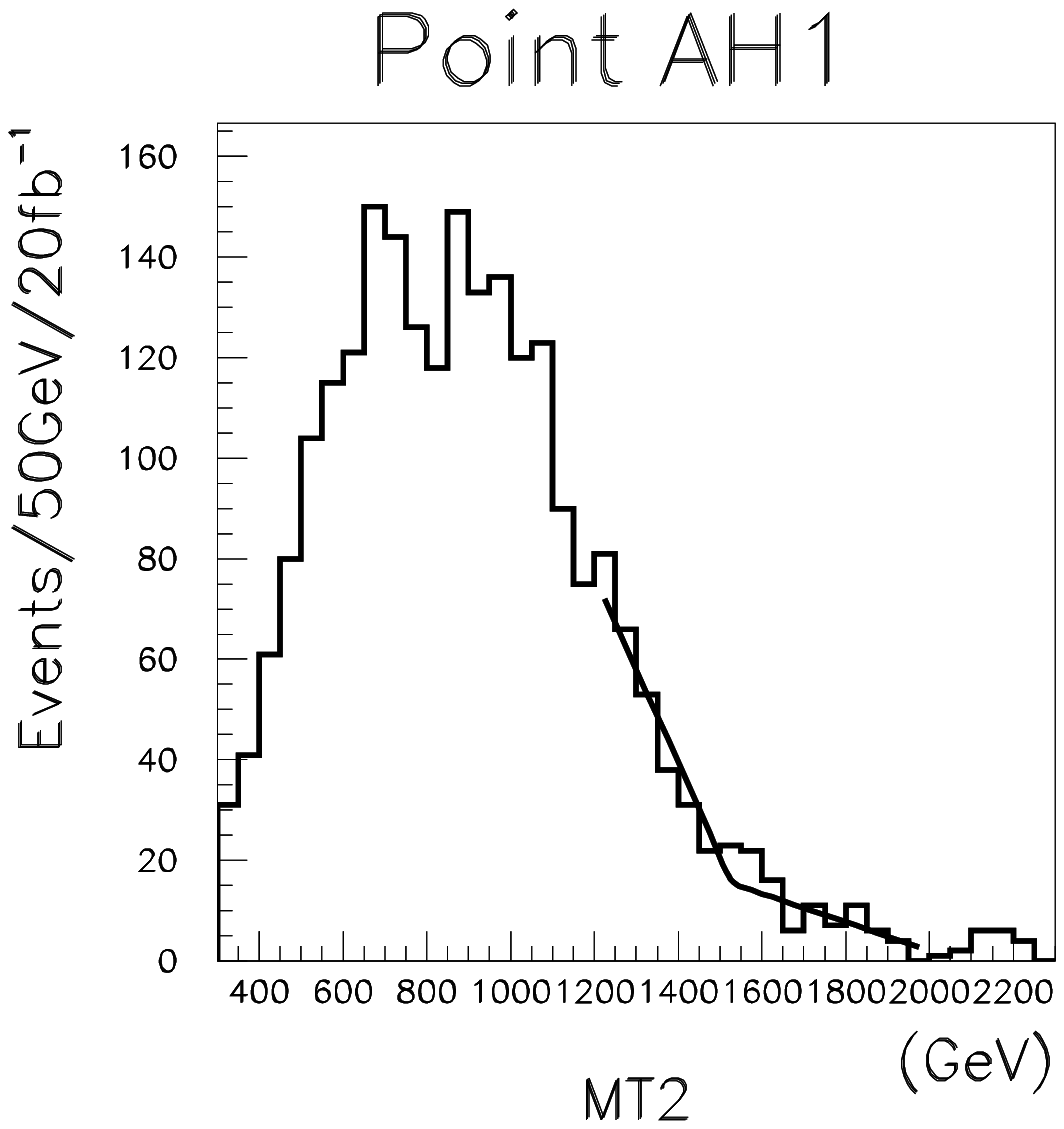}
\end{minipage}
\begin{minipage}{0.36\hsize}
\includegraphics[width=5.2cm,clip]{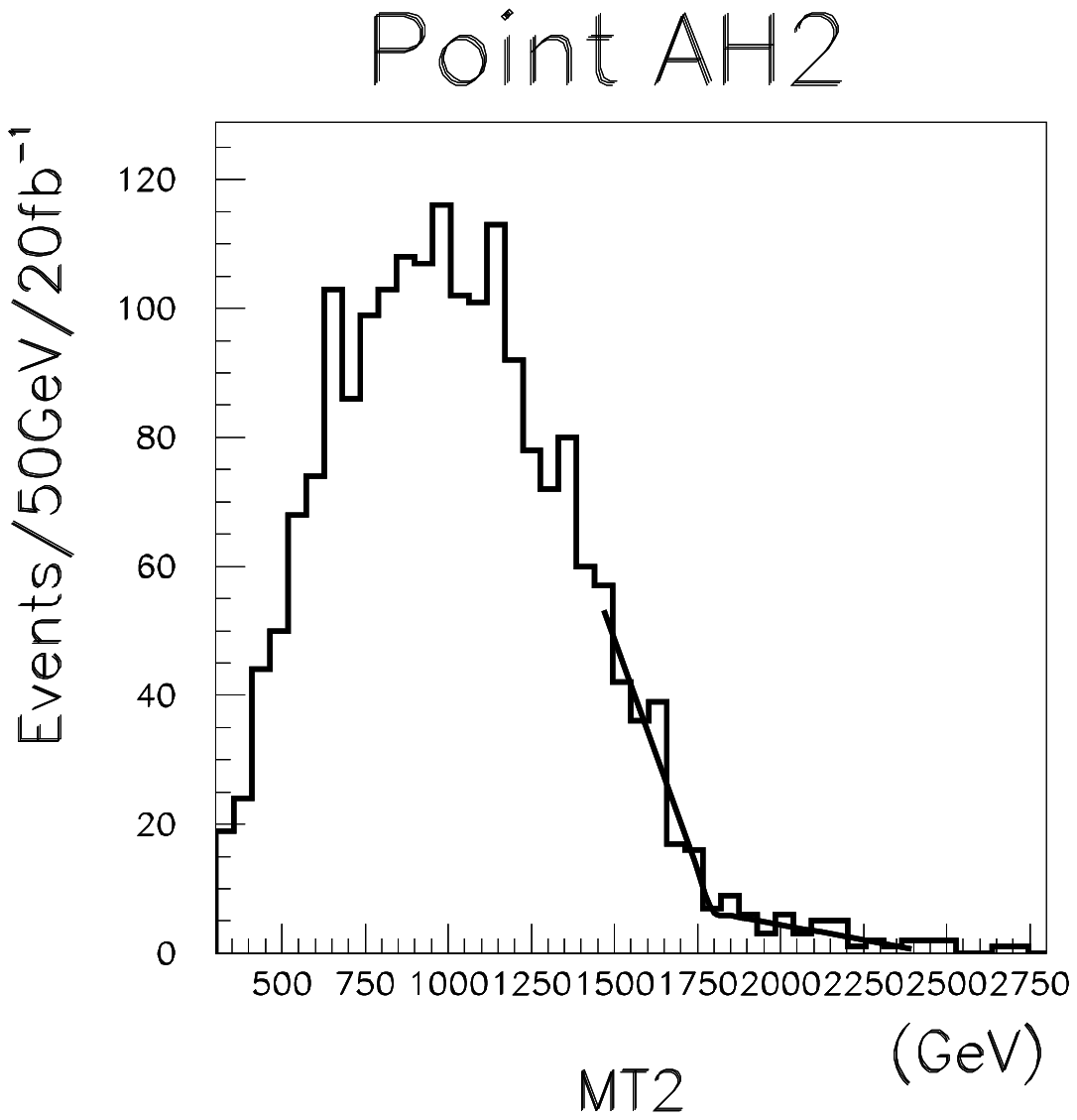}
\end{minipage}
\end{tabular}
\caption{\footnotesize{
The distributions of the $M_{T2}$ ($=\min\{ M^{(1)}_{T2}, M^{(2)}_{T2}\}$) at AH1 and  AH2.
The results of the endpoint fit are $M^{\rm end}_{T2} = 1530\pm31$ and $1798\pm21$\,GeV 
for Points AH1 and AH2, respectively.
The first two generation squark masses are $m_{\ti q} \simeq 1500$ and 1780\,GeV 
for Points AH1 and AH2, respectively.
}}
\label{mt2next}
\end{center}
\end{figure}

The distributions of the $M_{T2}$ calculated from such procedure at Points AH1 and AH2
are shown in Fig.\,\ref{mt2next}.
Here we adopt the same cuts as in Fig.\,\ref{mt2}. 
The distributions have robust endpoint structures near the correct squark masses:
1500 and 1780\,GeV for Points AH1 and AH2, respectively.
We fit the distributions by the fitting function in Eq.\,(\ref{fit}).
The results are $M_{T2}^{\rm end}= 1530\pm31$ and $1798\pm21$\,GeV 
for Points AH1 and AH2, respectively.
They are consistent with the input squark masses.

\begin{figure}[t!]
\begin{center}
\begin{tabular}{cc}
\begin{minipage}{0.36\hsize}
\includegraphics[width=5.4cm,clip]{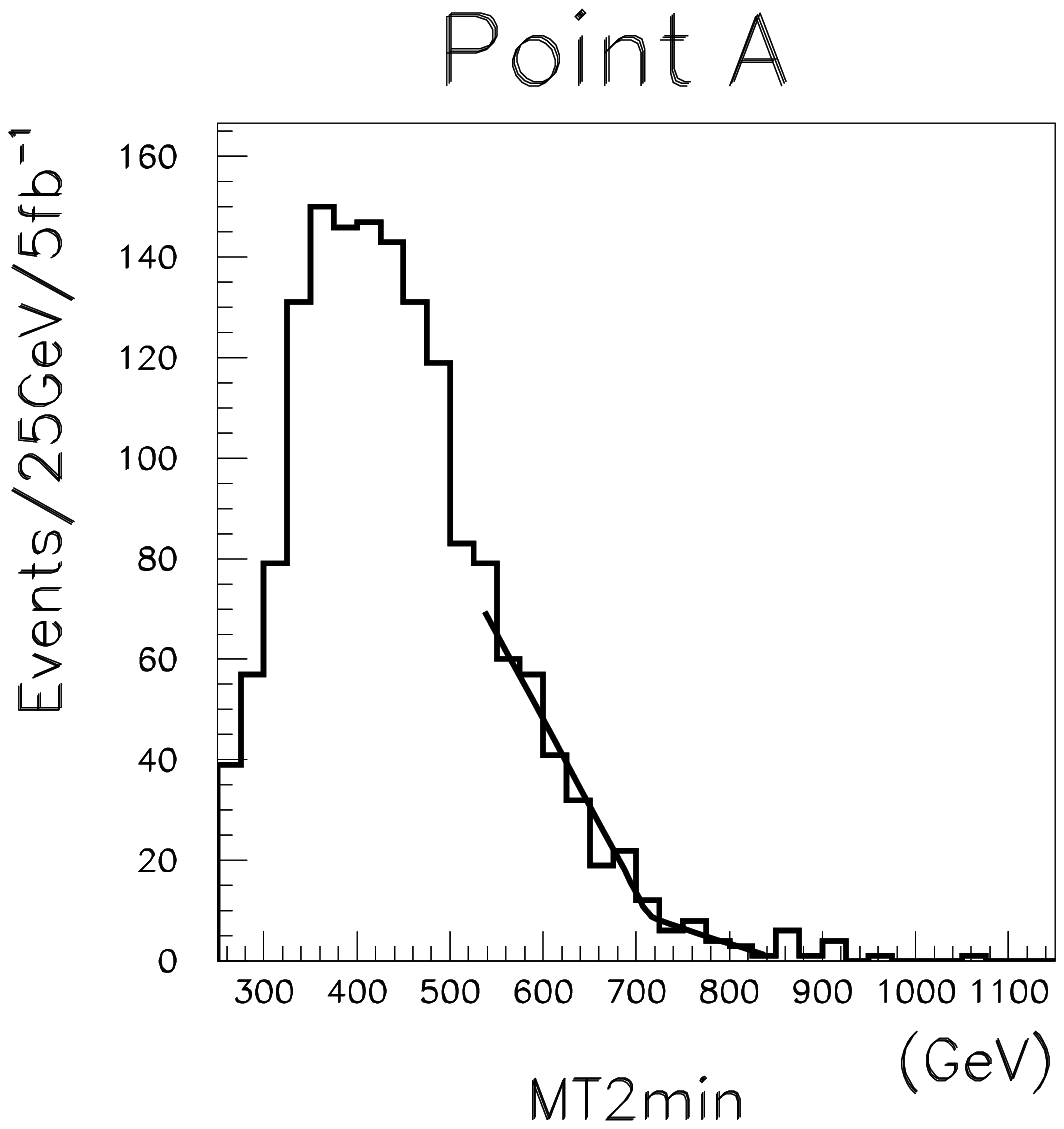}
\end{minipage}
\begin{minipage}{0.36\hsize}
\includegraphics[width=5.4cm,clip]{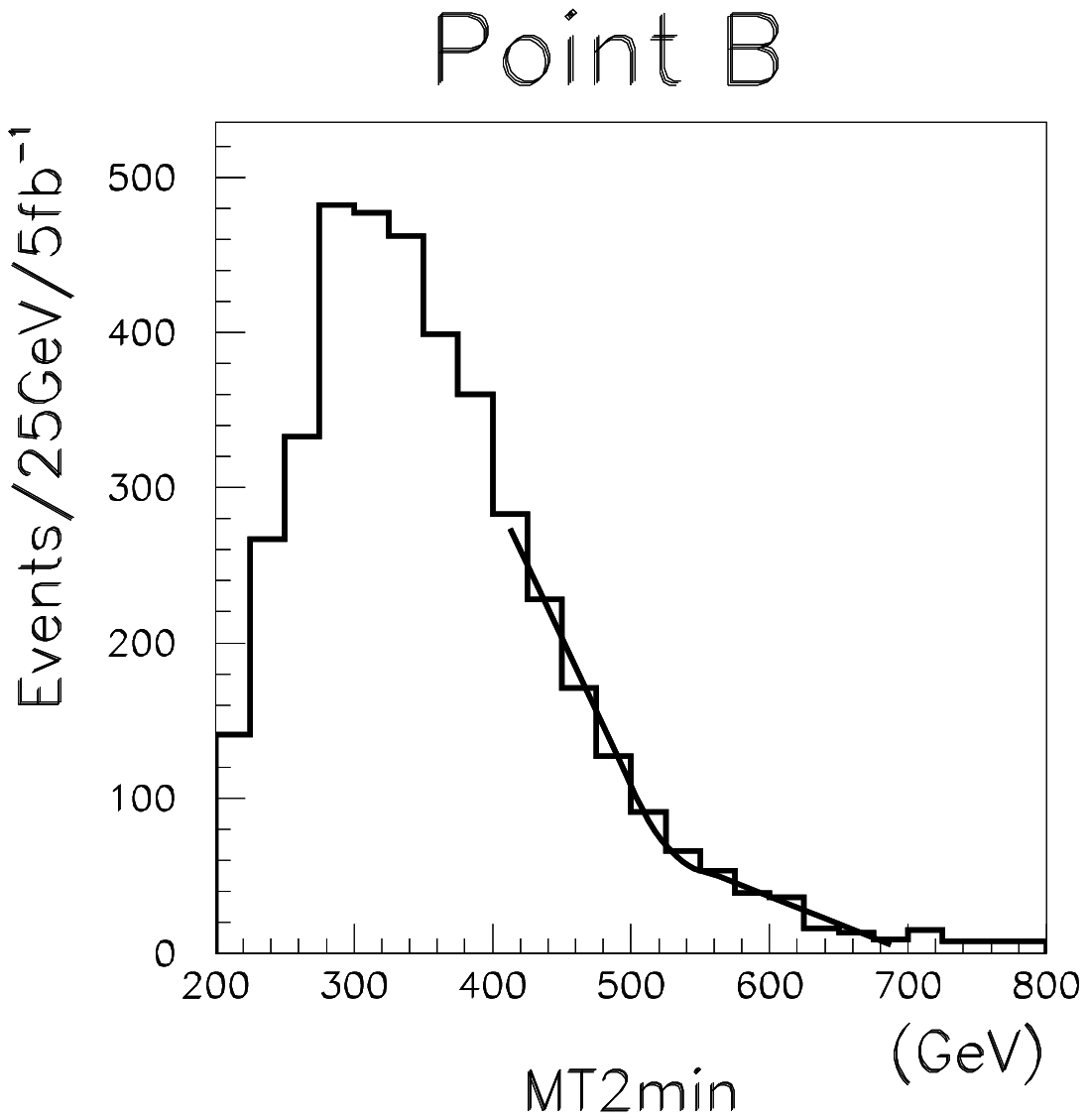}
\end{minipage}
\end{tabular}
\caption{\footnotesize{
The $M^{\rm min}_{T2}$ distributions at Points A and B.
The results of the endpoint fit are $(M^{\rm min}_{T2})^{\rm end} = 715\pm14$ and $524\pm8$\,GeV 
for Points A and B, respectively.
The gluino masses are $m_{\ti g} \simeq 697$ and 544\,GeV 
for Points A and B, respectively.
}}
\label{mt2sub}
\end{center}
\end{figure}
Next, we show the $M_{T2}^{\rm min}$ distributions for Points A and B in Fig.\,\ref{mt2sub}.\footnote {
We do not show   the distributions for Points AH1 and AH2 because
their gluino mass are almost the same as that of Point A. } 
Here, we require $N^{\rm jets}_{300} = 0$ to reduce the $\ti q$-$\ti q(\ti g)$ production events. 
The numbers of events corresponds to
$\int {\cal L}=5\,{\rm fb^{-1}}$ for each model points.
We fit the $f(x)$ in Eq.\,(\ref{fit}) to the $M_{T2}^{\rm min}$ distributions. 
We obtain $(M_{T2}^{\rm min})^{\rm end} = 715\pm14$ and $524\pm8$\,GeV
at Points A and B, respectively.
They are roughly consistent with input gluino masses: $m_{\ti g}=697$ and 544\,GeV 
for Points A and B, respectively.

If one assume GUT relation among the gaugino masses and gaugino dominance of the $\chi_1^{\pm}$,
one can roughly estimate the $\chi_1^{\pm}$ mass from the gluino mass.
Since the $tb$ endpoint is a function of $m_{\ti g}$,  $m_{\chi^{\pm}_1}$ and $m_{\ti t_1}$ ,
one can get the information of $m_{\ti t_1}$
from the the measured gluino mass and the $tb$ endpoint.

\section{Summary and Conclusion}

In this paper, we investigate the LHC signature of modified universal sfermion mass (MUSM) scenario.
In the scenario the sfermion mass matrices can be parametrized 
as in Eq.\,(\ref{nusm_matrix}) at the GUT scale. In this paper, we 
concentrate the region where $m_0 \gg m_{30}$, $m_{1/2}$, $m_{H_u}$, $\mu$ based on the considerations 
of naturalness and flavor and CP constraints.    


In  this scenario, gluino  decays entirely into the 3rd generation squarks. 
The SUSY events typically have 4 $b$ partons. 
The fraction of the SUSY events without 
$b$ tagged jets  is suppressed even if the $b$ tagging efficiency is 60\%.
This is a feature of models with the mass relation Eq.\,(\ref{mass_relation1}).


The 1st and 2nd generation squarks are much heavier than gluino and the 3rd generation squarks 
in  this scenario. 
The mass of the heavy squarks $m_0$ cannot be arbitrarily large from CCB constraint.
The heavy squarks can be observed at the LHC if $m_0 \lsim 1-2$\,TeV.
A quark jet from the decay $\ti q \to \ti g q$ 
tends to have large $p_T$ ($p_T \lsim m_{\ti q}/2$),
and they will be tagged  as the excess of  the non-$b$ tagged jets in the 
high $p_T$ range relative to the 
 $b$ tagged jets.  The excess indicates the mass relation Eq.(\ref{mass_relation2}).

The signature similar to that of MUSM scenario may be observed at some  CMSSM points with 
$|A_0|>m_0 \gg m_{1/2}$.
For MUSM scenario, gluino decay modes are 
$\ti g \to \ti t_1 t$ and $\ti g \to \ti b_1 b$,
while gluino decay is dominated by  $\ti g \to \ti t_1 t$ mode for the CMSSM region .
This difference may be seen in the $\ntwo \to \none Z$ channel.
If a  $\ntwo$ is originated from the $\ti b_1$ decay, the number of associated jets is 2 ($b \bar b$).
On the other hand if $\ntwo$ is originated from the $\ti t_1$ decay,
it is  6 ($b \bar b + 2 W(jj) $) when  both of the $W$ bosons decay into jets.
We demonstrate 
that we can discriminate these scenarios by 
investigating  the number of jets in the event with  2 leptons with $m_{ll} \sim m_Z$.

In MUSM scenario,  the events contain many jets arising from gluino decays. 
The mass reconstructions of the SUSY particles are  challenging due to the combinatorial background. 
However, we find that successful reconstructions are possible. 
The conventional endpoint analysis with OSSF lepton pair from a cascade decay chain ($\ref{sbdc}$)
is useful if  $m_{1/2}$ and $|A_0|$ are small enough  to satisfy conditions ($\ref{3cond}$).
We demonstrate that all sparticle masses arising from the decay chain ($\ref{sbdc}$) 
can be measured at the LHC in such case. 
On the other hand, if $m_{1/2}$ or $|A_0|$ is large enough, 
we have to use events without leptons for the exclusive analysis. 
We have succeeded to reduce the combinatorial background by searching for 
jet pair consistent with $W$ in the same hemisphere.  
Especially we can efficiently reconstruct the top quark arising from gluino decays into scalar top. 
Moreover we demonstrate that the endpoint in the $Wbj$ distribution can be seen even in the jet level analysis, 
which indicated the $tb$ endpoint of the $\ti g \to \ti t_1 t \to \chi_1^{\pm} b t$ mode.

The inclusive $M_{T2}$ and $M^{\rm min}_{T2}$ distribution 
is also useful for the mass determination of gluino and the heavy squarks.
The squark mass can be measured from the endpoint of the $M_{T2}$ distribution.
Moreover we reconstruct  the $\ti g$-$\ti g$ and/or $\ti g$-$\ti \chi_i$ ``sub-system'' 
by removing the $i$-th high $p_T$ jet before hemisphere reconstruction and calculating
subsystem $M_{T2}$ called $M_{T2}^{\rm sub}(i)$. The gluino mass can be measured from the endpoint of the 
$M^{\rm min}_{T2}$ ($\equiv \min_{i=1}^5 \big[ M_{T2}^{\rm sub} (i) \big]$) distribution.
By combining the measured gluino mass and $tb$ endpoint, 
one can get the information of the stop mass only from the analysis using jets.
Such inclusive analyses may tell us  both the mass scales of squarks and gluino 
in the early stage of the LHC such as $\int {\cal L} \simeq 5-20\,{\rm fb^{-1}}$.



\section*{Acknowledgements}
This work is supported by the World Premier International Research Cernter Initiative (WPI initiative) by MEXT, Japan.
S.K, N.M and K.I.N are supported in part by Grants-in-Aid for Scientific 
Research from the Ministry of Education, Culture, Sports, Science 
and Technology of Japan.

\newpage

\appendix

\section{Masses and Branching Ratios}
\label{masses}

\vspace{-5mm}
\begin{table}[!h]
\begin{center}
\caption{\footnotesize{
{\bf Left;} Mass spectra of sparticles and Higgs bosons 
for our model points in GeV.
{\bf Right;} Branching ratios of sparticles for our model points.
Here, $u$ and $d$ denote  both the first and the second generation up and down type quarks, respectively  }}
\label{mass_spectrum}
\begin{tabular}{cr}
\vspace{-5mm}
\begin{minipage}{0.5\hsize}
\vspace{-25mm}
\begin{center}

\renewcommand{\arraystretch}{1.2}
\begin{tabular}{|c||c|c|c|}
\multicolumn{4}{c}{}\cr
\hline
particle     & A & B & U \cr
\hline
\hline
$\gluino    $&    697&   537  & 707\cr
\hline
$\ti d_L       $&    1152&    1085 & 1150\cr
$\ti u_L       $&    1150&    1082 & 1147\cr
$\ti d_R       $&    1143&    1080 & 1142\cr
$\ti u_R       $&    1144&    1082 & 1142\cr
$\ti b_1       $&    540&     400  &  868\cr
$\ti t_1       $&    321&     296  &  484\cr
$\ti b_2       $&    1129&    1069 & 1077\cr
$\ti t_2       $&    612&     475  &  896\cr
\hline
$\ti e_L       $&    1016&    1010 & 1015\cr
$\ti e_R       $&    1005&    1000 & 1005\cr
$\ti {\nu}_e      $&    1013& 1007 & 1012\cr
$\ti {\tau}_1     $&    298&   183 &  932\cr
$\ti {\tau}_2    $&    1013&  1007 &  987\cr
$\ti {\nu}_{\tau}    $&  1011&  1006 &  979\cr
\hline
$\none      $&    110&    77 & 113\cr
$\ntwo      $&    210&   138 & 221\cr
$\nthre    $&     470&   258 & 742\cr
$\nfour     $&    486&   289 & 748\cr
$\ch_1^+    $&    211&    137 & 221\cr
$\ch_2^+    $&    486&    287 & 750\cr
\hline
$h^0      $&    115&     100 & 119\cr
$H^0      $&    560&     115 & 1153\cr
$A^0      $&   557&      105 & 1146\cr
$H^{\pm}   $&    567&     135 & 1157\cr
\hline
\end{tabular}

\end{center}
\end{minipage}
\vspace{-20mm}
\begin{minipage}{0.5\hsize}
\vspace{5mm}
\begin{center}

\renewcommand{\arraystretch}{1.1}
\begin{tabular}{|cl||c|c|c|}
\hline
 \multicolumn{2}{|c||}{mode} & \multicolumn{3}{c|}{BR(\%)}\\
\cline{3-5}
 &  & \multicolumn{1}{c|}{ A} & \multicolumn{1}{c|}{ B} & \multicolumn{1}{c|}{ U} \\
\hline
\hline
$ \ti u_L    $ & $ \to \gluino u$ & $67$ &74 &66\\
$            $ & $ \to \cpl_1 d$ & $21$ &15 &22\\
$            $ & $ \to \ntwo  u$ & $10$ &7 &11\\
\cline{1-5}
$ \ti d_L    $ & $ \to \gluino d $ & $68$ &74 &67\\
$            $ & $ \to \cm_1 u   $ & $20$ &12 &22\\
$            $ & $ \to \ntwo d   $ & $10$ &6 &11\\
\cline{1-5}
$ \ti u_R    $ & $ \to \gluino u $& $92$ &94 &91\\
$            $ & $ \to \none u   $ & $8$ &6 & 9\\
\cline{1-5}
$ \ti d_R    $ & $  \to \gluino d  $& $98$ &98 &98\\
$            $ & $  \to \none d   $ & $2$ &2 &  2\\
\cline{1-5}
$ \gluino    $ & $ \to \ti t_1 \bar{t} \,(\ti t_1^\ast t) $ & $64$ &30 &100\\
$            $ & $ \to \ti b_1 \bar{b} \,(\ti b_1^\ast b) $ & $36$ &70 &0\\
\hline
$ \ti t_1 $ & $ \to \cpl_1 b   $ & $73$ &91 &30\\
$         $ & $  \to \ntwo t   $ & $0$ &0 & 9\\
$         $ & $  \to \none t   $ & $27$ &9 &61\\
\cline{1-5}
$ \ti b_1 $ & $\to \ti t_1 W^-$ & $63$ &15 & 41\\  
$         $ & $  \to \cm_1 t   $ & $20$ &35 &22\\
$         $ & $  \to \ntwo b   $ & $16$ &39 &12\\
$         $ & $ \to \none b    $ & $1 $ &6 &  1\\
$         $ & $ \to \ti g b    $ & $0 $ &0 & 24\\
\cline{1-5}
$ \cm_1   $ & $ \to \none W^-              $ & $100$ &0 & 100\\
$         $ & $\to \none \bar{u} d          $ & $0$ &67 & 0\\
$         $ & $\to \none l^- \bar{\nu}_l     $ & $0$ &22 & 0\\
$         $ & $\to \none \tau^- \bar{\nu}_\tau $ & $0$ &11 & 0\\
\cline{1-5}
$ \ntwo   $ & $ \to \none Z^0       $ & $100$ &0 & 100\\
$         $ & $ \to \none u \bar{u}(d \bar{d}) $ & $0$ &24 & 0\\
$         $ & $ \to \none b \bar{b} $ & $0$ &31 & 0\\
$         $ & $ \to \none l^+ l^-   $ & $0$ &5 & 0\\
$         $ & $ \to \none \tau^+ \tau^- $ & $0$ &6 & 0\\
$         $ & $ \to \none \nu \bar{\nu} $ &$0$ &15 & 0\\
\hline
\end{tabular}

\end{center}
\end{minipage}
\end{tabular}
\end{center}
\end{table}

\newpage

\section{Comparison with SPS benchmark points}
\label{sps_appen}

For comparison of MUSM scenario with other scenarios with universal sfermion masses,  
we show the distribution of number of $b$ tagged jet and the $p_T^{(1)}$ distributions
at several benchmark model points called snowmass points and slopes (SPS) \cite{Allanach:2002nj}.
Here, the same cut as in Section\,\ref{charact} is adopted.
The number of generated events correspond to $\int {\cal L} dt =1\,{\rm fb}^{-1}$.

\vspace{10mm}
{\large {\bf SPS1a}} ``Typical point''
\begin{eqnarray}
m_0=100\,{\rm GeV},~~m_{1/2}=250\,{\rm GeV},~~A_0=-100\,{\rm GeV},~~\tan\beta=10,~~\sgn(\mu)=+
\nonumber
\end{eqnarray}

\vspace{-5mm}
\begin{figure}[h!]
\begin{center}
\begin{tabular}{ccc}
\begin{minipage}{0.3\hsize}
\includegraphics[width=3.8cm,clip]{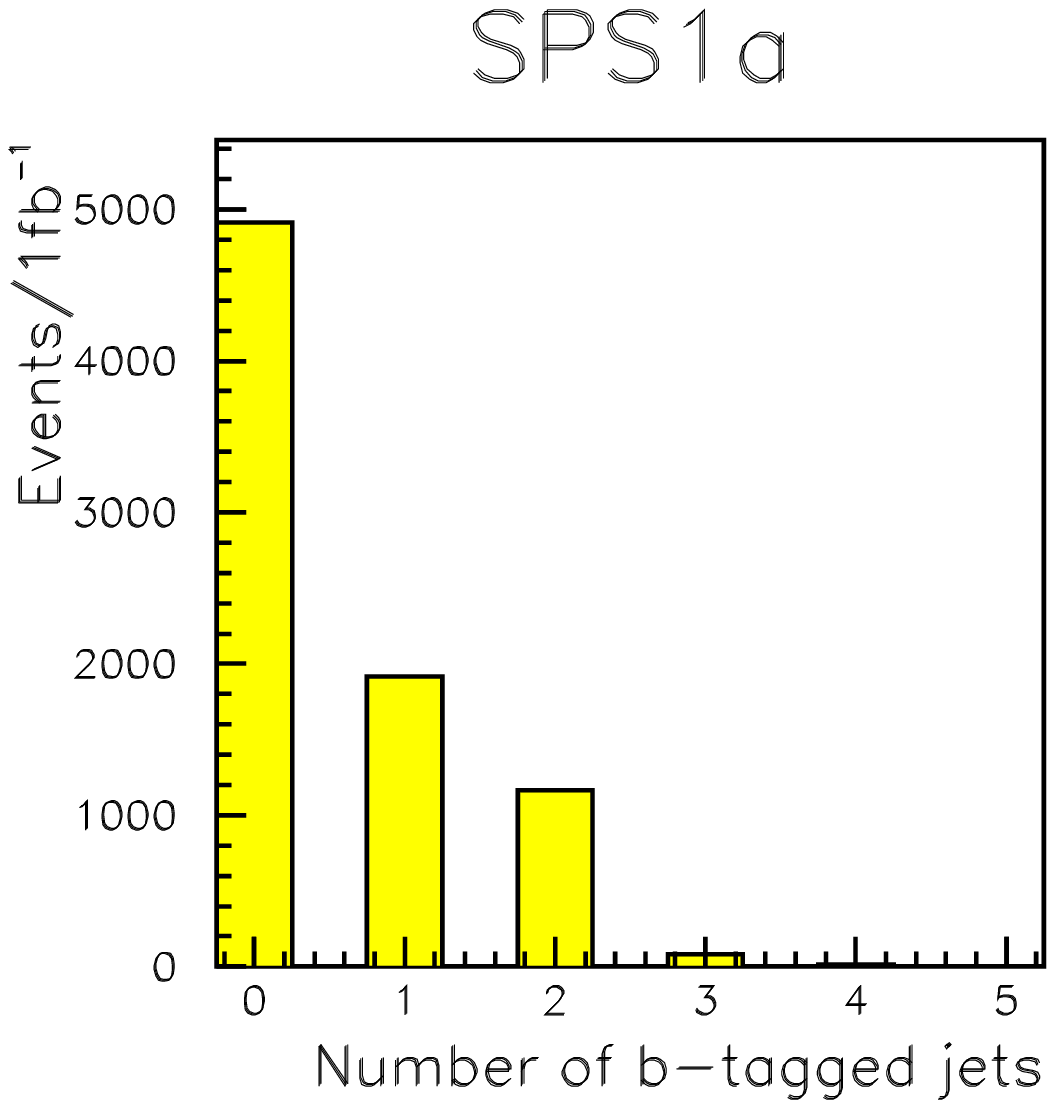}
\end{minipage}
\begin{minipage}{0.3\hsize}
\includegraphics[width=3.8cm,clip]{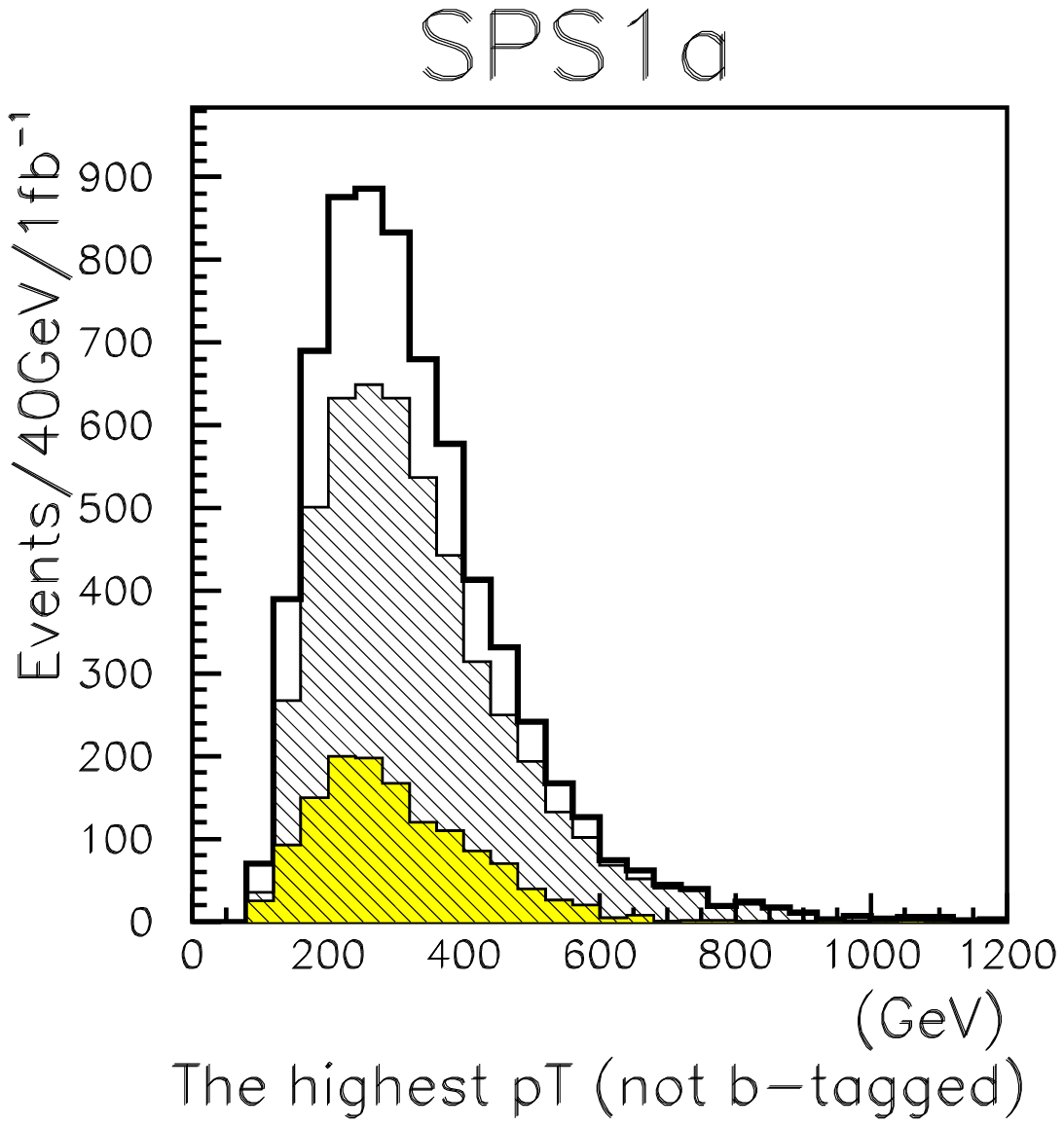}
\end{minipage}
\begin{minipage}{0.3\hsize}
\includegraphics[width=3.8cm,clip]{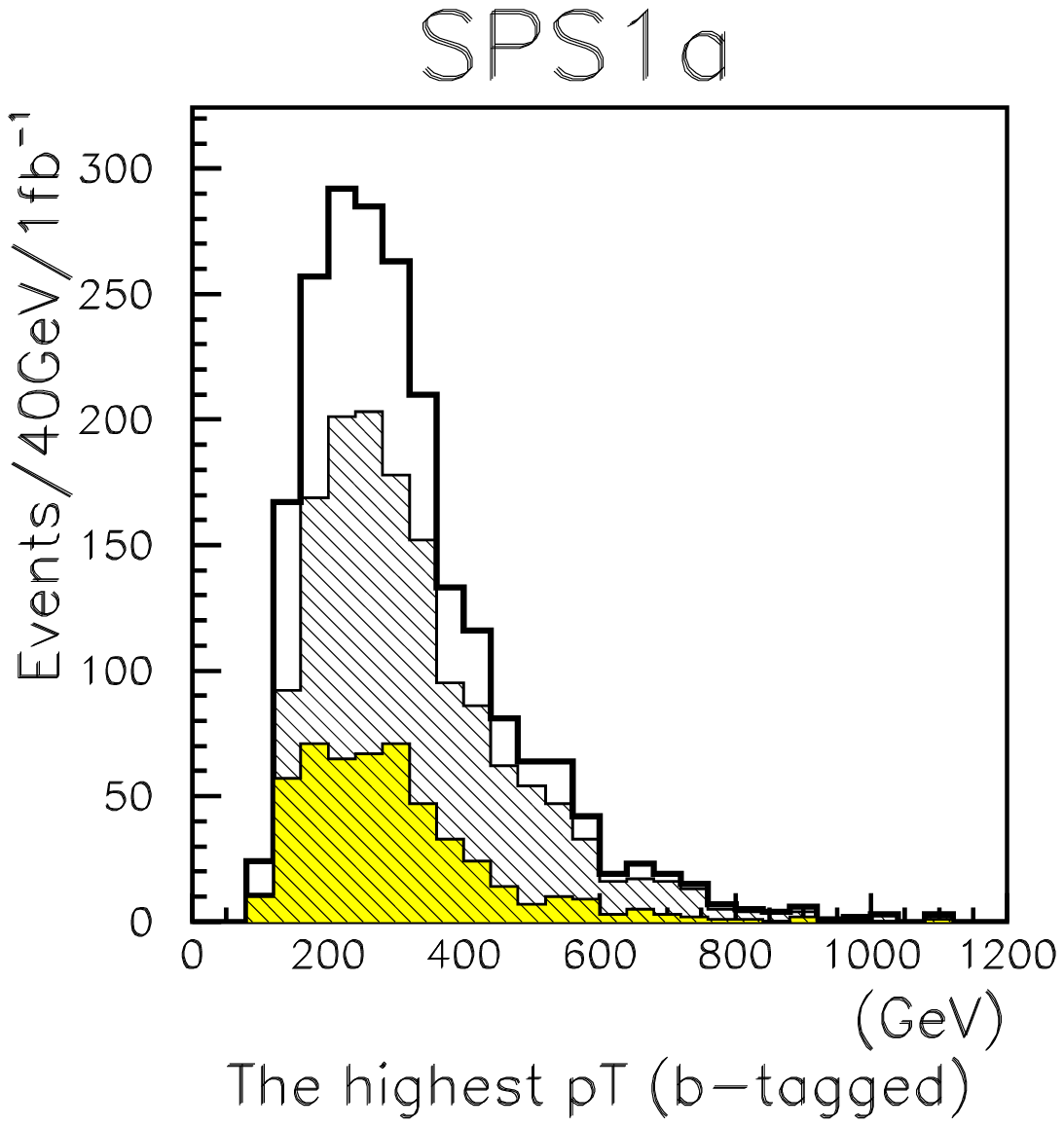}
\end{minipage}
\end{tabular}
\end{center}
\end{figure}


\vspace{10mm}
{\large {\bf SPS2}} ``Focus point''
\begin{eqnarray}
m_0=1450\,{\rm GeV},~~m_{1/2}=300\,{\rm GeV},~~A_0=0\,{\rm GeV},~~\tan\beta=10,~~\sgn(\mu)=+
\nonumber
\end{eqnarray}

\vspace{-5mm}
\begin{figure}[h!]
\begin{center}
\begin{tabular}{ccc}
\begin{minipage}{0.3\hsize}
\includegraphics[width=3.8cm,clip]{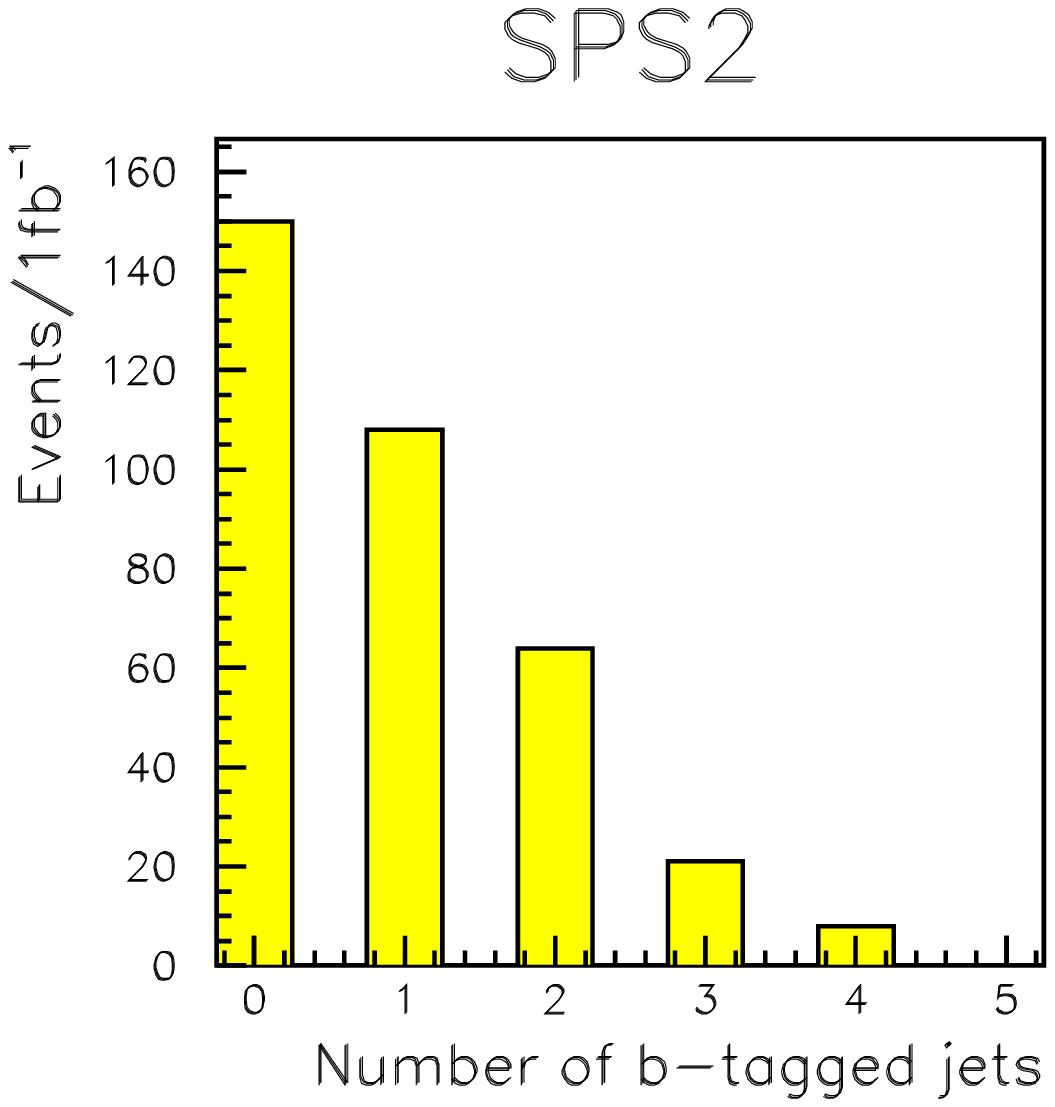}
\end{minipage}
\begin{minipage}{0.3\hsize}
\includegraphics[width=3.8cm,clip]{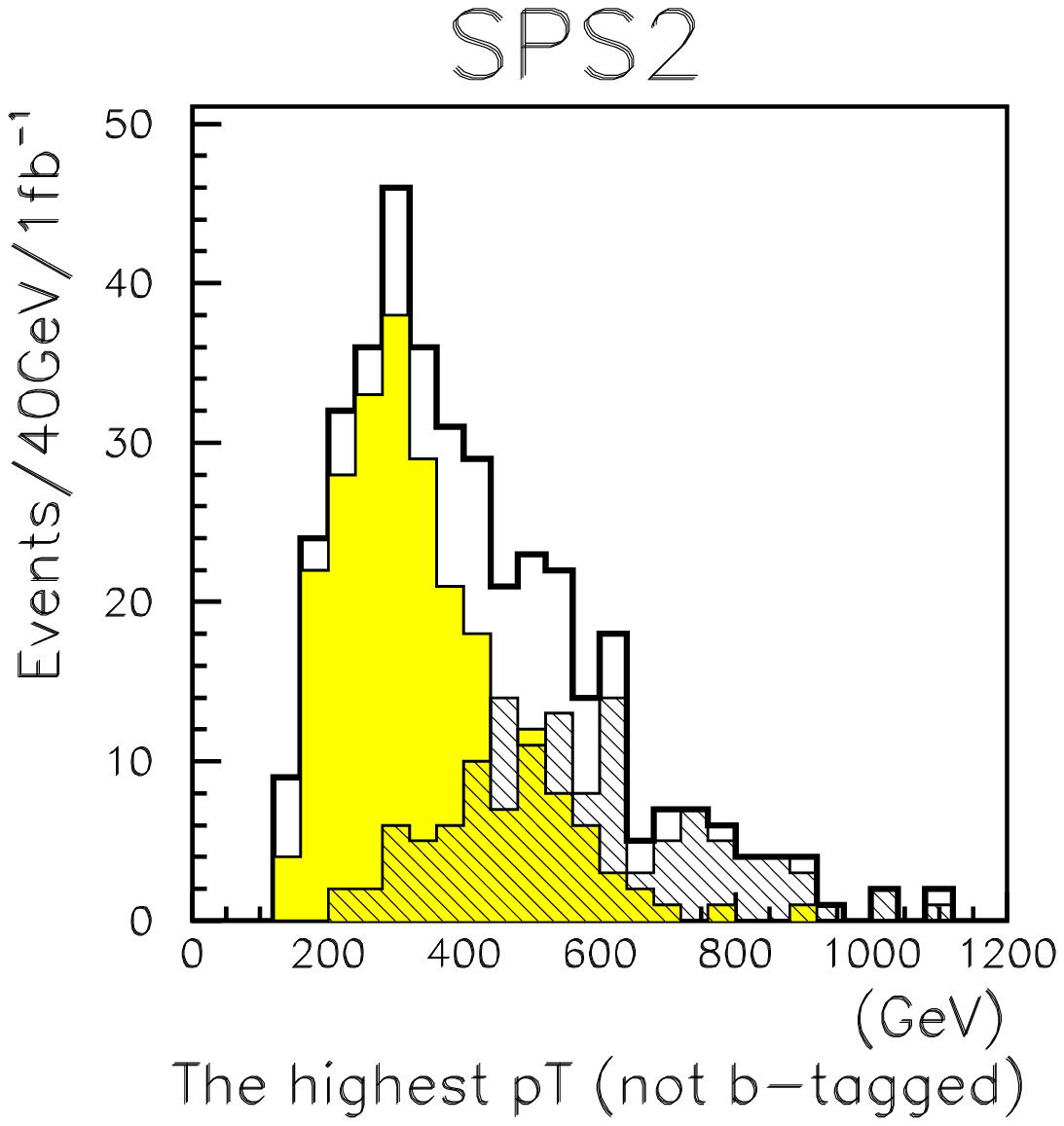}
\end{minipage}
\begin{minipage}{0.3\hsize}
\includegraphics[width=3.8cm,clip]{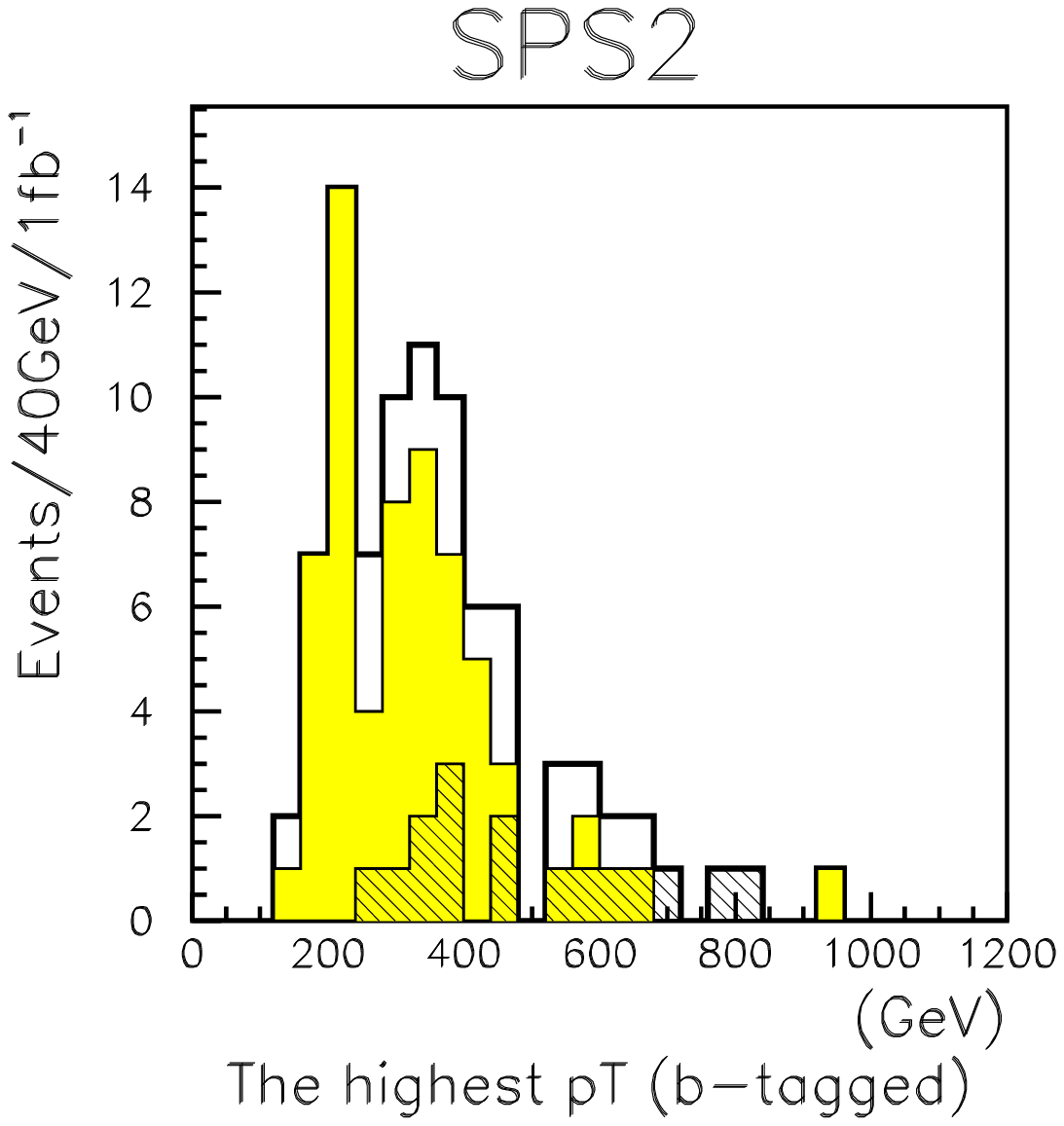}
\end{minipage}
\end{tabular}
\end{center}
\end{figure}

\newpage

\vspace{10mm}
{\large {\bf SPS3}} ``Coannihilation region''
\begin{eqnarray}
m_0=90\,{\rm GeV},~~m_{1/2}=400\,{\rm GeV},~~A_0=0\,{\rm GeV},~~\tan\beta=10,~~\sgn(\mu)=+
\nonumber
\end{eqnarray}

\vspace{-5mm}
\begin{figure}[h!]
\begin{center}
\begin{tabular}{ccc}
\begin{minipage}{0.30\hsize}
\includegraphics[width=3.8cm,clip]{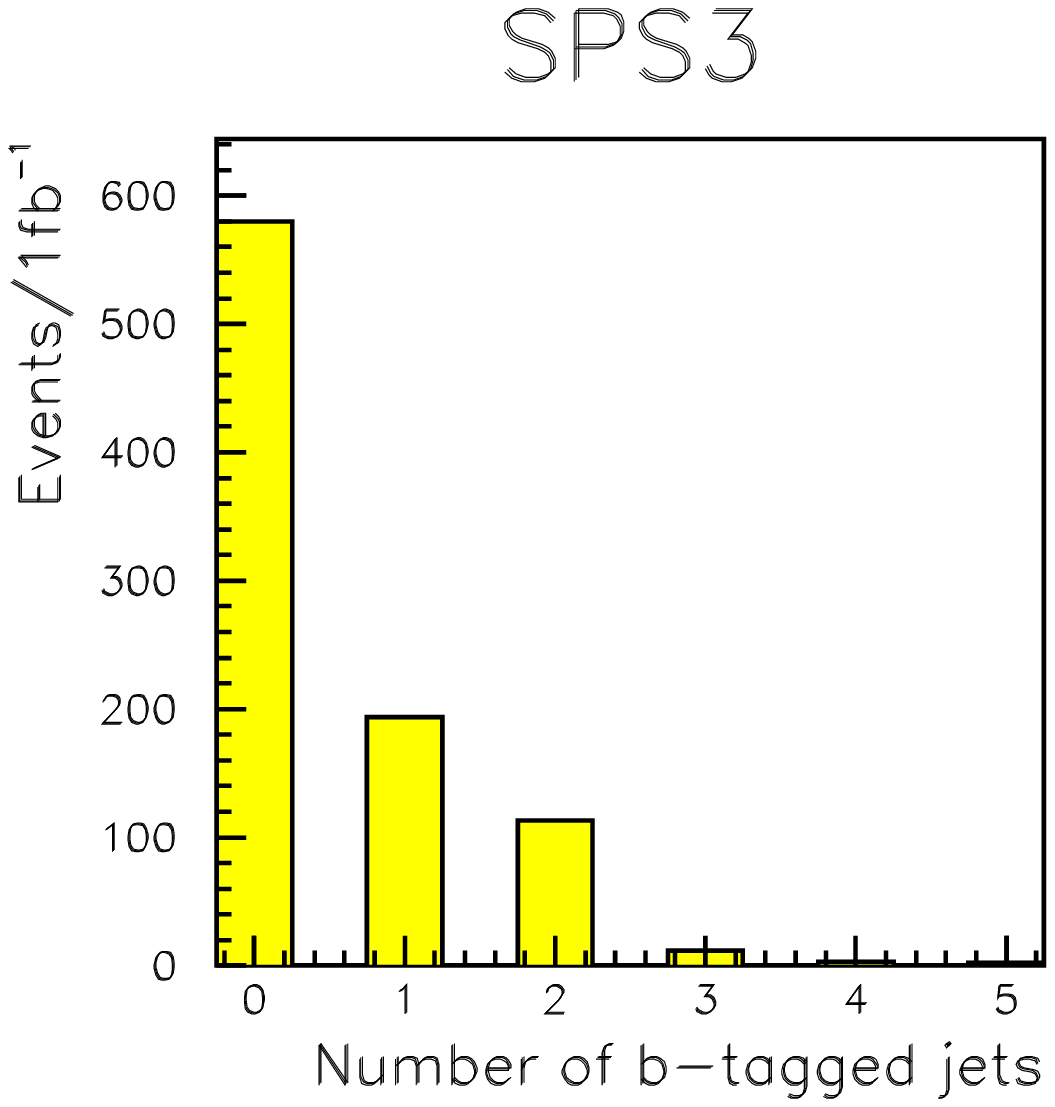}
\end{minipage}
\begin{minipage}{0.30\hsize}
\includegraphics[width=3.8cm,clip]{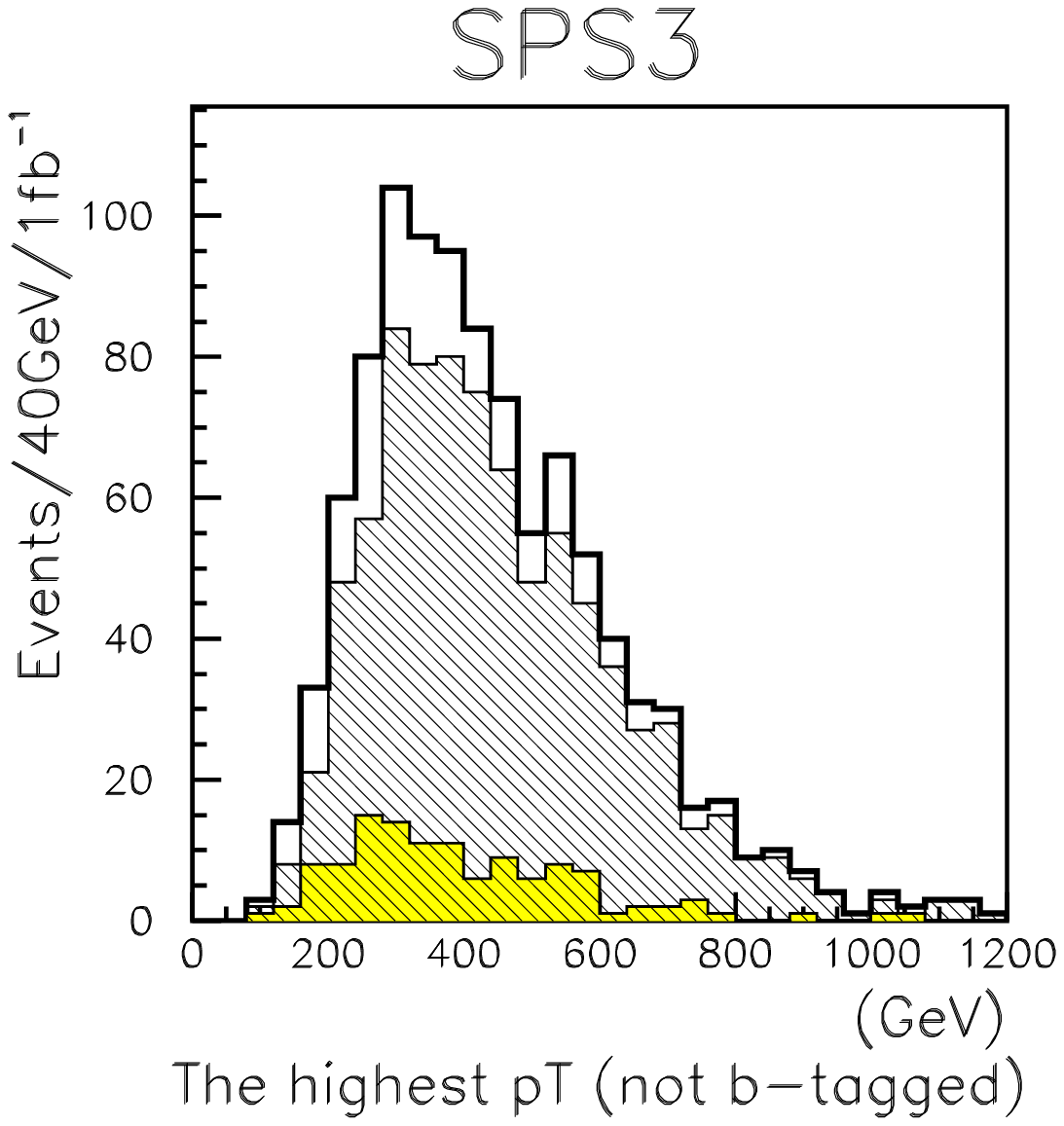}
\end{minipage}
\begin{minipage}{0.30\hsize}
\includegraphics[width=3.8cm,clip]{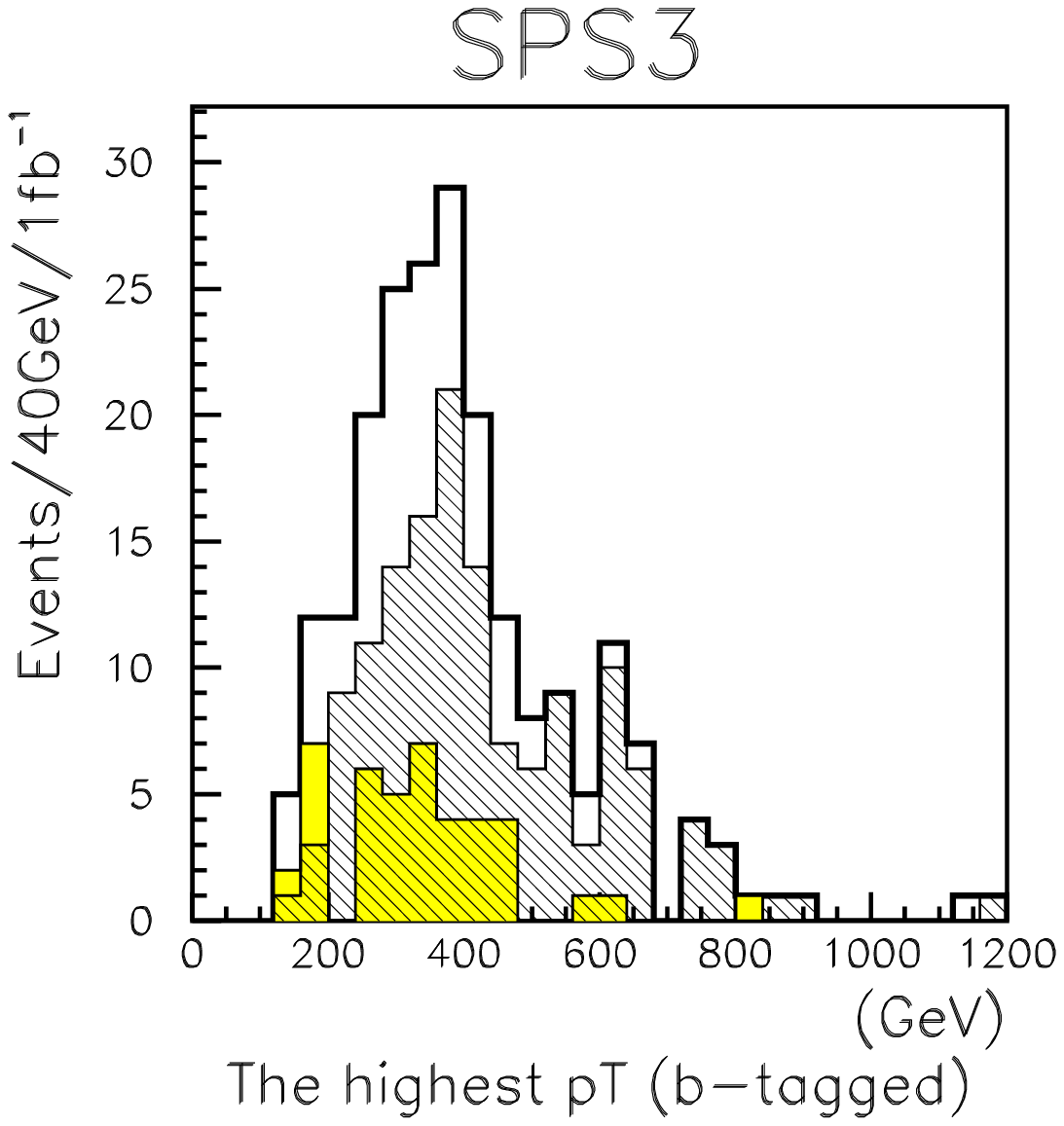}
\end{minipage}
\end{tabular}
\end{center}
\end{figure}


\vspace{10mm}
{\large {\bf SPS4}} ``Large $\tan\beta$''
\begin{eqnarray}
m_0=400\,{\rm GeV},~~m_{1/2}=300\,{\rm GeV},~~A_0=0\,{\rm GeV},~~\tan\beta=50,~~\sgn(\mu)=+
\nonumber
\end{eqnarray}

\vspace{-5mm}
\begin{figure}[h!]
\begin{center}
\begin{tabular}{ccc}
\begin{minipage}{0.3\hsize}
\includegraphics[width=3.8cm,clip]{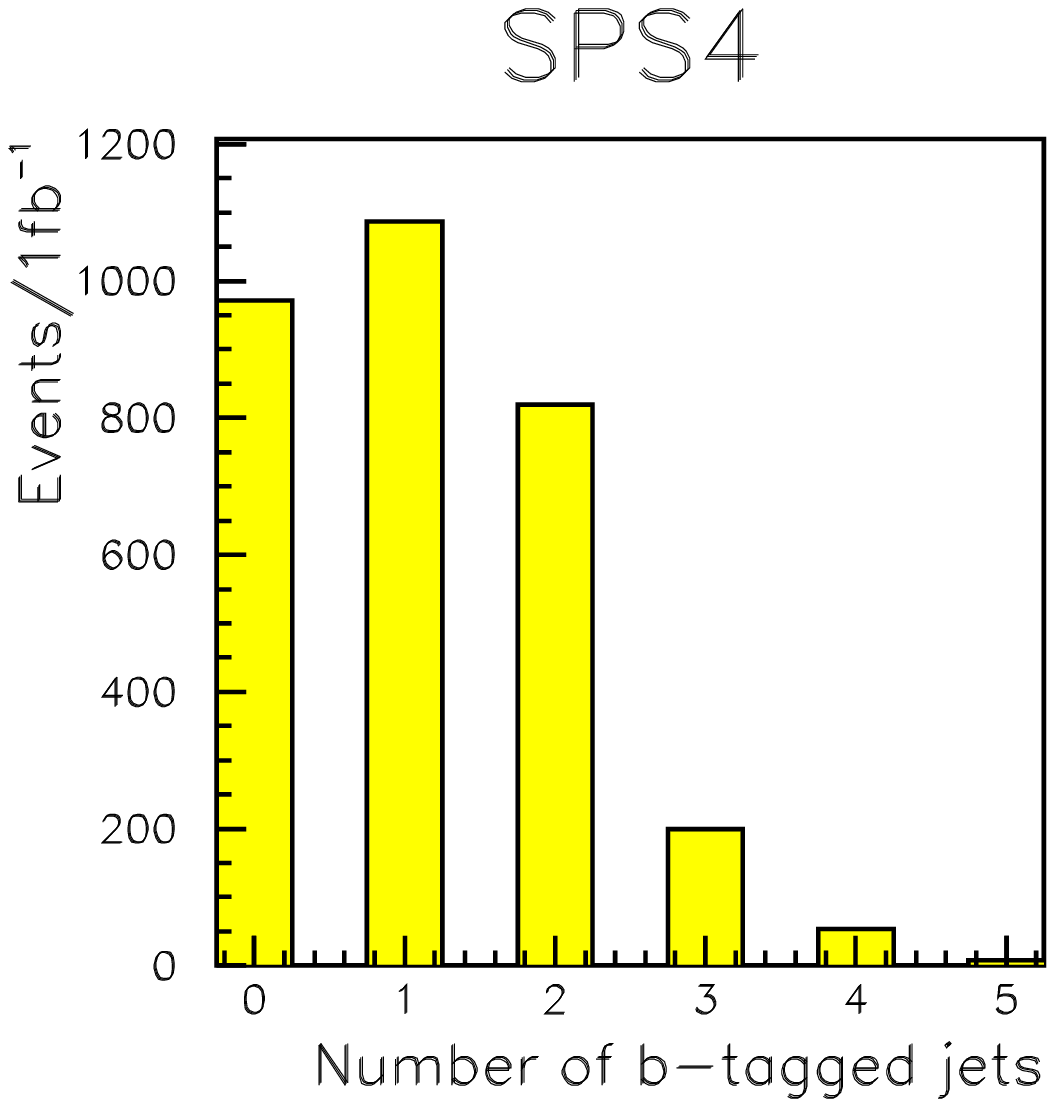}
\end{minipage}
\begin{minipage}{0.3\hsize}
\includegraphics[width=3.8cm,clip]{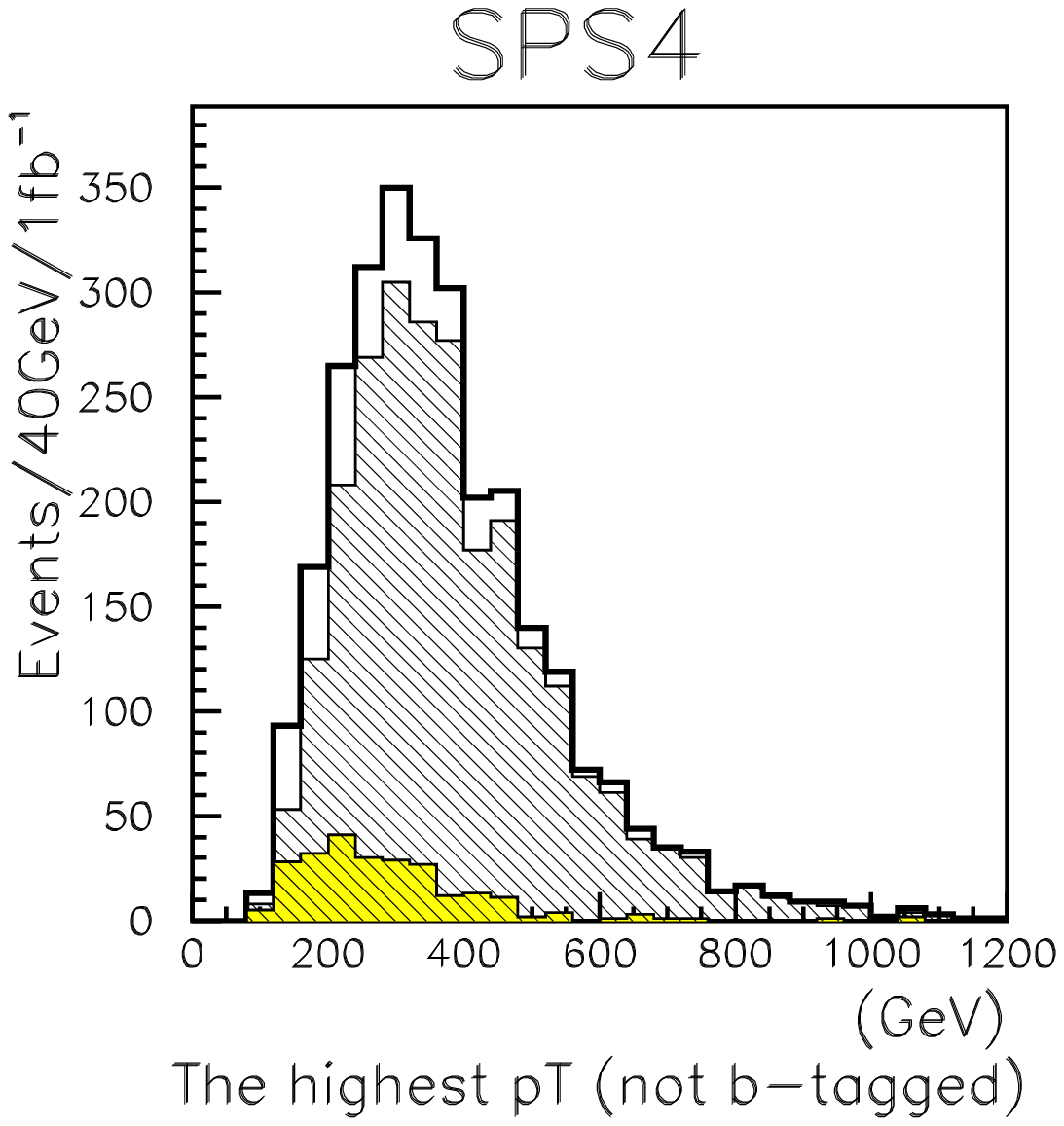}
\end{minipage}
\begin{minipage}{0.3\hsize}
\includegraphics[width=3.8cm,clip]{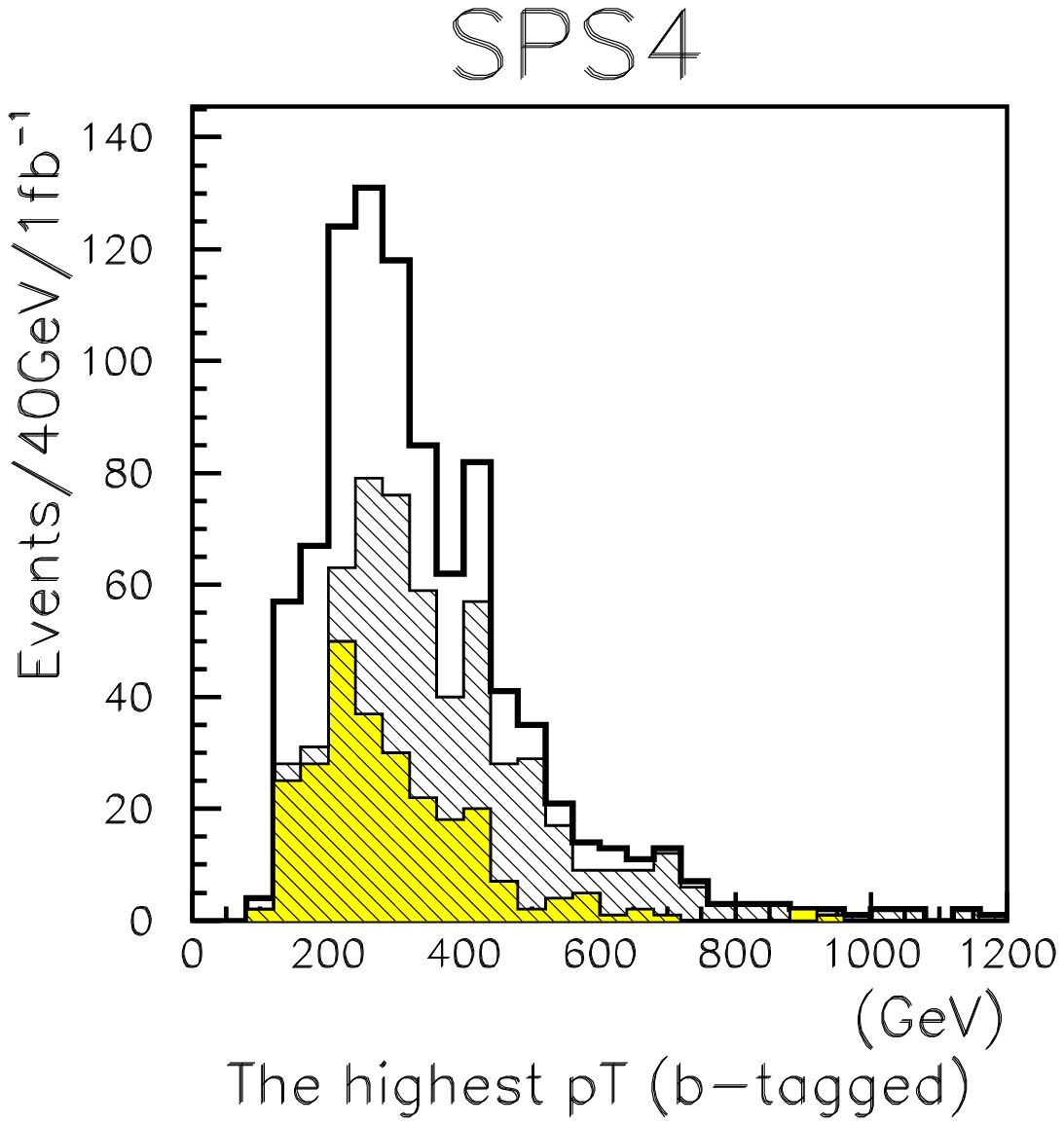}
\end{minipage}
\end{tabular}
\end{center}
\end{figure}


\vspace{10mm}
{\large {\bf SPS5}} ``Light stop''
\begin{eqnarray}
m_0=150\,{\rm GeV},~~m_{1/2}=300\,{\rm GeV},~~A_0=-1000\,{\rm GeV},~~\tan\beta=5,~~\sgn(\mu)=+
\nonumber
\end{eqnarray}

\vspace{-5mm}
\begin{figure}[h!]
\begin{center}
\begin{tabular}{ccc}
\begin{minipage}{0.3\hsize}
\includegraphics[width=3.8cm,clip]{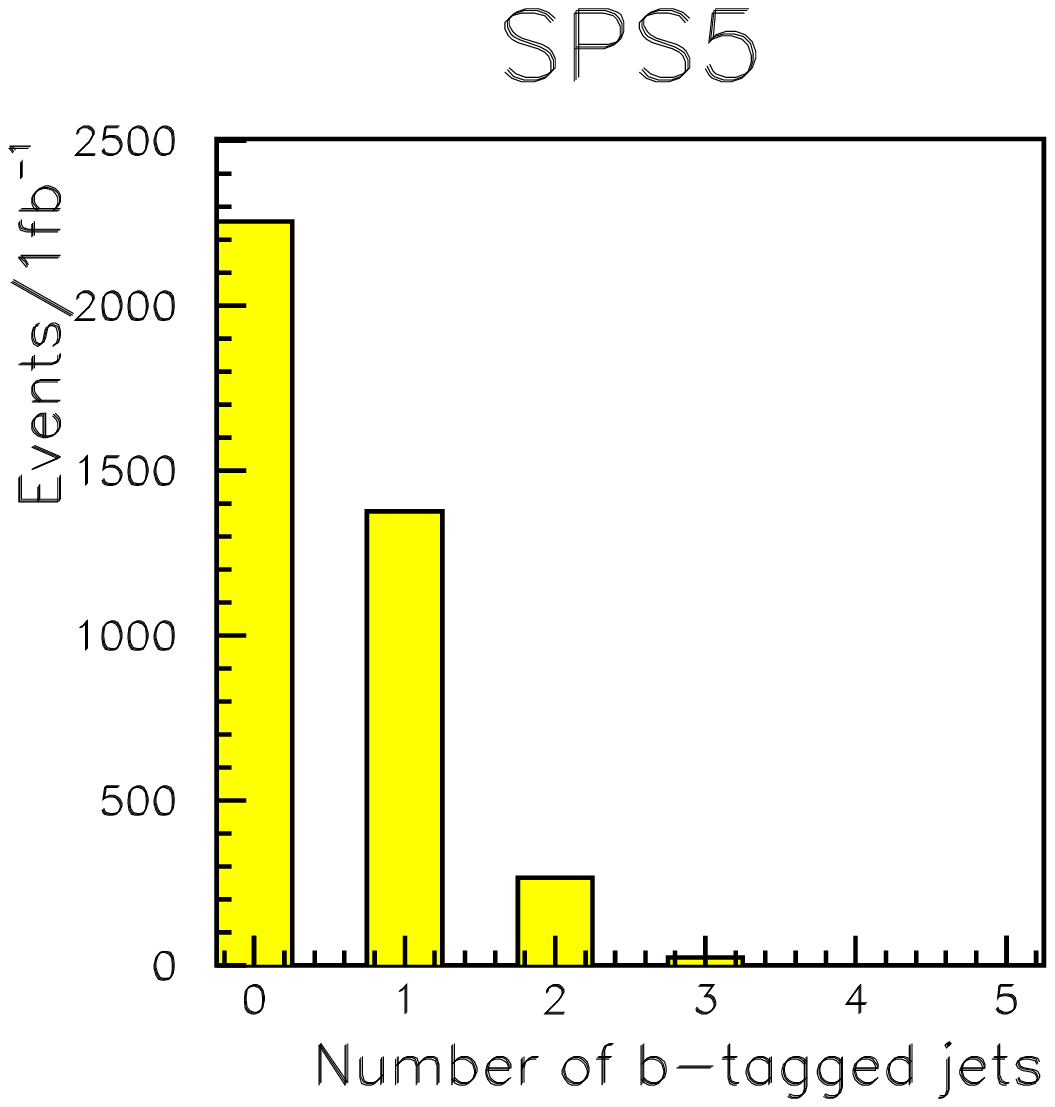}
\end{minipage}
\begin{minipage}{0.3\hsize}
\includegraphics[width=3.8cm,clip]{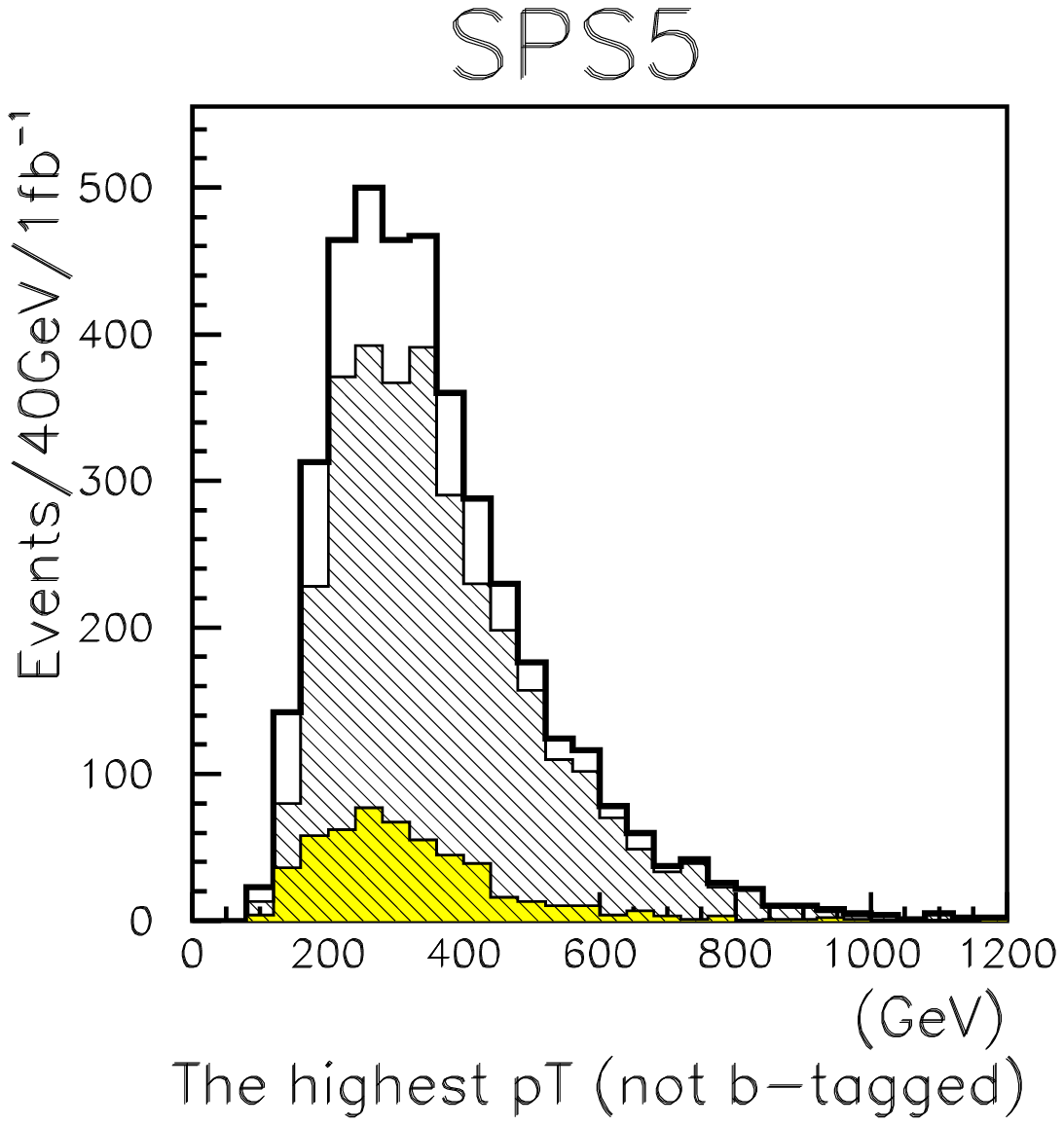}
\end{minipage}
\begin{minipage}{0.3\hsize}
\includegraphics[width=3.8cm,clip]{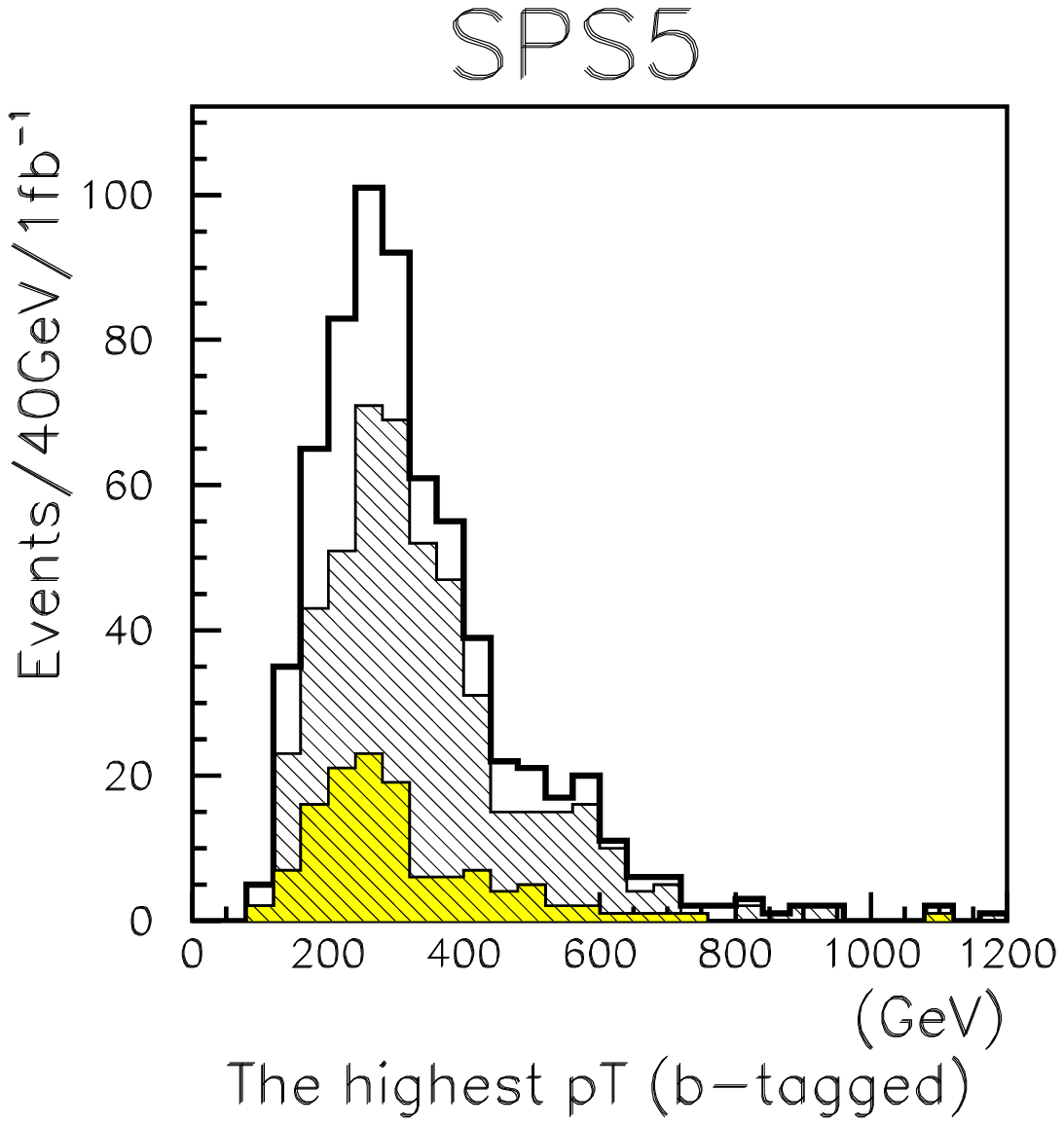}
\end{minipage}
\end{tabular}
\end{center}
\end{figure}

\newpage

\vspace{10mm}
{\large {\bf SPS6}} ``Non-universal gaugino''
\begin{eqnarray}
m_0=150\,{\rm GeV},~M_1=480\,{\rm GeV},~M_2=M_3=300\,{\rm GeV},~A_0=0\,{\rm GeV},
~\tan\beta=10,~\sgn(\mu)=+
\nonumber
\end{eqnarray}

\vspace{-5mm}
\begin{figure}[h!]
\begin{center}
\begin{tabular}{ccc}
\begin{minipage}{0.3\hsize}
\includegraphics[width=3.8cm,clip]{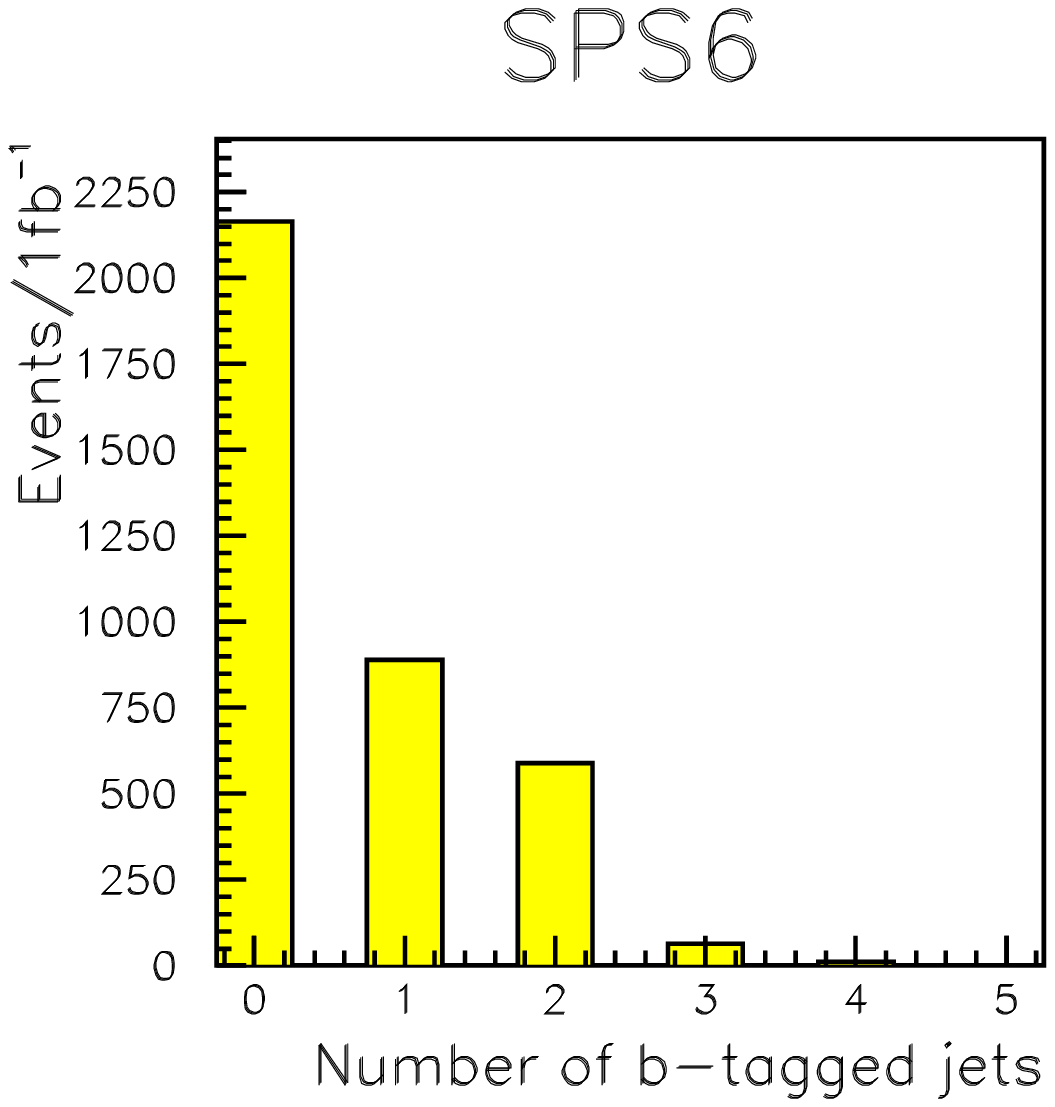}
\end{minipage}
\begin{minipage}{0.3\hsize}
\includegraphics[width=3.8cm,clip]{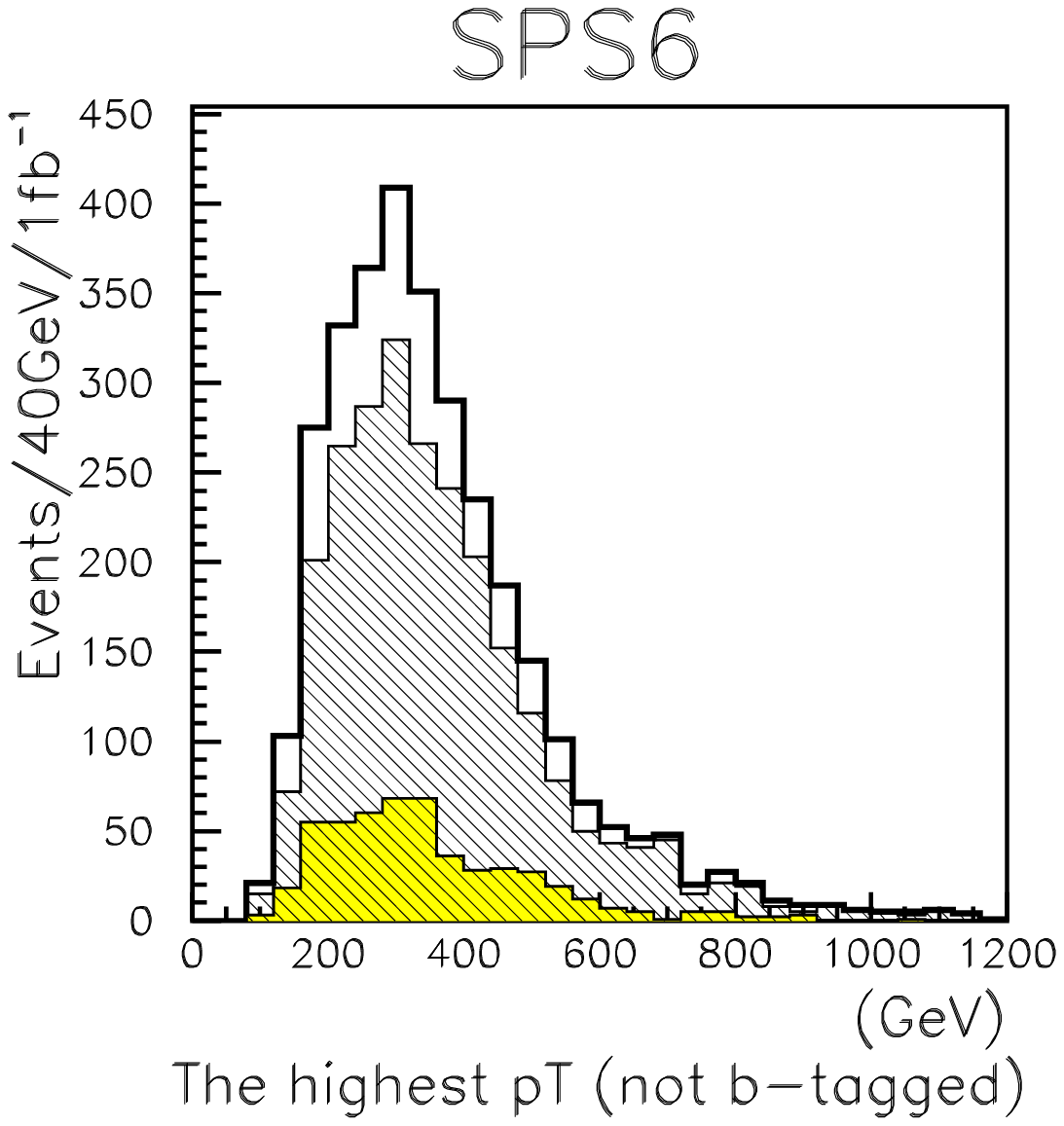}
\end{minipage}
\begin{minipage}{0.3\hsize}
\includegraphics[width=3.8cm,clip]{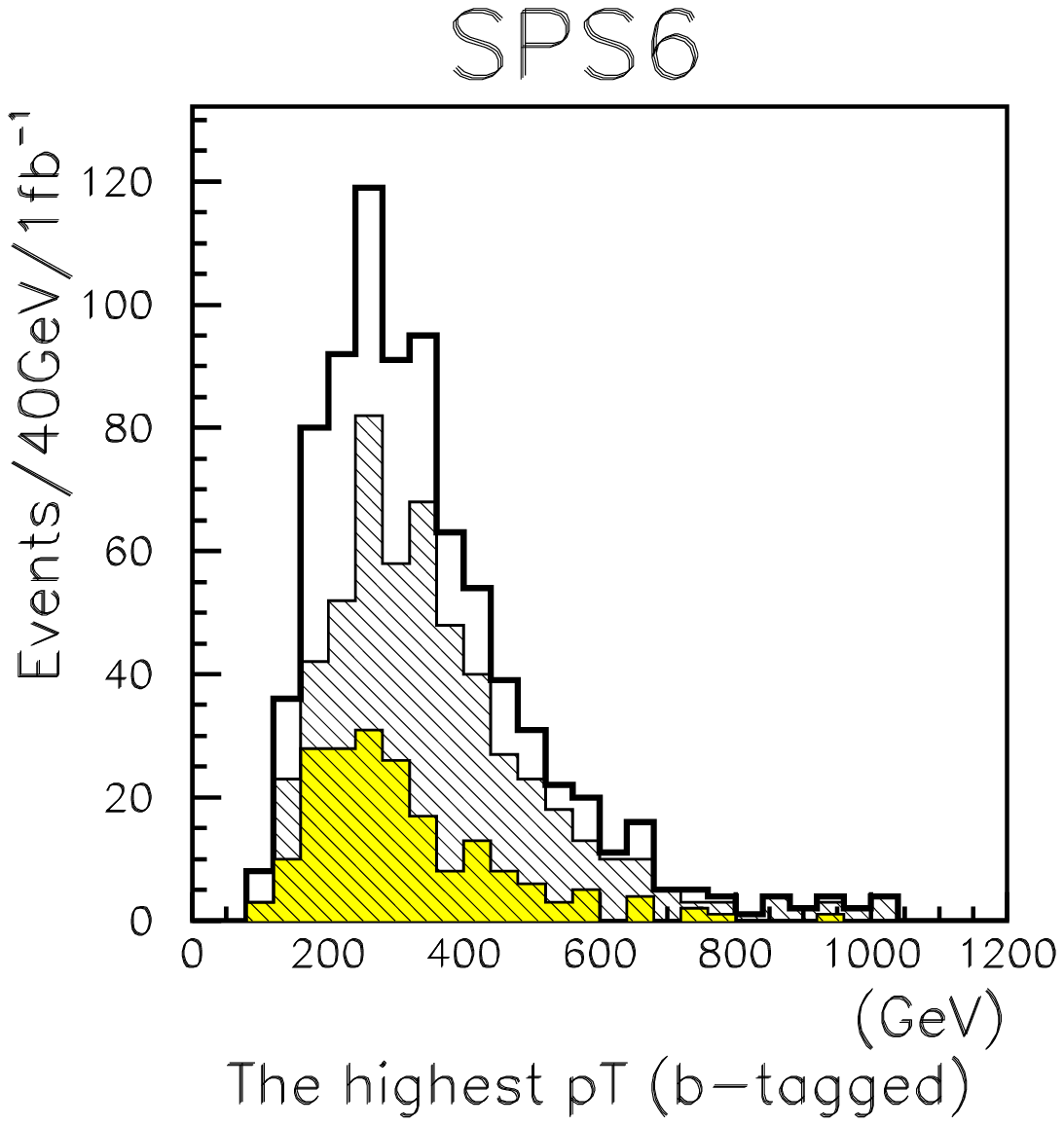}
\end{minipage}
\end{tabular}
\end{center}
\end{figure}


\vspace{10mm}
{\large {\bf SPS8}} ``GMSB scenario''
\begin{eqnarray}
\Lambda=40\,{\rm TeV},~~M_{mess}=80\,{\rm TeV},~~N_{mess}=1,~~\tan\beta=15,~~\sgn(\mu)=+
\nonumber
\end{eqnarray}

\vspace{-5mm}
\begin{figure}[h!]
\begin{center}
\begin{tabular}{ccc}
\begin{minipage}{0.3\hsize}
\includegraphics[width=3.8cm,clip]{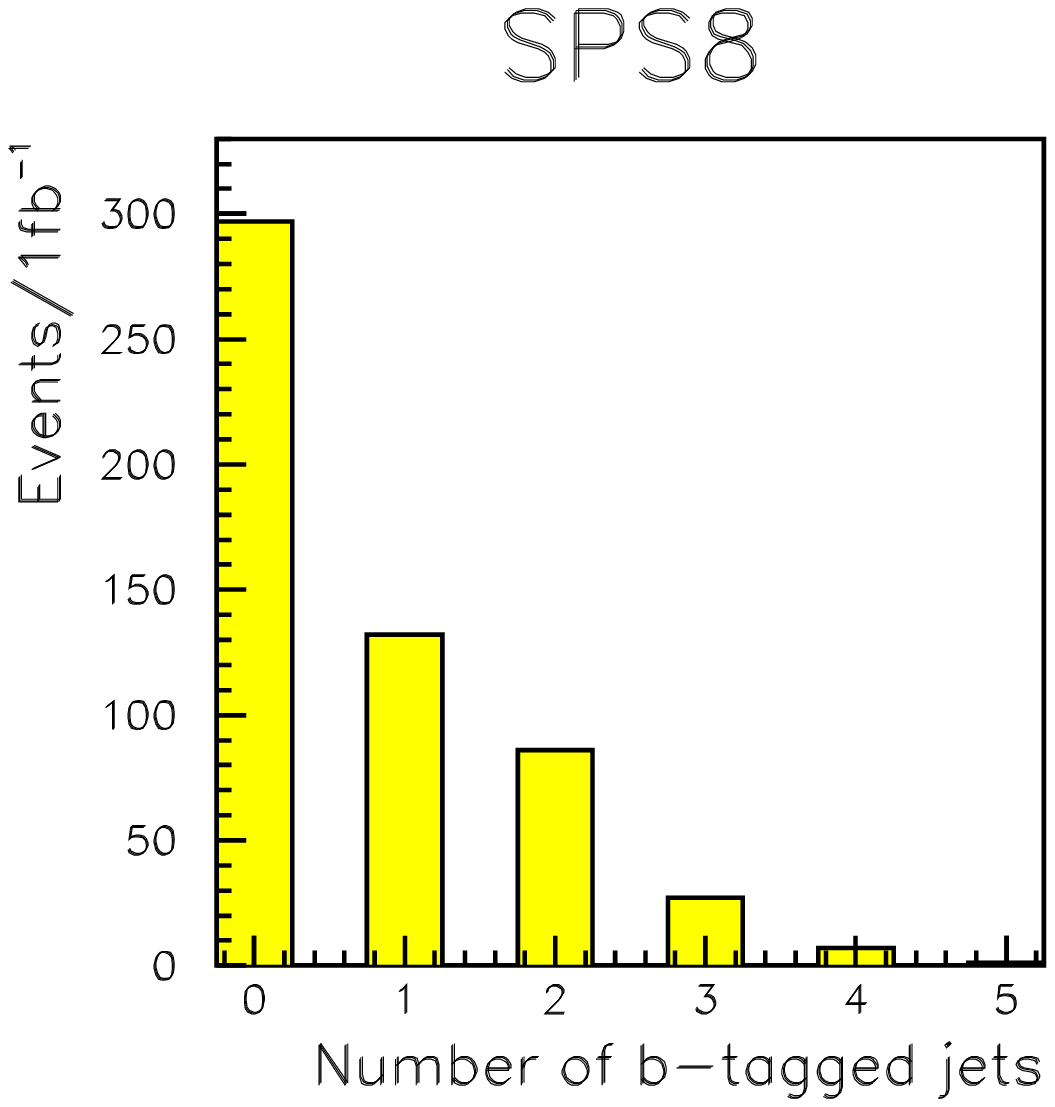}
\end{minipage}
\begin{minipage}{0.3\hsize}
\includegraphics[width=3.8cm,clip]{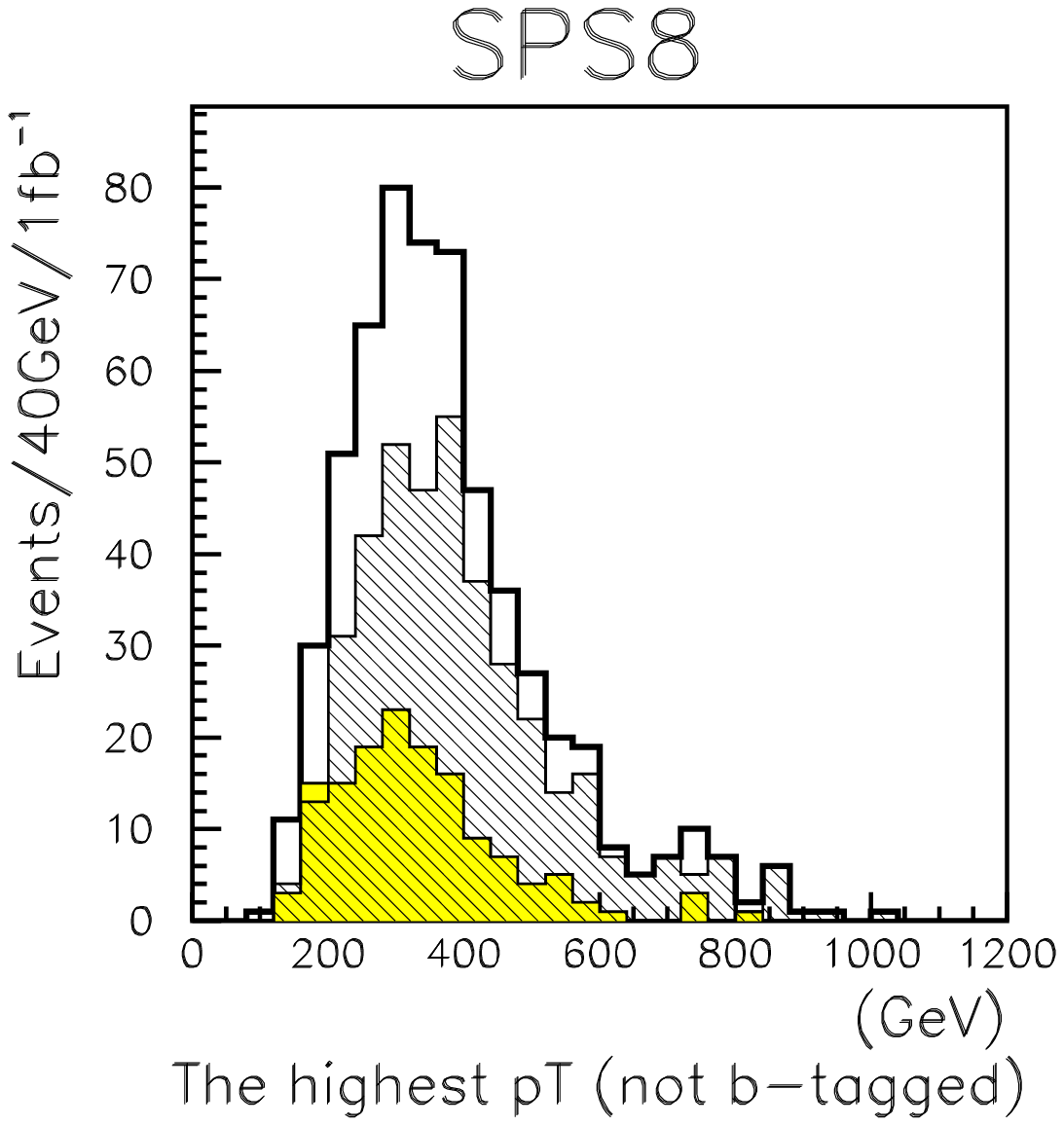}
\end{minipage}
\begin{minipage}{0.3\hsize}
\includegraphics[width=3.8cm,clip]{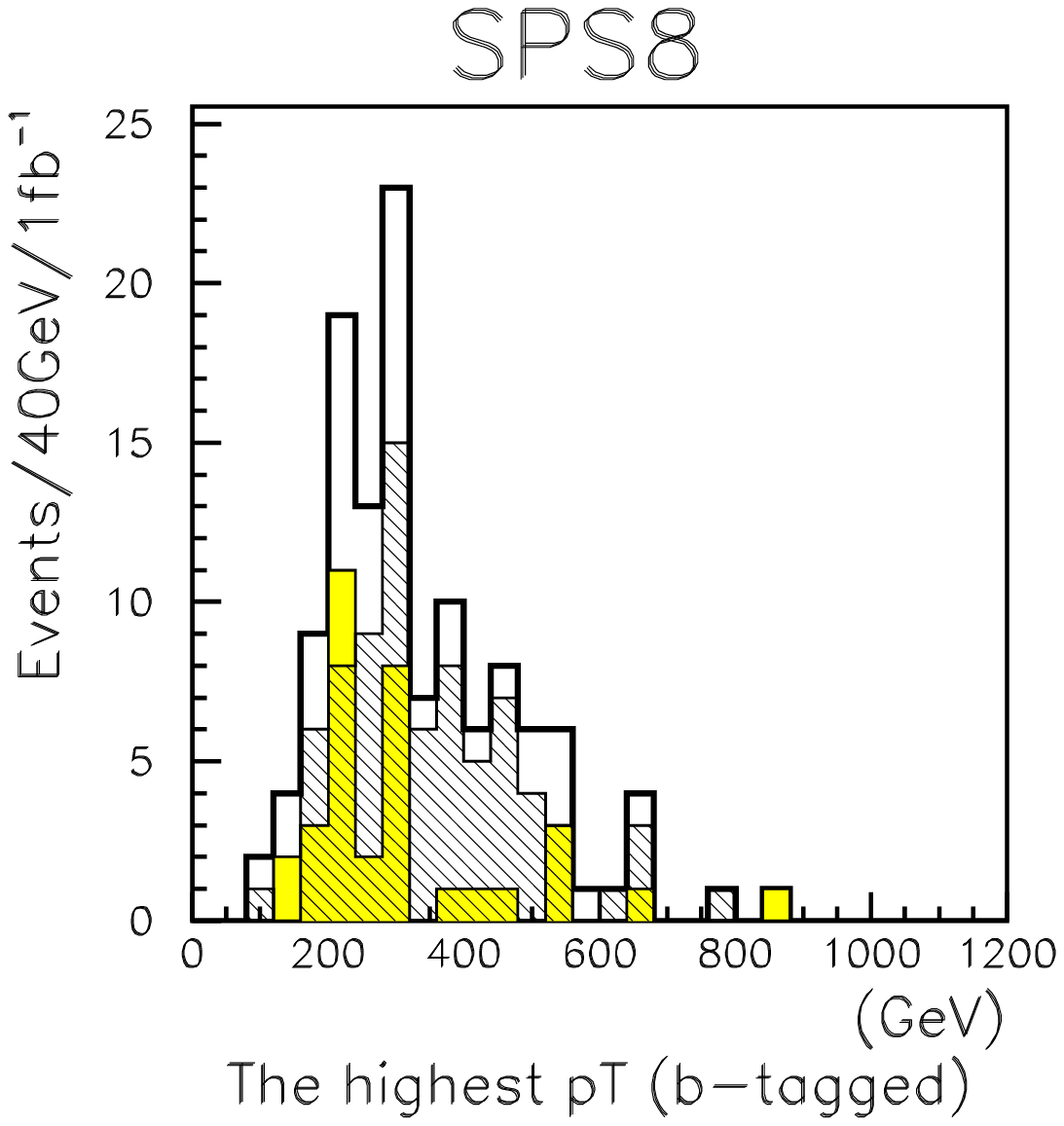}
\end{minipage}
\end{tabular}
\end{center}
\end{figure}


\vspace{10mm}
{\large {\bf SPS9}} ``AMSB scenario''
\begin{eqnarray}
m_0=400\,{\rm GeV},~~m_{3/2}=60\,{\rm TeV},~~\tan\beta=10,~~\sgn(\mu)=+
\nonumber
\end{eqnarray}

\vspace{-5mm}
\begin{figure}[h!]
\begin{center}
\begin{tabular}{ccc}
\begin{minipage}{0.3\hsize}
\includegraphics[width=3.8cm,clip]{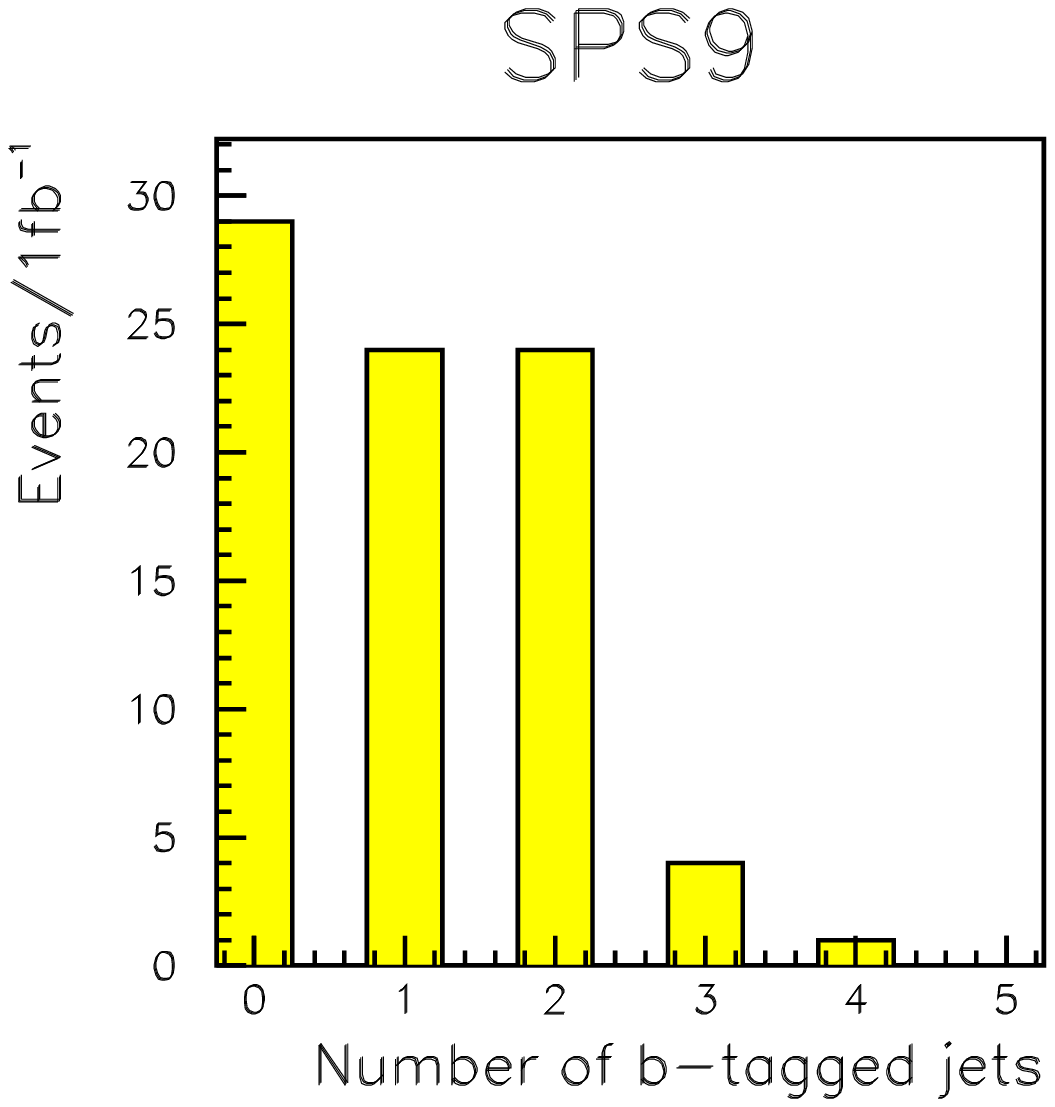}
\end{minipage}
\begin{minipage}{0.3\hsize}
\includegraphics[width=3.8cm,clip]{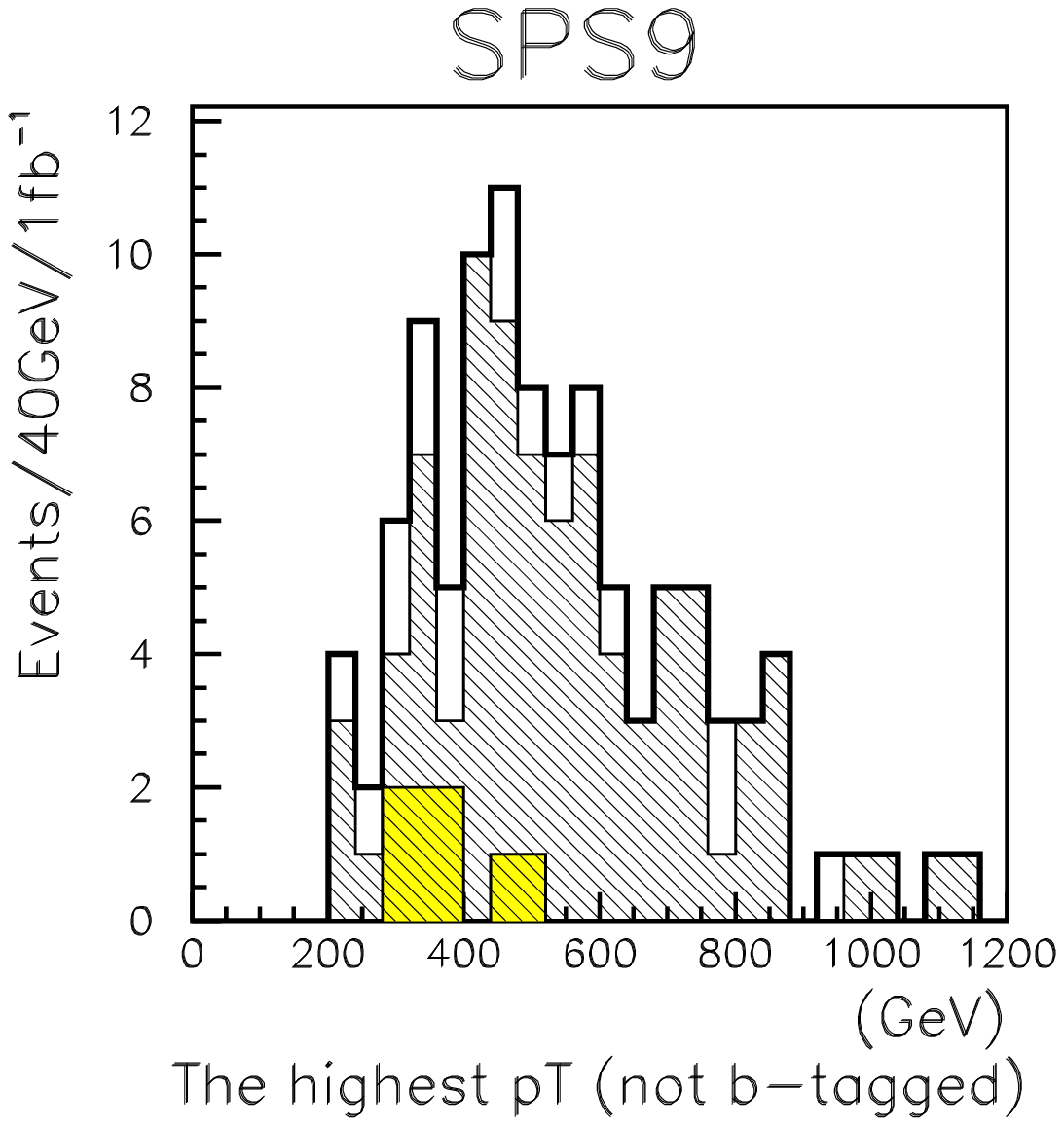}
\end{minipage}
\begin{minipage}{0.3\hsize}
\includegraphics[width=3.8cm,clip]{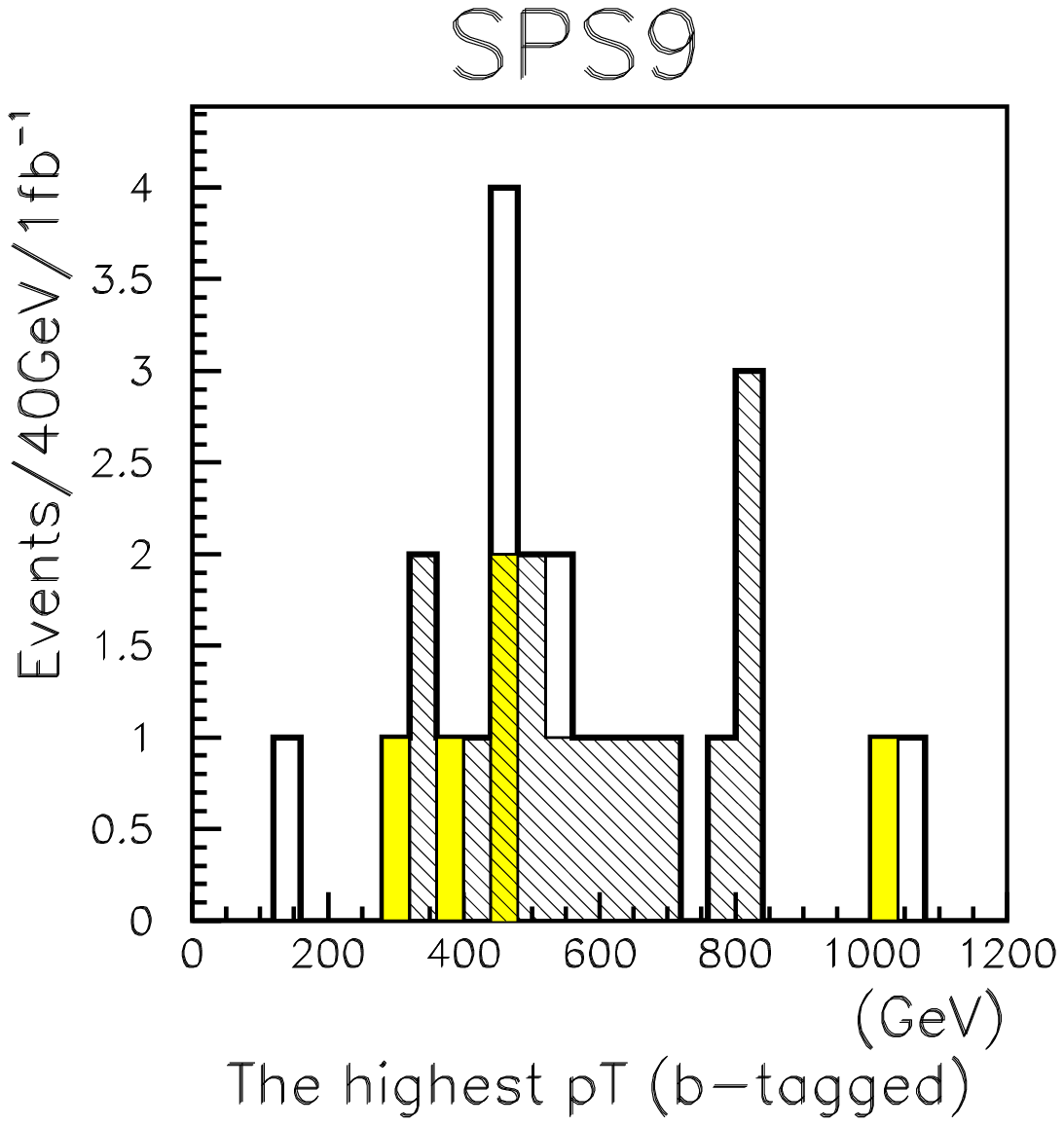}
\end{minipage}
\end{tabular}
\end{center}
\end{figure}

\newpage

\end{document}